%% file: FuturesSpot_Paper_Print.tex
\documentclass[opre]{informs3} 

\usepackage{endnotes}
\let\footnote=\endnote

\usepackage{natbib}
 \bibpunct[, ]{(}{)}{,}{a}{}{,}%
 \def\bibfont{\small}%
\TheoremsNumberedThrough     
\ECRepeatTheorems

\EquationsNumberedThrough    

\usepackage{multirow}

\input{Section-Def-OR}

\begin{document}


\RUNAUTHOR{Gao, Shou, Chen, and Huang}

\RUNTITLE{A Hybrid Market Approach to Dynamic Spectrum Access}

\TITLE{Combining Spot and Futures Markets: A Hybrid Market Approach to Dynamic Spectrum Access}

\ARTICLEAUTHORS{%
\AUTHOR{Lin~Gao}
\AFF{School of Electronic and Information Engineering, Harbin Institute of Technology (Shenzhen), China,
\\
and
Department of Management Sciences, City University of Hong Kong,
\EMAIL{gaolin@hitsz.edu.cn}}
\AUTHOR{Biying~Shou}
\AFF{Department of Management Sciences, City University of Hong Kong, \EMAIL{biying.shou@cityu.edu.hk}}
\AUTHOR{Ying-Ju~Chen}
\AFF{School of Business and Management and School of Engineering, The Hong Kong University of Science and Technology,
\EMAIL{imchen@ust.hk}}
\AUTHOR{Jianwei~Huang}
\AFF{Department of Information Engineering, The Chinese University of Hong Kong, \EMAIL{jwhuang@ie.cuhk.edu.hk}}
} 

\ABSTRACT{%
Dynamic spectrum access is a new paradigm of secondary  spectrum utilization and sharing.
It allows unlicensed secondary users (SUs) to exploit opportunistically the under-utilized licensed spectrum.
Market mechanism is a widely-used promising means to regulate the consuming behaviours of users and, hence, achieve the efficient allocation and consumption of limited resources.
In this paper, we propose and study a \textit{hybrid} secondary spectrum market consisting of both the \emph{futures market} and the \textit{spot market}, in which SUs (buyers) purchase under-utilized licensed spectrum from a  spectrum regulator, either through predefined contracts via the futures market, or through spot transactions via the spot market.
We focus on the optimal spectrum allocation among SUs in an exogenous hybrid market that maximizes the secondary spectrum utilization efficiency.
The problem is challenging due to the stochasticity and asymmetry of network information.
To solve this problem, we first derive an off-line optimal allocation policy that maximizes the ex-ante expected spectrum utilization efficiency based on the stochastic distribution of network information.
We then propose an on-line Vickrey¨CClarke¨CGroves (VCG) auction that determines the real-time allocation and pricing of every spectrum based on the realized network information and the pre-derived off-line policy.
We further show that with the spatial frequency reuse, the proposed VCG auction is NP-hard; hence, it is not suitable for on-line implementation, especially in a large-scale market.
To this end, we propose a heuristics approach based on an on-line VCG-like mechanism with polynomial-time complexity, and further characterize the corresponding performance loss bound analytically.
We finally provide extensive numerical results to evaluate the performance of the proposed solutions.
}%


\SUBJECTCLASS{Games/Group Decisions; Communications}
\AREAOFREVIEW{Games, Information, and Networks}
\HISTORY{Received October 2012; revisions received June 2014, May 2015; accepted March 2016}

\maketitle

%

\input{Section0-Intro-OR}

\input{Section8-Relat-OR}
\input{Section1-Model-OR}
\input{Section2-ProbFormu-OR}

\input{Section3-Symmetry-OR}

\input{Section4-VCG-OR}
\input{Section5-Loss-OR}
\input{Section6-Simu-OR}

\input{Section9-Conclusion-OR}

\ACKNOWLEDGMENT{The authors would like to thank Prof.~Asuman Ozdaglar and the anonymous review team for their valuable comments and suggestions.
This work is supported by the HK General Research Fund (Project Numbers: CityU 9041836, CUHK 412713, HKUST 616613 and 16206814)
established under the University Grant Committee of the Hong Kong Special Administrative Region, China. This work is also supported by HK RGC Theme-based Research Scheme No. T32-101/15-R.}

\input{Section-Ref-OR}

\input{Section-Appendix}


\end{document}

%% file: Section-Def-OR.tex
\newtheorem{mechanism}{Mechanism}

\newtheorem{prob}{Problem}

\ifodd 12
\newcommand{\reee}[1]{{\color{blue}#1}}
\newcommand{\cooo}[1]{\textbf{\color{red} (COMMENT: #1) }} \newcommand{\reve}[1]{{\color{blue}#1}}
\newcommand{\revf}[1]{{\color{red}#1}}
\newcommand{\yyy}[1]{{\color{blue}#1}}
\newcommand{\rrr}[1]{#1} 
\newcommand{\rrd}[1]{#1}
\newcommand{\rev}[1]{#1}
\newcommand{\revv}[1]{#1}
\newcommand{\revh}[1]{#1}    
\newcommand{\com}[1]{\textbf{\color{red} (COMMENT: #1) }} 
\newcommand{\comg}[1]{\textbf{\color{green} (COMMENT: #1)}}
\newcommand{\response}[1]{\textbf{\color{green} (RESPONSE: #1)}} 
\else
\newcommand{\reee}[1]{#1}
\newcommand{\cooo}[1]{}
\newcommand{\reve}[1]{#1}
\newcommand{\revf}[1]{#1}
\newcommand{\yyy}[1]{#1}
\newcommand{\rrr}[1]{#1}
\newcommand{\rrd}[1]{#1}
\newcommand{\revv}[1]{#1}
\newcommand{\rev}[1]{#1}
\newcommand{\revh}[1]{#1}
\newcommand{\com}[1]{}
\newcommand{\comg}[1]{}
\newcommand{\response}[1]{}
\fi
\newcommand{\revth}[1]{{#1}}

\def\ch{\textsc{ch}}    
\def\CH{\mathcal{Q}}    
\def\XI{\mathcal{X}}    
\def\THETA{\boldsymbol{\Theta}}    
\def\thxi{\theta, \xi}
\def\thxikt{\theta_{kt}, \xi_{kt}}
\def\M{\mathcal{M}}     
\def\N{\mathcal{N}}     
\def\CO{\mathcal{B}}    
\def\Cn{\textsc{Ctr}}     
\def\G{\textsc{G}}
\def\Gs{\textsc{G}^{\textsc{s}}}
\def\Gc{\textsc{G}^{\textsc{c}}}
\def\V{\mathcal{V}}
\def\Vs{\mathcal{V}^{\textsc{s}}}
\def\Vc{\mathcal{V}^{\textsc{c}}}
\def\E{\mathcal{E}}
\def\Es{\mathcal{E}^{\textsc{s}}}
\def\Ec{\mathcal{E}^{\textsc{c}}}
\def\CQ{\mathcal{C}}
\def\IS{\mathcal{I}}
\def\ISs{\IS^{\textsc{s}}}
\def\ISc{\IS^{\textsc{c}}}
\def\Ic{I_{\textsc{c}}}
\def\Adet{\boldsymbol{A}_{\textsc{det}}}

\def\Ais{\boldsymbol{A}_{\textsc{IS}}}
\def\Aisi{\boldsymbol{A}_{\textsc{IS,1}}}

\def\Aic{\boldsymbol{{A}}_{\textsc{ISc}}}
\def\Aicx{\boldsymbol{{A}}_{\textsc{ISc'}}}
\def\veca{\boldsymbol{a}}
\def\vecais{\boldsymbol{a}_{\textsc{IS}}}
\def\vecaic{\boldsymbol{a}_{\textsc{ISc}}}
\def\vecaicx{\boldsymbol{a}_{\textsc{ISc'}}}
\def\as{a^{\textsc{s}}}
\def\ac{a^{\textsc{c}}}
\def\ais{a^{\textsc{IS}}}
\def\aic{a^{\textsc{ISc}}}
\def\MW{\boldsymbol{\dot{w}}}	
\def\IMW{\boldsymbol{\dot{w}}^{\textsc{In}}}	
\def\CMW{\boldsymbol{\dot{w}}^{\textsc{Co}}}	
\def\CMWx{\boldsymbol{\widetilde{w}}^{\textsc{Co}}}	
\def\CMWmax{{\dot{W}_{\mathrm{1st}}}}
\def\CMWsec{{\dot{W}_{\mathrm{2nd}}}}

\def\ws{w^{\textsc{s}}}
\def\wc{w^{\textsc{c}}}
\def\wcx{\widetilde{w}^{\textsc{c}}}

\def\eq{\triangleq}

%% file: Section0-Intro-OR.tex

\section{Introduction}\label{sec:intro}

\subsection{Background and Motivations}

Radio electromagnetic spectrum is a limited resource for wireless   communications, and is becoming more congested and scarce with the explosive development of wireless services and networks. Nonetheless, extensive studies (e.g., \cite{FCC2, McHenry}) have shown that many spectrum bands are heavily \textit{under-utilized} due to the significant variations in spectrum occupying time and space. This suggests that the traditional static licensing approach, i.e., granting exclusive spectrum usage right to a single commercial or government department over a long time and in a large area, is no longer  efficient  or suitable for future wireless networks. 
\rrd{Dynamic spectrum access (DSA)  has been recently recognized as a novel and promising approach to increase spectrum utilization efficiency and alleviate spectrum scarcity.}
The key idea is to enable unlicensed secondary users (SUs) to exploit the under-utilized licensed spectrum  in an opportunistic manner
(see \cite{Haykin-survey,Akyildiz-survey,buddhikot-sur,Zhao-sur}).\footnote{This idea has been embraced by many government spectrum regulatory bodies (e.g., FCC in the U.S. and Ofcom in the UK), standardization organizations (e.g., IEEE and ECC), and major industry players (e.g., Google, Microsoft, Motorola, and Huawei).
The most recent progress along this frontier is the finalization of the IEEE 802.22 Wireless Regional Area Network standard (see \cite{IEEE80222}) \rrr{on the opening and secondary utilization of under-utilized spectrum in the UHF band (between around 500MHz and 800MHz) licensed for broadcast TV.}}
The long-term success of DSA requires many innovations in technology, economics, and policy.
In particular, it is essential to design a secondary spectrum sharing mechanism that achieves a high  secondary spectrum utilization efficiency, and meanwhile offers necessary incentives for licensees to open their licensed spectrum for secondary sharing and unlicensed access.\footnote{For example, when FCC in the U.S. decided to open up the TV band for secondary unlicensed access, the National Association of the Broadcasters and Association for Maximum Service Television sued the FCC as they do not have incentives for such spectrum sharing (see \cite{NABSuit}).}

{Market mechanism is a widely-used promising means to regulate and coordinate the consuming behaviours of users and, hence, achieve the efficient allocation and consumption of limited resource.}
\revf{Therefore, market-driven secondary spectrum trading is a natural choice to achieve the efficient spectrum utilization, and also a promising approach to address the incentive issue (see \cite{spectrum-trading2,Weiss}).}
\reee{With secondary spectrum trading, SUs temporarily \textit{purchase} the under-utilized licensed spectrum from a spectrum regulator (SR), such as an agent representing the Federal Communications Commission (FCC) in the United States or Ofcom in the United Kingdom.}\footnote{\reee{A real-world secondary spectrum sharing example (that involves the interactions of SR and unlicensed SUs) is the TV
white space network (see \cite{FCC3}).
In a TV white space network,  SUs request and exploit the under-utilized UHF/VHF radio spectrum licensed for broadcast television service (called TV white space) from a SR which is designated by the FCC in the United States to operate a white space database.}}
{The secondary spectrum market differs from conventional markets} (e.g., paintings, bonds, and electricity markets) in the following two aspects.
First, \textit{radio spectrum is spatially reusable}. That is, a spectrum can be potentially used by multiple SUs simultaneously, as long as these SUs do not generate mutual interferences (see Section \ref{sec:model:reuse} for details).
Second, \textit{the availability of spectrum is not deterministic.} To protect the subscribed primary users (PUs) benefits, a spectrum is only available to SUs when it is not occupied by any PU (i.e., when it is \textit{idle}).
Due to the stochastic nature of PUs' activities, the spectrum availability is often a random variable (see Section \ref{sec:model:1} for details).
Based on the above two facts, the mechanism design in such a secondary spectrum market needs to overcome a different set of challenges than in a conventional market.

\begin{figure}[t]
   \centering
    \includegraphics[scale=.55]{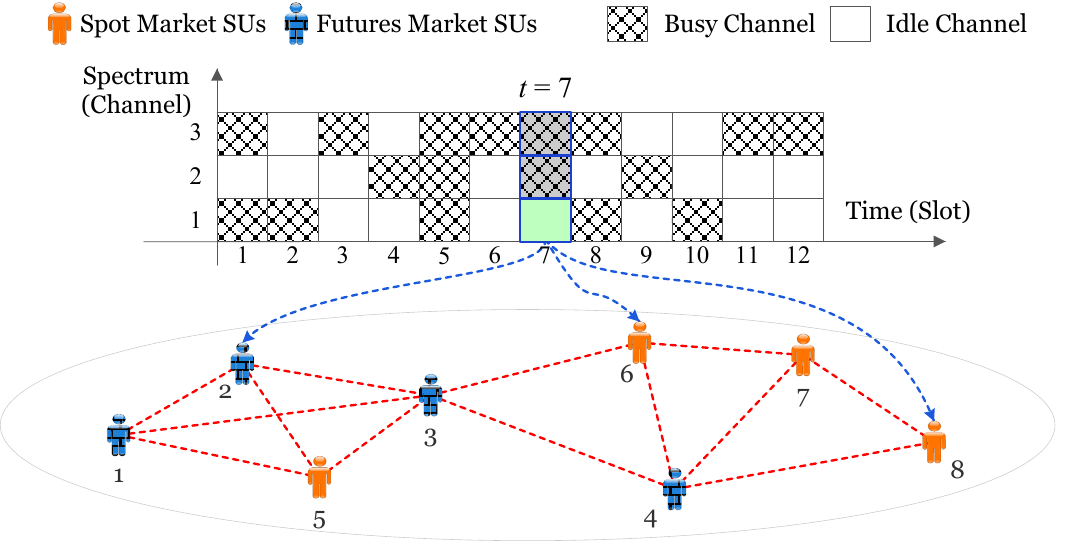}
    \caption{Illustration of a time-slotted secondary spectrum market.
The SUs not connected by an edge are interference-free, and can use the same (idle) spectrum at the same time without mutual interference.
For example, the idle (green) spectrum at time $t=7$ is allocated to SUs \{2, 6, 8\} simultaneously.}
\label{system-illu}
\vspace{-3mm}
\end{figure}

\subsection{Market Model and Problem Formulation}

\reee{In this paper, we consider the secondary spectrum trading between one SR (seller) and multiple SUs (buyers) for all idle spectrum in a given time period which consists of multiple time slots.\footnote{\revth{Detailed discussions regarding time period and time slot can be referred to Section \ref{sec:model:1}.}} Moreover, we consider the \textit{short-term} spectrum trading, i.e., the idle spectrum is traded on a slot-by-slot basis.}
Figure \ref{system-illu} illustrates such a time-slotted secondary spectrum market within a period of $12$ time slots.
The blank and shadowed squares denote the idle spectrum and busy spectrum in each time slot, respectively.
The edge connecting two SUs denotes the interference relationship between them.
In other words, any two SUs without a connecting edge (e.g., SUs \{4, 5\} or SUs \{6, 8\}) are interference-free (also called \textit{independent}), and thus can use the same (idle) spectrum concurrently without any mutual interference.
In this example, the idle spectrum at time $t=7$  is allocated to SUs \{2, 6, 8\} simultaneously, as these SUs are independent of each other. 

\reee{More specifically, we consider a \textit{hybrid} secondary spectrum market consisting of both the \textit{futures market} and the \textit{spot market}. In such a market, SUs can purchase spectrum either through predefined futures contracts via the futures market, or through spot transactions via the spot market.}
In the futures market, each SU reaches an agreement, called a futures contract or {guaranteed contract}, directly with the SR.
The contract specifies not only the SU's total demand and payment in the given period, but also the SR's penalty if violating the contract.\footnote{In this sense, the futures market in this paper can be seen as a generalized version of those in stock markets (see \cite{future-survey}). In futures stock markets, the seller is obligated to deliver the commodities at an explicit future date; whereas, in our market model, the seller is allowed to deliver the commodities at any time before the due date, and even to violate the contract with certain penalty.}
In the spot market, SUs buy spectrum in a real-time and on-demand manner through spot transactions, and multiple SUs may compete with each other for limited spectrum resource (e.g., through an auction).
In the example of Fig.~\ref{system-illu}, we denote the contract users (blue) by 1, 2, 3, and 4, each requesting a pre-specific number of spectrum.
Likewise, we denote the spot market users (red) by 5, 6, 7, and 8, each requesting and competing for spectrum in an on-demand manner.
It is easy to see that the futures market insures buyers (sellers) against uncertainties of future supply (demand) through predefined contracts, while the spot market allows buyers to compete for limited resource based on their real-time demands and preferences.\footnote{\rrr{Guaranteed contract has been widely adopted by wireless service providers for different kinds of services, e.g., monthly subscription of data traffic (e.g., \$25 for 2G wireless iPad data) or voice traffic (e.g., \$35 for 180 minutes of voice communication in Hong Kong). On the other hand, with the explosive development of technological innovations, especially the emergence of DSA and cognitive radio, spot transaction, e.g., auction (see \cite{JianweiHuang,SorabhGandhi,lixy,eBay,TRUST}, and \cite{WangXuXu2010}), has grown in prominence in secondary spectrum markets.}}
Therefore, such a hybrid market has both the reliability of the futures market and the flexibility of the spot market; hence, it is highly desirable for Quality of Service (QoS) differentiations in secondary spectrum utilization.
Specifically, an SU with elastic traffic (e.g., file transferring or FTP downloading) may be more interested in spot transactions to achieve a flexible resource-price tradeoff; whereas, an SU with inelastic traffic requiring specific data rates (e.g., Netflix video streaming or
VoIP) may prefer the certainty of contracts in the futures market. 

\revth{We consider a quasi-static market scenario, where  both the network topology and the hybrid market structure are exogenously given and fixed in each time period, but may change across different time periods.}\footnote{Here the network topology regards the interference relationships among all SUs, and the  market structure regards the SUs in each market and the contract of each SU in the futures market.}
\revth{In particular, we focus on the optimal spectrum allocation in a particular time period with a particular \emph{exogenous} and fixed hybrid market, i.e., with given contract
users (in the futures market) and given spot transaction users (in the spot market).}\footnote{Namely, we do not consider the endogenous market formulation and evolution in this work.
In a more general case, we can consider the problem of an  endogenous hybrid spectrum market structure, in which an SU can choose which market(s) to join, and what contract to accept if he or she is in the futures market.
Our study in this paper provides important insights into the
exogenous market with fixed user-market associations, and serves as an important first step for understanding the endogenous market.}
Namely, we want to answer the following question: how to optimally allocate spectrum among the given contract users and spot market users to maximize the secondary spectrum utilization efficiency in that time period.
Here, the secondary spectrum utilization efficiency (also called \textit{spectrum efficiency}) is defined as the total social welfare generated by all SUs from utilizing idle spectrum in the whole time period (see Section \ref{sec:model:utility} for the detailed definition).
\reee{This problem is challenging due to the \textit{stochasticity} and \textit{asymmetry} of network information.
Here, network information mainly refers to the benefit (utility) of each SU from using each (idle) spectrum in each time slot.
Stochasticity means that no one can observe the network information in the future, as the network changes randomly over time.
Asymmetry means that the currently realized information of an SU is the private information, and cannot be observed by other entities, such as the SR and other SUs.}
Nevertheless, in the hybrid spectrum market, the allocation of spectrum in the whole time period must be jointly optimized due to the time-coupling constraint on the contract user demand (e.g., each contract user requires a specific number of spectrum in the whole period).
This implies that solving the optimal spectrum allocation problem directly would require the complete network information in all time slots.
As mentioned above, however, in practice the SR only has partial knowledge (e.g., the stochastic distribution) about the future network information because of the stochasticity of information. Moreover, it cannot observe all of the realized current network information, especially the private information of SUs, due to the asymmetry of information.
Therefore, the key research problem becomes the following:
\begin{prob}
How should the SR optimally allocate the idle spectrum in the given period among the contract users and spot market users to maximize the spectrum efficiency, taking into consideration the spatial spectrum reuse, information stochasticity, and information asymmetry?
\end{prob}



\subsection{Solutions and Contributions}

\reee{To solve this problem, we first derive an \textit{off-line  optimal policy} that maximizes the ex-ante expected spectrum efficiency based on the stochastic distribution of network information.
We then design an \textit{on-line Vickrey¨CClarke¨CGroves (VCG) auction} that elicits
SUs' private information realized in every time slot.
Based on the elicited network information and the derived off-line policy, the VCG auction determines the real-time allocation and pricing of every spectrum.
Such a solution technique (i.e., off-line policy and on-line auction) allows us to optimally allocate every spectrum in an on-line manner under stochastic and asymmetric information.
We further show that with the spatial spectrum reuse, the proposed VCG auction relies on solving the maximum weight independent set (MWIS) problem, which is well-known to be NP-hard.
Thus, it is not suitable for   on-line implementation, especially in a large-scale market.
This motivates us to further study low-complexity sub-optimal solutions.
To this end, we propose a heuristics approach based on an on-line VCG-like mechanism with polynomial-time complexity, and further characterize the corresponding performance loss bound analytically.
Our numerical results indicate that the heuristics approach exhibits good and robust performance (e.g., reaches at least 70\% of the optimal efficiency in our simulations).}

\begin{table*}[t]
\small
\centering
\caption{Key results in this paper}
\label{major-work}
\vspace{1mm}
\begin{tabular}{|c||c|c|c|c|}
\hline
\textbf{Problem} & \textbf{\rev{Solution}} & \textbf{Performance} &  {\textbf{\rev{Complexity}}} &  \textbf{Section No.} \\
\hline
\hline
E-SEM: Expected Spectrum &
Policy + VCG & Optimal & Exponential & \ref{sec:symm}, \ref{sec:vcg} \\
\cline{2-5}
Efficiency Maximization & Policy + VCG-like & Sub-optimal & Polynomial & \ref{sec:symm}, \ref{sec:vcg}, \ref{sec:pl} \\
\hline
\end{tabular}
\end{table*}

In summary, we list the key results and the corresponding section numbers in Table \ref{major-work}.
\reee{It is important to note that the main contribution of this work is not the development of new auction theory, but rather the formulation of the hybrid spectrum market and the solution techniques (including the use of auction theory) to optimize the spectrum utilization in a given hybrid market.}
Specifically, the main contributions of this paper are as follows:

\begin{itemize} 

 \item
\textit{New modeling and solution technique}:
\reee{We propose and study a hybrid spectrum market, which has both the reliability of futures market and the flexibility of spot market. Hence, it is highly desirable for QoS differentiations in secondary spectrum utilization.
To the best of our knowledge, this is the first paper to study such a hybrid
spectrum market with spatial spectrum reuse.}~~~~

\item
\textit{Optimal solution under stochastic and asymmetric information}:
\reee{We analyze the optimal spectrum allocation in an exogenous hybrid market (in a particular time  period) under stochastic and asymmetric information systematically.
Our proposed solution consists of two parts: (i) an off-line allocation {policy} that maximizes the ex-ante expected spectrum efficiency based on the stochastic network information;
and (ii) an on-line VCG auction that determines the real-time allocation of every (idle) spectrum based on the realized network  information and the pre-derived policy.
Such a solution technique allows us to optimally allocate every spectrum in an on-line manner.}

\item
\textit{Heuristic solution with polynomial-time complexity}:
\reee{We propose a heuristic approach based on an on-line VCG-like mechanism with polynomial-time complexity, and further characterize the corresponding performance loss bound analytically.
This polynomial-time solution is particularly useful for achieving the efficient spectrum utilization in a large-scale network.}

\item
\textit{Performance evaluation}:
\reee{We provide extensive numerical results to evaluate the performance of the proposed solutions.
Our numerical results show that: (i) the proposed optimal allocation significantly outperforms the traditional greedy allocations, e.g., with an average increase of 20\% in terms of the expected spectrum efficiency;
and (ii) the proposed heuristics approach exhibits good
and robust performance, e.g., reaching at least 70\% of the optimal efficiency in our simulations.}~~~~~~~~~~
\end{itemize}

The rest of this paper is organized as follows. After reviewing the literature in Section \ref{sec:relat}, we describe the system model in Section \ref{sec:model}, and present the problem formulation in Section \ref{sec:prob}. Then we derive the off-line optimal policy in Section \ref{sec:symm}, and design the on-line VCG mechanisms in Section \ref{sec:vcg}.
In Section \ref{sec:pl}, we analyze the performance loss in the low-complexity heuristic solution.
in Section \ref{sec:simu}, we provide the detailed simulation results. We finally conclude in Section \ref{sec:conclu}.
%
%

%% file: Section8-Relat-OR.tex

\section{Literature Review}\label{sec:relat}

\subsection{Secondary Spectrum Trading for DSA}

A major motivation of this work is to establish economic incentives and improve spectrum utilization efficiency in dynamic spectrum access (DSA) and cognitive radio networks (CRNs).
There are several comprehensive surveys on the technical aspects of DSA and CRNs (see \cite{Haykin-survey,Akyildiz-survey,buddhikot-sur,Zhao-sur}).

\subsubsection{Secondary Spectrum Trading in Pure Spectrum Markets.}

Recent years have witnessed a growing body of literature on the
economic analysis (in particular the incentive issues) of DSA (see, e.g., \cite{spectrum-trading2,Weiss}).
Market-driven \emph{secondary spectrum trading} is a promising paradigm to address the incentive issue and achieve high spectrum efficiency in DSA.
The literature on secondary spectrum trading often considers \emph{pricing}, \emph{contract}, and \emph{auction}. 
Pricing and Contract are generally adopted by
the \emph{futures market}, wherein sellers and buyers enter into certain agreements, specifying the
price, quality and demand in advance.
In particular, Pricing is often used under information symmetry, where sellers and buyers
possess the same market information (e.g., utility and quality) or hold the same degree of uncertainty about the market information (see, e.g., \cite{pricing1,pricing4,pricing5}).
Contract has been widely used in supply chain models, with a main focus on the supply chain coordination (see
\cite{kok}~for a comprehensive survey). Contract is effective in a market under information asymmetry, where
sellers and buyers possess different information. The seller (or buyer) can offer a selling (purchasing)
contract to extract the buyer surplus (seller surplus) as much as possible (see, e.g.,  \cite{gao1,duan,contract2}).
In contrast to Pricing and Contracts,
{Auctions} are one of the most common mechanisms used by the \emph{spot market}, wherein buyers compete to obtain commodities by offering different bids.
Auctions are particularly suitable for the information asymmetry scenario. With a well-designed auction, buyers (bidders) have the incentive to bid for the resource in a truthful manner, and the seller can allocate the resource efficiently without knowing the buyers' private information in advance (see, e.g., \cite{JianweiHuang,lixy,SorabhGandhi,WangXuXu2010,eBay,TRUST}).
\yyy{However, the above work considers the secondary spectrum trading in a pure spectrum market (i.e., either spot or futures market).
Instead, our setup incorporates the short-term secondary spectrum trading in a \emph{hybrid} spectrum market.}

\subsubsection{Secondary Spectrum Trading in Hybrid Spectrum Markets}

\cite{contract3} and \cite{Muthusamy}~considered the secondary spectrum
trading in a hybrid market. In their settings, primary sellers offer two types of contracts: the guaranteed-bandwidth
contract and the opportunistic-access contract.
{The main difference  between these two prior papers and our paper  lies in the formulation of the guaranteed contract}.
Specifically, in \cite{contract3} and \cite{Muthusamy}, the guaranteed-bandwidth contract provides guaranteed access to a certain amount of bandwidth at \emph{every} time slot.
In our model, the guaranteed-delivery contract provides guaranteed access to a total amount of bandwidth in one
time period; nevertheless, the bandwidth delivery at every time slot can be different, depending on the PUs' own demand.
The main advantage of our approach  is its
flexibility in shifting
secondary demand across time slots (to comply with the PUs' random demand). That is, it enables opportunistic delivery of a small (or large) bandwidth to SUs in those time slots that the PUs' own demand is high (or low).
Additionally, our model is also more practically relevant to a wide range of applications, which do not require fixed data delivery per time slot, but demand a guaranteed average data rate over each time period.
Furthermore, {the underlying market models
are also different}. \cite{contract3} and \cite{Muthusamy}~assumed that the demand (supply) markets have infinite liquidity. That is, any bandwidth amount supplied by the seller can be sold out (any bandwidth amount demanded
by the buyer can be bought from the market) at an ``outside fixed price''.
In this sense, their market models are closely related to the ideal competitive market.
We assume that the market price is
endogenously determined by the associated seller and buyer (through, for example, an VCG mechanism).
Thus, we essentially consider the monopoly market.

\begin{table}[t]
\footnotesize
\renewcommand{\arraystretch}{1.5}
\centering
\caption{\normalsize A summary of secondary spectrum trading literature}
\label{summary-trading}
\vspace{1mm}
\begin{tabular}{c||l}
\hline
\textbf{Market Type}&~~~~~~~~~~~~~~~~~~~~~~~~~~~~~~~~~~~~\textbf{Related Work} \\
\hline
\hline
 \vspace{-2mm}
 \multirow{2}*{\textbf{Futures Market}} & 	Pricing: \cite{pricing1,pricing4,pricing5}. \\
 & Contract: \cite{gao1,duan,contract2}.  \\
\hline
\vspace{-2mm}
 \multirow{2}*{\textbf{Spot Market}} & Auction: \cite{JianweiHuang,SorabhGandhi,lixy, WangXuXu2010}, \\
& ~~~~~~~~~~~ \cite{eBay,TRUST}. \\
\hline
\textbf{Hybrid Market} & \cite{contract3,Abhishek-2012,Muthusamy,gao2}\\
\hline
\end{tabular}
\vspace{-3mm}
\end{table}

\reee{\cite{Abhishek-2012} also considered a hybrid market, in which a cloud service provider sells its service to users via two different pricing schemes: pay-as-you-go (PAYG) and spot pricing.
Under the PAYG, users are charged a fixed price per unit time.
Under the spot pricing, users compete for services via using an auction. They focused on the optimal market formulation,
that is, the service provider selects different PAYG prices
such that different users will choose different pricing schemes or market types; consequently, different hybrid markets will be formulated.
In our work, we focused on the optimal spectrum allocation in a \textit{given} hybrid market.
In other words, each SU is associated and fixed in a particular market (based on his or her application type), and the SR determines the optimal spectrum allocation among the given contract users and spot market users.}

In \cite{gao2}, we studied the secondary spectrum
trading in the same hybrid market. However, \cite{gao2}~did not consider spatial spectrum reuse, which is the key contribution of this paper.
With spatial spectrum reuse, the same spectrum can be potentially used by multiple SUs simultaneously; thus, the total spectrum efficiency can be greatly improved. This new coupling in the spatial dimension creates many challenges (e.g., solving MWIS problems) in the optimal mechanism design, and makes the problem significantly different from those without spatial spectrum reuse.
Precise mathematical modeling and understanding of the spatial coupling are often lacking in the wireless literature. One of the main contributions of this paper is to propose a low-complexity heuristic algorithm to tackle this issue, and to quantify the performance bound of the proposed algorithm. %
For convenience, we summarize the key literature in Table \ref{summary-trading}.

%

\subsection{Dual Sourcing in Supply Chain Management}

The theoretic model used in our work is related to the \emph{dual sourcing} problem in supply chain management. Specifically, with   dual sourcing, a firm can procure a single commodity from a supplier via a long-term contract and/or from a spot market via short-term purchases (see \cite{klei2003}). In such a context, the long-term contract and short-term purchase in the supply chain model corresponds to the guaranteed contract and spot transaction in our model, respectively.
Note that the firm in the supply chain model buys commodities through   dual sourcing, while the SR in our model sells commodities through guaranteed contracts and spot transactions.
{The long-term/guaranteed contract ensures consistency over time availability of commodities with guaranteed quality at a predetermined price;
whereas, the short-term purchase/spot transaction provides high inventory flexibility, allowing firms to buy and sell commodities at any quantity with zero lead time, but at a random market price.}
The main advantage of dual sourcing is to hedge the future uncertainties in the supplier's commodity supply and the end users' demand, so as to hedgy the financial risk in the supply chain (see \cite{klei2008a,klei2008b}).

\rev{The integration of long-term contracts and short-term purchases is of particular interest to the firm in a supply chain, due to the following two reasons: (i) sourcing competition can keep purchasing prices under control by the firm; and (ii) a wider supply base can mitigate the risk induced by the uncertainty at one supplier.}
\cite{Lee2002}~were the first to integrate, after sales, the spot market considerations within a newsvendor ordering framework. \cite{Peleg2002}~studied the long-term and short-term integrated sourcing using a stylized two period model.
\cite{Yi2003}~studied the nature
of the optimal inventory policy when such a dual sourcing is used, in the presence of a fixed cost for the spot market participation.
\cite{Wu2005}~proposed a general framework with integrated long-term  and short-term contract decisions
for non-storable commodities. \cite{Kouvelis2009}~studied the problem of dual sourcing with financial hedging for storable commodities.

\rrr{Our work differs from the above work in the following aspects. First, the spectrum in our model can be potentially used by multiple SUs simultaneously, while a traditional commodity can usually be used by one user only. Second, the availability of spectrum is stochastic, while the availability of a traditional commodity is usually deterministic.
Third, we focus on the spectrum utilization  efficiency maximization, rather than the financial risk hedging.
Fourth, instead of designing the optimal contracts, we treat contracts as exogenously given, and focus on the problem of  how to fulfill these contracts and cope  with additional demand from spot markets. Finally, we consider stochastic and asymmetric network information.}

%% file: Section1-Model-OR.tex

\section{System Model}\label{sec:model}

\subsection{System Description}\label{sec:model:1}

We consider a DSA system with one SR and multiple SUs.
\reee{The SR has certain licensed frequency band, which  is divided into $K$ orthogonal \textit{channels}
using channelization methods such as frequency  division and code division.}
In Fig.~\ref{system-illu}, for example, we have $K=3$.
The frequency band is licensed to a set of subscribed users (here we call primary users, PUs), who access channels with the slotted transmission protocol which is widely-used in today's wireless communication systems (e.g., GSM, WCDMA, and LTE).
That is, the total time is divided into fixed-time intervals, called \textit{time slots}, and each PU transmits over one or multiple channel(s) according to a synchronous time slot structure.
Depending on the activities of PUs, some channels may be not used by any PU (i.e., idle) in some time slots, which can be potentially assigned for the secondary utilization of SUs.

We consider the \textit{market-driven} dynamic spectrum access, also called secondary spectrum trading, in which SUs temporarily purchase the idle channels from the SR.
More specifically, we consider the \textit{short-term} secondary spectrum trading, in which the idle channel is traded on a slot-by-slot basis.
\yyy{Namely, the basic resource unit for trading is ``a particular channel at a particular time slot'', referred to as a {spectrum opportunity} or \textit{spectrum}.
The main motivation for considering the \textit{short-term} spectrum trading is as follows.
The spectrum availability changes frequently and randomly over time due to the stochasticity
of PUs' activities; thus, a channel that is idle in a particular time slot may not always be idle in the future.}

\yyy{Let~~$\ch_{kt}$~~denote the spectrum on the $k$-th channel at time slot $t$. Let $\xi_{kt} \in \{0,1\}$ denote the state (availability) of spectrum $\ch_{kt}$, with $\xi_{kt} = 1$ indicating that the spectrum  is not used by any PU  and thus is available for SUs, and $\xi_{kt} = 0$ otherwise.
We consider the operations in a given period  of $T$ time slots.
The total amount of spectrums in the period is referred to as the SR's \textit{spectrum supply}, denoted by
$\textstyle
\CH \triangleq  \{\ch_{kt} \}_{k\in\{1,...,K\},t\in\{1,...,T\}.}
$
The size of total supply is denoted by $S = |\CH| \triangleq K\cdot T$.
The states of all spectrums in $\CH$ are referred to as the \textit{spectrum availability}, denoted by
$\textstyle
\XI \triangleq  \{\xi_{kt}\}_{k\in\{1,...,K\},t\in\{1,...,T\}.}
$

Due to the uncertainty of PUs' activities, the spectrum availability changes randomly across both time and frequency.
Fig.~\ref{system-illu} illustrates an example of such a system, where $K=3$, $T=12$, and $S=36$. At the time slot $t=1$, the spectrum $\ch_{21}$ is idle, while the spectrum $\ch_{11}$ and $\ch_{31}$ are busy, i.e., $\xi_{21} = 1$ and $\xi_{11} = \xi_{31} = 0$.}
\revth{Note that in practical wireless communication systems, $T$ is usually very large.
This is mainly due to two reasons.
First, the length of each time slot in a wireless
system is often quite small, e.g., in milliseconds or even in microseconds, which corresponds to the
typical length of frame in many wireless communication systems (such as 4G LTE
and Wi-Fi). In fact, the physical limit of choosing the time sloth length is the so called
\emph{coherence time}, which is the time within which the channel condition does not change.
In wireless
communications, such coherence time is usually very small due to fast small scale multi-path fading.
Second, the time scale of each time period is relatively large, e.g., in minutes or even hours, which
corresponds to the validity period of contracts (to be defined in Section \ref{sec:model:contract}).
Hence, the number of slots in each time period
(i.e., $T$) is very large, e.g., $T = 6 \cdot 10^5$ when the length of slot is 1 millisecond and the length of period is 10 minutes.}

Following are our assumptions on the spectrum availability  and spectrum  usage.

\begin{assumption}
 The spectrum availability is independently and identically distributed ($\textrm{i.i.d.} $) across time and frequency.\footnote{We will show in \cite{Techrpt}, that our results can be generalized to the  $\textrm{non-i.i.d.}$ spectrum availability case with minor modifications.}
 \revth{Moreover, the spectrum availability is ergodic.}
\end{assumption}

  \begin{assumption}
Each SU can transmit over multiple channels simultaneously.\footnote{This is supported by most physical layer access technologies (e.g., OFDMA) even with just one transmitter antenna per SU, and is possible with other technologies when each SU has multiple transceivers.}
\end{assumption}

\revth{For convenience, we denote the  availability of   spectrum as a random variable  $\boldsymbol{\xi} \in \{0,1\}$, and
 the probability mass function of $\boldsymbol{\xi}$ as $f_{\boldsymbol{\xi}}(\cdot)$.\footnote{We will use $f_{\boldsymbol{X}}(\cdot)$ and $F_{\boldsymbol{X}}(\cdot)$ to denote the probability distribution function (PDF) and cumulative distribution function (CDF) of a continuous random variable or vector $\boldsymbol{X}$.
For notational convenience, we use the same notation $f_{\boldsymbol{X}}(\cdot)$ and $F_{\boldsymbol{X}}(\cdot)$ to denote the probability mass function and CDF of a discrete random variable $\boldsymbol{X}$.}
Suppose $\rho $ is the idle probability of  spectrum $\ch_{kt}$. Then, $f_{\boldsymbol{\xi}}(\xi_{kt}) = \rho $ if $\xi_{kt}=1$, and  $f_{\boldsymbol{\xi}}(\xi_{kt}) = 1- \rho $ if $\xi_{kt}=0$.
}

\subsection{Futures Market and Spot Market}\label{sec:model:futurespot}

Wireless applications can be broadly categorized as either \textit{elastic}  or \textit{inelastic} depending on Quality-of-Service (QoS) requirements.\footnote{Note the meaning of ``elastic/inelastic demand'' in wireless networks (in this paper) is a bit different from that in microeconomics, where it is mainly used to characterize whether a consumer's demand changes with the market price.
In wireless networks, however, it is mainly used to characterize an SU's inherent demand for spectrums.}
With elastic traffic, SUs have elastic demands for spectrum, in the sense that the tasks are not urgent in terms of time; thus, the QoS will not be significantly affected, even if the spectrum resource is limited and the transmission rate is low
for a substantial amount of time.
With inelastic traffic, SUs have inelastic demands for spectrum.
In other words, the tasks can only function well when the data rate is guaranteed to be above certain thresholds such that the delay requirements are met; otherwise they will suffer significant performance loss.
It is important to note that the
information of the application type is usually explicitly represented in the headers of data packets,
and can be easily extracted by the network operator through deep packet inspection.
Hence, it is
reasonable to assume that an SU cannot fake his or her application~type arbitrarily.
Examples of elastic traffic include FTP downloading, data backup, and cloud synchronization, and examples of inelastic traffic include VoIP, video streaming,
and real-time data collection.


To accommodate the various requirements of elastic traffic and inelastic traffic (so as to achieve desirable QoS differentiations), we propose a \textit{hybrid spectrum market} combining both the {futures market} and the {spot market}.
Next we define the futures market and spot market formally.

 \textit{Futures Market.} In the futures market, each SU enters into an agreement, called a futures contract or \textit{guaranteed contract}, directly with the SR. The contract specifies the key elements in trading, e.g., the SU's total demand and payment in the given period.
Once the SR accepts a contract, it is committed to deliver the specified number of spectrum to the SU.  {If the SR fails to do so, it needs to pay certain \textit{penalty} for compensating the SU's potential welfare loss.}
This penalty can be a unit price paid for every undelivered spectrum (called a {soft contract}) or simply a total payment for the violation of contract (called a {hard contract}).

\textit{Spot Market.} In the spot market, each SU purchases spectrums in a real-time and on-demand manner through spot transactions. That is, an SU initiates a purchasing request only when he or she needs spectrum, and multiple SUs requesting the same spectrum would compete with each other for the spectrum, e.g., through an auction.
The spectrum is delivered immediately to the winner at a real-time market price, which depends on both the SUs' preferences and competitions.
The winner's payment
is also dependent on the SUs' preferences and competitions. 


\yyy{Although, in practice, it is more desirable to allow SUs to have the flexibility to choose which market to join, in this work we assume that {the SUs with inelastic traffic always choose the futures market, and the SUs with elastic traffic always choose the spot market}.\footnote{\reee{As mentioned previously, one key motivation for this assumption is that SUs cannot fake their applications arbitrarily.
In \cite{Techrpt}, we will further show that even if an SU has the capability of faking his or her application type, he or she does not
have the incentive to do so, if the hybrid market is properly designed.}}
In other words, we will leave out the SU's market selection problem, and instead focus exclusively on
the spectrum allocation in the given hybrid markets in each time period.
This setup helps us to concentrate on the technical challenges brought by  spatial spectrum reuse and  information stochasticity \& asymmetry.
}

Next we provide the key assumptions on the hybrid market structure.

\begin{assumption}
\revth{We assume a quasi-static market scenario, where the hybrid spectrum market is exogenously given and fixed in each time period, but may change across different time periods.}\footnote{This means that the user-market association is fixed, and the contract of each futures market user is also fixed during the whole period of interest. \revth{The changes of hybrid market in different time periods can be caused by the changes of user parameters (e.g., user service types) as well as the changes of the contract selection of the SR at the beginning of each period.}}
\end{assumption}

\begin{assumption}
We assume that each SU has one application, and all SUs with inelastic traffic (elastic traffic) are in the futures market (spot market).\footnote{Note that the assumption can be easily generalized: For an SU having multiple applications with different QoS requirements, we can simply divide the SU into multiple virtual SUs, each associated with one application.}
\end{assumption}

\yyy{Based on these assumptions, we can divide SUs into two disjoint sets, each associated with one market. Let $\N\eq\{1,2,...,N\}$ and $\M\eq\{1,2,...,M\}$ denote the sets of SUs in the futures market and the spot market, respectively.
For convenience, we will use the notation ``$n$'' to denote a contract user, and  ``$m$'' to denote a spot market user.
When needed, we will use the superscripts ``${\textsc{c}}$'' and ``$\textsc{s}$'' to indicate variables related to the futures market and spot market, respectively.}

\revth{For analytical convenience, we further assume that the size of futures market is much smaller than that of the spot market, i.e., $\N \ll \M$. This assumption is used to facilitate the computation of independent contract user sets in the later analysis.\footnote{\revth{Although this is an assumption, it can be justified by the fact that the SR can actually control the size of the futures market, via intelligently choosing a limited set of contract users at the beginning of each period. Namely, to balance the performance and the complexity, the SR can intelligently choose a proper set of contract users to serve in a particular time period.
Hence, from the system perspective, the size of the futures market is controllable, whereas the size of spot market is generally random and uncontrollable.}}}

\subsection{{Futures Contract Definition}}\label{sec:model:contract}

A contract could be quite complicated, depending on specific requirements of a contract user's application.
In this work, we focus on a basic contract form, which consists of two parts: (i) the \rev{SU's} demand and payment for spectrum in each period; and (ii) the \rev{SR's penalty when} not delivering the demanded spectrums to the SU.
Formally, we write the contract of an SU $n\in \N$ as:
\begin{equation}\label{eq:contract}
\Cn_n  \triangleq  \big\{B_n , D_n,  \boldsymbol{J}_n \big\},
\end{equation}
where $B_n$ is the SU's payment, $D_n$ is the SU's demand in one period, and $\boldsymbol{J}_n = (J_n, \{\widehat{P}_n, \widehat{B}_n\})$ is the SR's penalty scheme.
Here, $J_n\in\{0,1\}$ \rev{indicates the contract type (soft or hard), $\widehat{P}_n$ is the unit penalty for a soft contract ($J_n=0$), and $\widehat{B}_n$ is the total penalty for a hard contract ($J_n=1$).}
Thus, the \textit{actual payment} of contract user $n$  (denoted by $R_n$) depends on the number of spectrums he or she actually obtains (denoted by $d_n$), that is:
\begin{equation}\label{eq:ERn-xxxx}
 R_n = B_n  - \mathbb{P}( d_n  ,D_n),
\end{equation}
where  $\mathbb{P}(d_n  ,D_n)$ is the SR's penalty if violating the contract, given by:
\begin{equation}\label{eq:penalty-xxxx}
\mathbb{P}(d_n  ,D_n) \triangleq \left\{
\begin{aligned}
&[D_n-d_n ]^+\cdot \widehat{P}_n , &&  J_n=0
\\
& \mathbf{1} {( D_n - d_n )} \cdot \widehat{B}_n  ,\qquad &&  J_n=1
\end{aligned}
\right.
\end{equation}
where $[x]^+ = \max\{0, x\}$ and \revth{$\mathbf{1} {(x)} = \max\{0, \frac{x}{|x|}\}$}. That is,
$[x]^+ = \mathbf{1} {(x)} = 0$ if $x\leq 0$, and
$[x]^+ = x$ and $\mathbf{1} {(x)} = 1$ if $x > 0$.
For convenience, we denote the set of all contracts in the futures market by $\CO \triangleq \{\Cn_n\}_{n\in \N}$.
As each contract user is guaranteed to obtain the desired number of spectrum (otherwise he can get the SR's penalty as compensation), we refer to such a contract as the \textit{guaranteed-delivery contract}.

The important assumptions on the guaranteed contract are listed below.


\begin{assumption}
We assume that contract users care about the expected number of spectrums they obtain (hence the SR's penalty is based on the expected number of allocated spectrums).
\end{assumption}


\reee{The main practical motivation for such an assumption is that most wireless applications in
practice require an expected/average data rate during a certain time period.
For example, video streaming concerns the average downloading rate in every minute, and VoIP concerns the average downloading/uploading rate in every second.
When the actual data rate is occasionally less than the average
rate, various coding and error concealment technologies can be employed, so that the SUs will
not feel significant performance degradation.
Thus, with a proper choice of the length of allocation
period (e.g., one minute for video streaming or one second for VoIP), such a contract
with the expected demand is suitable for most wireless applications.
In
fact, with the uncertainties of spectrum availability and the stochasticity of SU utility on each spectrum,
it is practically impossible to guarantee a strict spectrum supply for each contract user. To the best of our knowledge, no
such guarantee has been provided even in the latest communication standard.
To make our analysis more complete, we also provide detailed
theoretical analysis and numerical evaluation for the impact of this assumption on the performance in \cite{Techrpt}.}~~~~~~~

\subsection{Spatial Spectrum Reuse}\label{sec:model:reuse}

As mentioned previously, an important characteristic of a spectrum market is that radio spectrum is \textit{spatially reusable} subject to certain spatial interference constraints (see \cite{wirelessbook}). For example, a spectrum can be used by multiple SUs simultaneously if they are far enough apart and do not interfere with each other.
This leads to a fundamental difference between the spectrum market and conventional markets: the same spectrum can be potentially sold to multiple SUs.~~~~~~~~

In this work, we adopt a widely-used physical model to capture the interference relationships between SUs, namely, the \textit{protocol  interference model} (see \cite{Gupta2000}). Under the protocol interference model, multiple SUs can use the same spectrum simultaneously without mutual interference, if and only if none of the SUs falls inside the interference ranges of others (e.g., they are sufficiently far away from each other).
The network under a protocol interference model can usually be represented by a \textit{conflict graph}, with each vertex denoting an SU, and each edge indicating that two associated SUs fall inside the interference range of each other.
Thus, any two SUs not connected by an edge (also called \textit{independent} users) can use the same spectrum simultaneously, while any two connecting SUs cannot.

\begin{figure}
\centering
    \includegraphics[scale=.4]{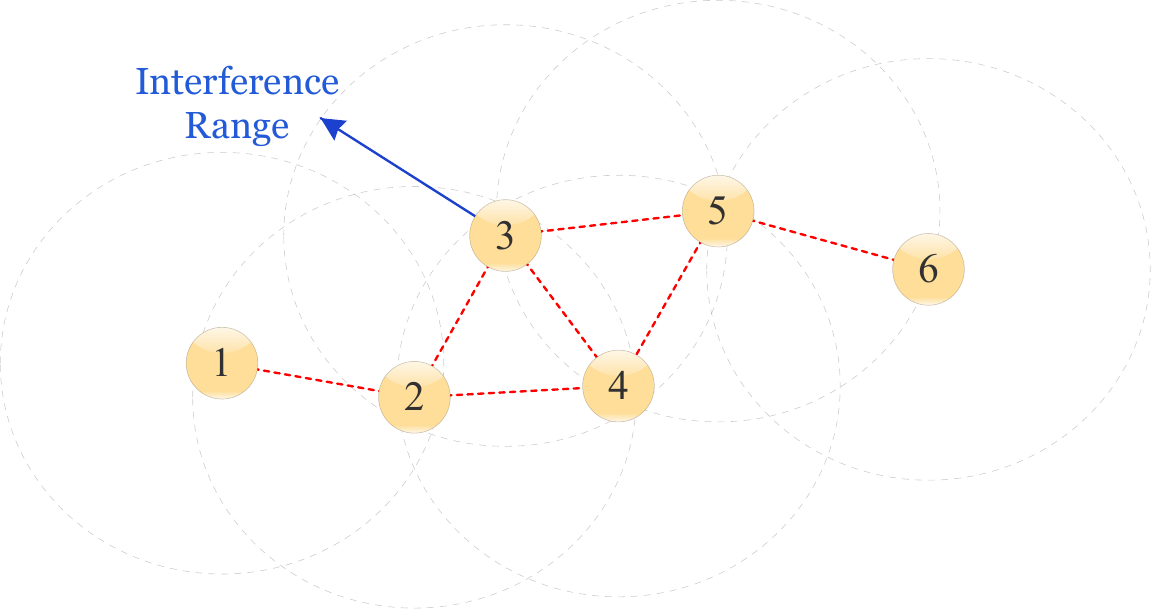}
    \caption{Conflict graph representation.} \label{conflict-graph}
\end{figure}

\yyy{Fig.~\ref{conflict-graph} illustrates the conflict graph representation for a network with 6 SUs, in which each dash circle denotes the interference range of the associated SU.
{In this example, SUs 1 and 3 are independent of each other (and therefore can  use the same spectrum simultaneously), while SUs 1 and 2 are not.}
Fig.~\ref{system-illu} also illustrates a similar conflict graph, where SUs 2, 6, and 8 are independent of each other, whereas SUs 1, 2, and 3 are not.}

\yyy{Let $ \G \triangleq (\V,\E)$ denote the graph consisting of all SUs (and the associated edges, similarly hereafter), where $\V \triangleq  \M \bigcup \N $ is the set of vertices (SUs), and $\E \triangleq \{\mathtt{e}_{ij}\ |\ i,j\in \V \}$ is the set of edges. Let $\Gs \triangleq (\Vs,\Es)$ denote the \textit{spot market subgraph} consisting of all spot market users, and $\Gc \triangleq (\Vc,\Ec)$ denote the \textit{futures market subgraph} consisting of all contract users, that is, $\Vs \triangleq  \M$, $\Vc \triangleq  \N$, $\Es \triangleq \{\mathtt{e}_{ij}\in \E\ |\ i,j\in \M\}$, and $\Ec \triangleq \{\mathtt{e}_{ij}\in \E\ |\ i,j\in \N\}$.
For presentation convenience, we further introduce the concept of side market.
%

 \begin{definition}[{Side Market}]\label{def:subg-side}
The side market (subgraph) of a contract user set $\boldsymbol{n}  \subseteq \N$, denoted by  $\Gs_{\boldsymbol{n}}=(\Vs_{\boldsymbol{n} },\Es_{\boldsymbol{n} })$, is a subgraph of the spot market, consisting of all spot market users \textit{except} the neighbors of contract users $\boldsymbol{n} $.
 \end{definition}

Here, the ``neighbors'' of $\boldsymbol{n}$ refer to spot market users connecting to at least one contract user in the set $\boldsymbol{n} $.
It is easy to see that $\Vs_{\boldsymbol{n}_1 } \subseteq \Vs_{\boldsymbol{n}_2 }\subseteq \Vs$ for any $ \boldsymbol{n}_2 \subseteq \boldsymbol{n}_1 \subseteq \N$, since a neighbor of $\boldsymbol{n}_2$ must be a neighbor of $\boldsymbol{n}_1$ if $\boldsymbol{n}_2 \subseteq \boldsymbol{n}_1$.
In Fig.~\ref{system-illu},
we have  $\Vc = \{1,2,3,4\}$, $\Vs = \{\mbox{5,6,7,8}\}$, and examples of side markets include $\Vs_{\{1\}} = $ $ \{\mbox{6,7,8}\}$ and $\Vs_{\{1,2,3\}} = \{\mbox{7,8}\}$. Obviously,  $\Vs_{\{1,2,3\}}  \subseteq \Vs_{\{1\}}  \subseteq \Vs$.}


\revth{Let  $\IS(\G )$ and $\CQ(\G )$ denote the sets of all independent sets and all cliques of a graph $\G $, respectively.}\footnote{The detailed definitions for independent set and clique can be referred to a standard textbook, e.g., \cite{graph2001}.}
\yyy{Then, the spatial interference constraint can be defined as follows.
\begin{definition}[{Spatial Interference Constraint -- SIC}] \label{def:spatial-interference-cond}
Any two SUs in a clique of $\G $ cannot use the same spectrum simultaneously; on the other hand, all SUs in an independent set of $\G $ can use the same spectrum simultaneously.
\end{definition}}

Finally, we list the key assumption on the network topology or   graph $\G $ below.

\begin{assumption}
We assume that the network topology $\G $ is exogenously given, and remains unchanged during the whole period of interest.
\end{assumption}

{This corresponds to a static or quasi-static network scenario, in which SUs do not move or move slowly. With this assumption, it is reasonable to assume that the network topology is  public information to the SR.}

%
%
%

\subsection{Valuation}\label{sec:model:utility}

The \revth{\textit{valuation}}
of an SU $i$ over a spectrum $\ch_{kt}$ reflects
the SU's preference for the transmission rate (capacity) achieved on spectrum $\ch_{kt}$.
Specifically, it is determined by two factors: (i) the transmission rate of SU $i$ over spectrum $\ch_{kt}$ (denoted by $r_{i,kt}$), reflecting how efficiently SU $i$ utilizes spectrum $\ch_{kt}$; and (ii) the user preference of SU $i$  (denoted by $\alpha_{i,kt}$), reflecting how eagerly SU $i$ desires for spectrum $\ch_{kt}$.
To avoid confusion, we use notation ${v}_{m,kt}$  and ${u}_{n,kt} $ to denote the valuations of spot market user $m$ and contract user $n$ over spectrum $\ch_{kt}$, respectively.
Formally, we can write the valuations ${v}_{m,kt}$  and ${u}_{n,kt} $  as generic functions:
\begin{equation}\label{eq:utility}
{v}_{m,kt} \triangleq g_m\big( {r}_{m,kt},\ \alpha_{m,kt} \big) \mbox{~~and~~} {u}_{n,kt} \triangleq g_n\big( {r}_{n,kt},\ \alpha_{n,kt} \big),
\end{equation}
where $g_i(\cdot)$ is the generic valuation function of SU $i$, which generally increases in both the transmission rate and the user preference.
Typical examples of valuation function include: (a) linear functions, e.g., $ g_i(r, \alpha)   =\alpha  \cdot  r  $; (b) sigmoid functions, e.g., $ g_i(r, \alpha)   =  \alpha  \cdot \frac{1}{ 1 + e^{r_0-r}} $; and (c) step functions, e.g.,  $ g_i (r, \alpha)   =  \alpha $  if $ r $  exceeds a certain threshold $r_0$, and 0 otherwise.

The user preference $\alpha_{i,kt}$ is a \textit{subjective} feeling of SU $i$ for spectrum $\ch_{kt}$, and is   related to factors such as traffic state, urgency, and importance.
The transmission rate $r_{i,kt}$ is an \textit{objective} attribute of spectrum $\ch_{kt}$ for SU $i$, and is   related to a spectrum-specific \emph{quality} $q_{kt}$ and a user-specific channel coefficient ${h}_{i,kt}$.
The ${q}_{kt}$ is common for all SUs, regarding factors such as channel bandwidth, noise level, and power constraint. The ${h}_{i,kt}$ is usually independent between different SUs, regarding factors such as path loss, shadow fading, and small scale fading (or multiple-path effect). Typically, $r_{i,kt}$ is given by the Shannon-Hartley theorem (see \cite{Shannon2001}):
\begin{equation}\label{eq:capacity}
\textstyle
{r}_{i,kt} = B_{k} \cdot  \log\Big( 1 + \frac{P_i  \cdot  |{H}_{i,kt}|^2 }{ \sigma_{kt}^2 } \Big),
\end{equation}
where $B_{k}$ is the channel bandwidth, $P_i$ is the transmission power,
$\sigma_{kt}^2$ is the noise level over spectrum $\ch_{kt}$  (captured by the common spectrum quality $q_{kt}$), and $ |{H}_{i,kt}|^2 $  is the channel gain of SU $i$  over spectrum $\ch_{kt}$ (captured by the user-specific channel coefficient ${h}_{i,kt}$).


By (\ref{eq:utility}) and (\ref{eq:capacity}), an SU's valuation (over a spectrum) is essentially a joint function of the common spectrum quality, user-specific channel coefficient, and user preference.
For notational convenience,
we will write it as $
{v}_{m,kt} \triangleq g_m\big(q_{kt}, {h}_{m,kt}, \alpha_{m,kt} \big) $ and $ {u}_{n,kt} \triangleq g_n\big( q_{kt}, {h}_{n,kt}, \alpha_{n,kt} \big)$.
In general, the spectrum quality, user-specific channel coefficient, and user preference change randomly across time and frequency. For simplicity, the following assumptions are used in our work.

\begin{assumption}
The spectrum quality, user-specific channel coefficient  (of every SU), and user-preference  (of every SU) are all $ \textrm{i.i.d.} $ across time and channels.
\end{assumption}

Based on the above assumption, we can see that \textit{the valuation of each SU is also $ \textrm{i.i.d.} $ across time and channels}.\footnote{Note that in practice, an SU's valuation may not be $\textrm{i.i.d.} $ across time and channels, due to the $\textrm{non-i.i.d.} $ of spectrum quality, user-specific channel coefficient, and/or user preference. We will show in
\cite{Techrpt},
that our work can be easily applied to the scenario with $\textrm{non-i.i.d.} $ valuation across time or channels.}
Let random variable $\boldsymbol{q}$ denote the spectrum quality, ${\boldsymbol{h}}_i$ and ${\boldsymbol{\alpha}}_i$ denote the user-specific channel coefficient and preference  of SU $i$.
Let random variables $ {\boldsymbol{v}}_m \triangleq g_m\big(\boldsymbol{q}, {\boldsymbol{h}}_m , {\boldsymbol{\alpha}}_m \big) $ and $ {\boldsymbol{u}}_n\triangleq g_n\big(\boldsymbol{q},  {\boldsymbol{h}}_n , {\boldsymbol{\alpha}}_n \big)$ denote the valuations of spot market user $m$ and contract user $n $, respectively.  We can further see that {the valuations of different SUs (over the same spectrum) are {partially correlated} by the common spectrum quality $\boldsymbol{q}$}.
That is, \textit{the valuation of each spectrum is partially correlated across users}.
Such a valuation formulation generalizes both the case of  independent user valuation  (by removing the uncertainty of $\boldsymbol{q}$) and the case of  completely correlated user valuation (by removing the uncertainty of $ {\boldsymbol{h}}_i$ and ${\boldsymbol{\alpha}}_i$, $\forall i $).

Let random vector $\THETA \triangleq \left(\boldsymbol{v}_1 ,..., \boldsymbol{v}_M ; \boldsymbol{u}_1 ,...,\boldsymbol{u}_N \right)$ denote the {valuations (vector) of all SUs}, and $\theta_{kt} \triangleq \big( {v}_{1,kt} , $ $ ...,$ $ {v}_{M,kt} ; $ $  {u}_{1,kt} ,..., {u}_{N,kt} \big)$ denote a realization of $\THETA$ on   spectrum $\ch_{kt}$.
As will be shown later, in the spot market, the valuation $v_{m,kt}$ uniquely reflects the SU $m$'s maximum willingness-to-pay (or utility) for the spectrum $\ch_{kt}$.
\revth{In the futures market, however, the valuation  $u_{n,kt}$ only partially reflects the SU $n$'s maximum willingness-to-pay (or utility) for the spectrum $\ch_{kt}$, which depends also on the total number of spectrums that he obtains.}
\rrr{Since each spectrum $\ch_{kt}$ is fully characterized by the utility vector $\theta_{kt}$ and its availability $\xi_{kt}$, we also refer to $\{\thxikt\}$ as the \textit{information of spectrum $\ch_{kt}$}. Formally,}
\rrr{\begin{definition}[{Network Information}]\label{def:subg-info}
The information of each spectrum $\ch_{kt}$ consists of  its availability and  the valuations of all SUs over that spectrum, i.e., $\{ \thxikt\}$. The information of all spectrums in $\CH$ is referred to as the network information, i.e., $\{\thxikt\}_{k\in\{1,...,K\},t\in\{1,...,T\}}$.
\end{definition}}

For convenience, we will refer to a spectrum with information $\{\thxikt\}$ as  spectrum $\{\thxikt\}$.
\revth{Note that the private information of a contract user $n$ (or spot market user $m$) is defined as his or her valuation ${u}_{n,kt}$ (or $v_{m,kt}$) for each spectrum $\textsc{ch}_{kt}$.}
That is, we consider all other information of a contract user (e.g.,  the spectrum demand) as public information.
It is important to note that in a broad sense, we can generalize our current analytical framework by allowing multi-dimensional private information, as we adopt a general VCG framework in the later analysis.

\subsection{Spectrum Efficiency}\label{sec:model:eff}

When the SR leases an idle licensed spectrum to an SU for secondary utilization, the SU achieves a certain benefit, called \textit{SU-surplus}, which equals the difference between the total \revth{\textit{utility}} achieved from using these spectrums and the payment transferred to the related PU;
the PU also achieves a certain benefit, called \textit{PU-surplus}, which equals the difference between the payment from the SU and the cost of holding spectrum and sharing spectrum with unlicensed SUs.\footnote{\rrr{Such a cost may include  the spectrum license fee, management fee, maintenance fee, etc.}}
The \textit{social surplus} generated by such a process  is the sum of the SU-surplus and the PU-surplus, which equals   the difference between the  (SU)  utility generated and the (PU) cost involved. That is:
$$
\mathrm{Social~Surplus} = \mathrm{SU~Surplus} + \mathrm{PU~Surplus} = \mathrm{SU~Utility} - \mathrm{PU~Cost}.
$$
Notice that the PU's cost is a constant (sunk cost) independent of the secondary spectrum allocation.
Without loss of generality, we normalize it to zero.
As a result, the social surplus is equivalent to the  utility generated by the SU.
Next we define the \textit{utility} of each SU formally.


\textit{Utility of Spot Market User.}
In the spot market, based on the characteristics of elastic traffic, an SU achieves an immediate utility from any spectrum allocated, and such a utility is directly related to his or her valuation of the spectrum.
Thus, the total utility of a spot market user $m$ (denoted by $\ws_m$) can be directly defined as the total valuation achieved: \footnote{Our method can be directly applied to a general case, in which the SU's utility is a generic function of his valuation. In that case, we have $\ws_m = \sum_{k=1}^K\sum_{t=1}^T \as_{m,kt} \cdot y_m\big(v_{m,kt}\big) \cdot \xi_{kt}$, where $y_m(\cdot)$ is the valuation function of SU $m$.
}
\begin{equation}\label{eq:spot-welfare}
\textstyle
\ws_m = \sum_{k=1}^K\sum_{t=1}^T \as_{m,kt} \cdot v_{m,kt} \cdot \xi_{kt},
\end{equation}
where $\as_{m,kt} \in \{0,1\}$ indicates whether to allocate a spectrum $\ch_{kt}$ to SU $m$. The factor $\xi_{kt}$ means that only the idle spectrums can be used by SUs.

\textit{Utility of Contract User.}
\revth{In the futures market, however, an SU's utility is not only related to the valuation that he or she achieves, but also related to the total number of spectrums that he or she obtains.}
First, depending on the requirement of inelastic traffic, an SU achieves a fixed total utility if the number of allocated spectrums meets (or exceeds) the demand; otherwise the SU suffers certain utility loss.
\revth{Such a utility is usually referred to as the \emph{demand-related} utility,} which coincides with the SU's payment defined in the contract, i.e., Eq.~(\ref{eq:ERn-xxxx}).\footnote{It is worth noting that we are not contending that the contract user's actual payment in (\ref{eq:ERn-xxxx}) or the contract details in (\ref{eq:contract}) determines the user's (demand-related) utility definition; on the contrary, the user's (demand-related) utility determines the  type of contract that he or she will accept and the payment that he or she is willing to pay.}
\reee{For convenience, we will use the notation $D_n^{\ddag}$, $B_n^{\ddag}$, $ \widehat{P}_n^{\ddag}$, and $\widehat{B}_n^{\ddag} $ to denote the spectrum demand, the  fixed total utility when the demand is satisfied, the unit utility loss (for soft contract) and the fixed total utility loss (for hard contract) when the demand is not satisfied, respectively.
Since we do not address the problem of contract design in this work, we simply consider that each contract coincides perfectly with the contract user's utility definition, i.e., $D_n^{\ddag} = D_n$, $B_n^{\ddag} = B_n$, $ \widehat{P}_n^{\ddag} = \widehat{P}_n$, and $\widehat{B}_n^{\ddag} = \widehat{B}_n$.}
\revth{Thus, the total \emph{demand-related} utility of a contract user $n$ (denoted by $\wc_n$) has exactly the same form as his actual payment in the contract, that is:}
\begin{equation}\label{eq:contract-welfare}
\begin{aligned}
\textstyle
\wc_n  =   B_n  - \mathbb{P}( d_n  ,D_n) =  B_n  - \mathbb{P}\Big(\sum_{k=1}^K\sum_{t=1}^T \ac_{n,kt}\cdot \xi_{kt},\ D_n\Big)  ,
\end{aligned}
\end{equation}
where $\ac_{n,kt} \in \{0,1\}$ indicates whether to allocate a spectrum $\ch_{kt}$ to SU $n$, $d_n \triangleq \sum_{k=1}^K\sum_{t=1}^T \ac_{n,kt} \cdot \xi_{kt}$ is the total number of \textit{idle} spectrums allocated to the SU $n$, and $\mathbb{P}( d_n  ,D_n)$ is the potential utility loss of the SU caused by the unmet demand, that is:
\begin{equation}\label{eq:contract-welfare-loss}
\begin{aligned}
\mathbb{P}( d_n  ,D_n) = (1-J_n) \cdot [D_n-d_n]^+ \cdot \widehat{P}_n + J_n \cdot \mathbf{1}(D_n-d_n) \cdot \widehat{B}_n.
\end{aligned}
\end{equation}
which has the same form as the SR's potential penalty defined in  (\ref{eq:penalty-xxxx}).


\revth{Second, similar as spot market users, a contract user's utility is also related to the valuation that he or she can achieve on the allocated spectrum (or equivalently, the quality of the allocated spectrum).
The reason is following: to achieve
a desirable quality-of-service (QoS), a contract user needs to pay different levels of efforts (e.g.,
transmission powers) on spectrums with different qualities.
For example, he may achieve the desirable QoS with a lower effort on a spectrum with a higher quality, hence lead to a higher utility.
Such a utility is referred to as the \emph{quality-based} utility.
Similar as that of spot market users, the total \emph{quality-based} utility of a contract user $n$ (denoted by $\wcx_n$) can be defined as the total valuation achieved:}
\begin{equation}
\label{eq:contract-welfare-u}
\begin{aligned}
\textstyle
\wcx_n =
 \sum_{k=1}^K\sum_{t=1}^T \ac_{n,kt} \cdot u_{n,kt} \cdot \xi_{kt}.
\end{aligned}
\end{equation}

\revth{To capture the above two factors in a contract user's utility, we define the utility of contract user $n$ as a weighted sum of the demand-related utility $\wc_n$  and the quality-based utility $\wcx_n$, i.e.,}
$$
 \tau_n \cdot \wc_n  + (1-\tau_n) \cdot \wcx_n  ,
 $$
where $\tau_n \in [0,1]$ is the factor weighting $\wc_n $ and $\wcx_n$.
Such a utility definition is flexible
in capturing contract users' different preferences regarding spectrum demand and spectrum quality.

\revth{The {spectrum secondary utilization efficiency} (or \textit{spectrum efficiency}) is the aggregate utility of all SUs, also called \emph{social welfare}. Specifically, we have the following:}
\begin{definition}[{Spectrum Efficiency}]\label{def:spectrum-efficiency}
The {spectrum efficiency} (or social welfare) is
\begin{equation}\label{eq:spectrum-efficiency}
W \triangleq \sum_{m=1}^M \ws_m + \sum_{n=1}^N \big( \tau_n \cdot \wc_n  + (1-\tau_n) \cdot \wcx_n \big),
\end{equation}
given any spectrum allocation $\{a_{i,kt}\}_{i\in \M\bigcup \N, k\in\{1,...,K\}, t\in \{1,...,T\}}$ that is feasible.\footnote{A spectrum allocation is feasible if it satisfies the spatial interference constraint (SIC) in Definition \ref{def:spatial-interference-cond}.}
\end{definition}

Based on the above, the spectrum efficiency maximization (SEM) problem can be defined as follows: finding a feasible spectrum allocation $\{a_{i,kt}\}_{i\in \M\bigcup \N, k\in\{1,...,K\}, t\in \{1,...,T\}}$, such that the spectrum efficiency $W$ defined in (\ref{eq:spectrum-efficiency}) is maximized.

%% file: Section2-ProbFormu-OR.tex
\section{Problem Formulation}\label{sec:prob}

We first formulate the spectrum efficiency maximization (SEM) problem under deterministic (complete) network information (Section \ref{sec:prob-1}), which is essentially a \textit{matching} problem, and can be solved by many existing algorithms.
Then we formulate the expected spectrum efficiency maximization (E-SEM) problem under stochastic network information (Section \ref{sec:prob-2}), which is an \textit{infinite-dimensional} optimization problem.
We will focus on solving the E-SEM problem in Section \ref{sec:symm}.




\subsection{Problem Formulation under Deterministic Information}\label{sec:prob-1}

With complete and deterministic network information, i.e., $\{\thxikt \}_{k\in\{1,...,K\},t\in\{1,...,T\}}$, the spectrum efficiency maximization (SEM) problem can be formulated as a \textit{matching} between all spectrums and all SUs, and the optimal solution explicitly specifies the allocation of every spectrum to a particular set of (independent) SUs. 
Let $\as_{m,kt} \in \{0,1\}$ (or $\ac_{n,kt} \in \{0,1\}$)  indicate whether to allocate a spectrum $\ch_{kt}$ to spot market user $m$ (or contract user $n$).
Denote:
$$
\textstyle
\veca_{kt} \triangleq \big(\as_{1,kt},  ...,\as_{M,kt};\
\ac_{1,kt},...,\ac_{N,kt}
\big)
$$
as the allocation (vector) of spectrum $\ch_{kt}$, which contains the allocation of spectrum $\ch_{kt}$ to every SU.
An \textit{allocation strategy} consists of the allocations of all
spectrums in the whole time period, denoted by $\Adet \triangleq  \{\veca_{kt}  \}_{k\in\{1,...,K\},t\in\{1,...,T\}}$.
Intuitively, the allocation strategy $\Adet$ is a mapping from every particular spectrum $\ch_{kt}$ to an allocation vector $\veca_{kt} $.


By the Spatial Interference Constraint (SIC) given in Definition \ref{def:spatial-interference-cond}, an allocation strategy $\Adet $ is feasible, if and only if  the allocation $\veca_{kt}$ for every spectrum $\ch_{kt}$ satisfies:
\begin{equation}\label{eq:feasible-discrete}
\textstyle
\mathtt{SIC}:~~~~ \sum_{m \in \CQ_i\bigcap \M } \as_{m,kt}
+\sum_{n \in \CQ_i\bigcap \N } \ac_{n,kt}
\leq \xi_{kt}, \quad \forall \CQ_i \in \CQ(\G),
 \end{equation}
where $\CQ_i$ is the $i$-th clique of graph  $\G$, and $\CQ_{i}^1 \triangleq \CQ_i\bigcap \M$ and $\CQ_{i}^2 \triangleq \CQ_i\bigcap \N$ denote the spot market user set and contract user set in $\CQ_i$, respectively.
It is easy to see that $\as_{m,kt} = \ac_{n,kt} \equiv 0,\ \forall m,n$, if $\xi_{kt} = 0$ (i.e., $\ch_{kt}$ is not idle).
Intuitively, (\ref{eq:feasible-discrete}) states that (i) only idle spectrums can be used by SUs, and (ii) every idle spectrum can be allocated to, at most, one SU in a clique.

According to the definition of spectrum efficiency in Definition \ref{def:spectrum-efficiency} and the SIC in (\ref{eq:feasible-discrete}), the spectrum efficiency maximization (SEM) problem can be formally defined as follows:
\begin{equation}\label{eq:obj-comp-discrete}
\begin{aligned}
\mathtt{SEM}:~~~~ \Adet^* & = \arg \max_{\Adet}\ W = \arg \max_{\Adet}\ \sum_{m=1}^M \ws_m + \sum_{n=1}^N \Big( \tau_n \cdot \wc_n  + (1-\tau_n) \cdot \wcx_n \Big) \\
s.t. &\textstyle
\quad \as_{m,kt} \in \{0,1\},\ \ac_{n,kt} \in \{0,1\},\ _{ \forall m\in \M,\ \forall n\in \N,\ \forall k\in \{1,...,K\},\ \forall t \in \{1,...,T\},}\\
&\textstyle
\quad \mathtt{SIC} \mbox{ defined in } (\ref{eq:feasible-discrete}),\ _{\forall k\in \{1,...,K\},\ \forall t \in \{1,...,T\},}\\
& \textstyle
\quad D_n \geq d_n ,\ _{\forall n\in \N,}
\end{aligned}
\end{equation}
where  $d_n \triangleq \sum_{k=1}^K\sum_{t=1}^T \ac_{n,kt} \cdot \xi_{kt}$ is the number of spectrums allocated to contract user $n$. The last constraint in (\ref{eq:obj-comp-discrete}) states that none of the contract users will get spectrums more than his or her demand. The reason for this  is that over-allocation brings trivial welfare gain for contract users.
It is easy to see that the above problem is an 0-1 integer linear programming (also called matching problem),
and many algorithms, e.g., branch-and-bound algorithm and Kuhn-Munkres algorithm (see \cite{Munkres1957}), have been developed to solve a matching problem effectively. Due to space limitations, we omit the details here. 

\subsection{Problem Formulation under Stochastic Information} \label{sec:prob-2}

In practice, the assumption of complete and deterministic information is often unreasonable, as the network changes randomly over time.
Hence, in this section we consider the case in which the network information is stochastic and the SR only knows its distribution, but not realization.
In this case, we formulate the \rev{\textit{expected} spectrum efficiency maximization (E-SEM) problem} based on the \revv{stochastic} distribution (e.g., PDF or CDF) of network information.\footnote{\rev{This stochastic network information can be obtained empirically after counting the realized network information \rev{over a sufficiently long time period}. Note that in \cite{Techrpt}, 
we also study the problem when this stochastic distribution information is \textit{unavailable}. In that case, we propose a learning mechanism, which converges to the optimal solution without stochastic information.}}
The  solution specifies the allocation of any spectrum under every possible information realization (rather than the explicit allocation of every particular spectrum).
In this sense, it essentially defines a contingency plan (also called a policy) for allocating every spectrum under every possible information realization.~~~~~~

Based on the above, we can define the \textit{allocation strategy} as a mapping from every possible information realization (rather than every particular spectrum, as under deterministic information) to an allocation vector consisting of the allocation probability to every SU. However, as shown in \cite{Techrpt}, such a primitive definition is not appropriate due to the high complexity in determining the SIC constraint.

To this end, we propose a new definition for the allocation strategy.
The basic idea is to \textit{consider each independent SU set as a virtual player, and decide the spectrum allocation among the virtual players}.
That is, we transform the primitive allocation strategy, regarding the allocation probability (of every information realization) to every SU, into an {equivalent strategy}, regarding the allocation probability to every independent SU set.

Denote $\IS_i \in \IS(\G)$ as the $i$-th independent set of the graph $\G $, and $I \triangleq |\IS(\G )|$ as the total number of independent sets in the graph $\G $.
Let $ \ais_i(\thxi) \in [0,1]$ denote the allocation probability of a spectrum $\{\thxi\}$ to independent SU set $\IS_i$.
The allocation probability vector of spectrum $\{\thxi\}$ is:
$$
\textstyle
\vecais(\thxi) \triangleq \big(
\ais_{1}(\thxi),\ais_{2}(\thxi),...,\ais_{I}(\thxi) \big),
$$
and the new allocation strategy  can be formally defined as follows.
\begin{definition}[{Allocation Strategy}]\label{def:alloc-strategy-is}
An allocation strategy is a mapping from every information realization $\{\thxi\}$ to a vector $\vecais (\thxi)$ consisting of the allocation probability of a spectrum $\{\thxi\}$ to every independent SU set, denoted by $\Ais  \triangleq  \{ \vecais  (\thxi) \}_{  \theta \in \THETA, \xi\in\{0,1\} }$.
\end{definition}

Since the SUs in an independent set can use the same spectrum simultaneously, we can easily find that strategy $\Ais$ is feasible, if the allocation  $\vecais(\thxi)$ for every spectrum $\{\thxi\}$ satisfies:
\begin{equation}\label{eq:feasible-i}
\textstyle
  \sum_{i = 1}^I \ais_i (\thxi) \leq \xi.
\end{equation}
That is, the condition in (\ref{eq:feasible-i})
is \textit{sufficient} for a feasible allocation strategy.
The next question is whether it is able to represent all feasible allocations.
The answer is YES, as shown below.

\begin{proposition}\label{prop:feasible}
Any feasible allocation can be represented by an allocation strategy (defined in Definition \ref{def:alloc-strategy-is}) subject to the condition in (\ref{eq:feasible-i}).
\end{proposition}

For detailed proof, please refer to \cite{Techrpt}.
Proposition \ref{prop:feasible} implies that the condition in (\ref{eq:feasible-i}) is an effective SIC constraint.
%
From Proposition \ref{prop:feasible}, we can see that the main advantage of such a transformation is that we can systematically identify the necessary and sufficient conditions of a feasible allocation strategy (i.e., the SIC constraint), which is essential for the E-SEM problem formulation.
Certainly, this will bring some challenges, among which the most critical one is that the problem of finding all independent sets is NP-hard. This motivates our later study on the approximate algorithm and the associated performance loss analysis.

Next we can provide the E-SEM problem formulation based on the new allocation strategy.
Let $l_n^i\in \{0,1\}$ (or $l_m^i\in \{0,1\}$) indicate whether a contract user $n$ (or spot market user $m$) belongs to the $i$-th independent set, i.e., $l_n^i=1$ if $n \in \IS_i$, and $l_n^i=0$ otherwise. Then, the total allocation probability of spectrum $\{\thxi\}$ to contract users $n$ or spot market user $m$ can be written as:
\begin{equation}\label{eq:anam}
\textstyle
  \as_m (\thxi) = \sum_{i=1}^I l_m^i\cdot \ais_i(\thxi) \quad \mbox{ and } \quad \ac_n (\thxi) = \sum_{i=1}^I l_n^i\cdot \ais_i(\thxi).
\end{equation}
Given a feasible allocation strategy $\Ais $, the expected welfare generated by a spot market $m$ is:
\begin{equation}\label{eq:ER0}
\begin{aligned}
\mathbb{E}[\ws_m]  & = \sum_{k=1}^K \sum_{t=1}^T \sum_{\xi_{kt} = 0}^1 f_{\boldsymbol{\xi}} (\xi_{kt}) \cdot \xi_{kt}  \int_{\theta_{kt}}
  \as_m(\thxikt)  \cdot v_m(\theta_{kt} ) \cdot f_{\THETA}(\theta_{kt}) \mathrm{d} \theta_{kt}
  \\
& = S \cdot \sum_{\xi=0}^1 f_{\boldsymbol{\xi}} (\xi ) \cdot \xi \int_{\theta }
   \as_m(\thxi )  \cdot v_m(\theta ) \cdot  f_{\THETA}(\theta ) \mathrm{d} \theta  \\
 &= \rho   S\cdot \int_{\theta} \as_m(\theta ,1) \cdot v_m(\theta ) \cdot f_{\THETA}(\theta) \mathrm{d} \theta,
\end{aligned}
\end{equation}
where $ v_m(\theta ) $ is the spot market user $m$'s utility (i.e., the $m$-th element) in the utility vector $\theta = (v_1,...,v_M;$ $u_1,...,u_N)$,
and $f_{\THETA}(\theta) $ is the joint $\textrm{PDF}$ of $\THETA$, i.e., $f_{\THETA}(\theta)\triangleq  \prod_{m=1}^M f_{\boldsymbol{v}_m }(v_m )
\prod_{n=1}^N f_{\boldsymbol{u}_n }(u_n ) $.
The second equation   follows because  $\boldsymbol{\xi}$ and $\THETA$ are $ \textrm{i.i.d.} $ across   time and  channels. The third equation follows because (i) $f_{\boldsymbol{\xi}} (\xi) = 1-\rho$ if $\xi=0$ and (ii) $f_{\boldsymbol{\xi}} (\xi) = \rho$ if $\xi=1$. {Note that we write $ \mathrm{d}v_1  ...\mathrm{d}v_M \mathrm{d}u_1 ...\mathrm{d}u_N$ as $\mathrm{d}\theta$  and $\int_{v_1 }  ...\int_{v_M } \int_{u_1 } ...\int_{u_N }$ as $\int_{\theta}$ for convenience.}

The expected number of spectrums allocated to a contract user $n$ is:\footnote{\revth{Here we impliedly use the time-average value to approximate the expected value. This is available due to the ergodicity of network information and allocation strategy.}}
\begin{equation}\label{eq:EDn}
\begin{aligned}
\mathbb{E}[d_n] & = \sum_{k=1}^K \sum_{t=1}^T  \sum_{\xi_{kt} = 0}^1 f_{\boldsymbol{\xi}} (\xi_{kt})   \cdot \xi_{kt}   \int_{\theta_{kt}}
 \ac_n(\thxikt)   \cdot f_{\THETA}(\theta_{kt}) \mathrm{d} \theta_{kt} \\
& = \rho   S\cdot \int_{\theta} \ac_n(\theta,1 ) \cdot f_{\THETA}(\theta) \mathrm{d} \theta .
\end{aligned}
\end{equation}

Then, the expected demand-related utility generated by a contract user $n$ is:\begin{equation}\label{eq:ERn}
\mathbb{E}[ \wc_n ] = B_n  - \mathbb{P}(\mathbb{E}[d_n] ,D_n),
\end{equation}
where $\mathbb{P}(\mathbb{E}[d_n] ,D_n)$ is defined in (\ref{eq:contract-welfare-loss}).
Note that the above utility formula is based on the expected number of spectrums that the contract user $n$ obtains, rather than the explicit number as in (\ref{eq:contract-welfare}).
The major motivation for this assumption is  discussed in Section \ref{sec:model:contract}.

The expected quality-based utility generated by a contract user $n$ is:
\begin{equation}\label{eq:ERn-social}
\begin{aligned}
\mathbb{E}[\wcx_n]  &   = \sum_{k=1}^K \sum_{t=1}^T \sum_{\xi_{kt} = 0}^1 f_{\boldsymbol{\xi}} (\xi_{kt})\cdot \xi_{kt}   \int_{\theta_{kt}}
  \ac_n(\thxikt)  \cdot u_n(\theta_{kt} ) \cdot f_{\THETA}(\theta_{kt}) \mathrm{d} \theta_{kt} \\
& = \rho   S\cdot \int_{\theta} \ac_n(\theta,1 ) \cdot u_n(\theta ) \cdot f_{\THETA}(\theta) \mathrm{d} \theta,
\end{aligned}
\end{equation}
where $ u_n(\theta ) $ denotes the contract user $n$'s valuation in the vector $\theta= (v_1,...,v_M;$ $u_1,...,u_N)$.



The expected spectrum efficiency (expected social welfare) is the expected aggregate utility generated by all SUs, that is:
$$
\textstyle
\mathbb{E}[W] \eq  \sum_{m=1}^M \mathbb{E}[\ws_m] + \sum_{n=1}^N  \big( \tau_n \cdot \mathbb{E}[\wc_n] + (1-\tau_n) \cdot \mathbb{E}[\wcx_n]  \big).
$$
Therefore, we have the following expected spectrum efficiency maximization (E-SEM) problem:
\begin{equation}\label{eq:obj-incomp}
\begin{aligned}
\mathtt{E}\mbox{-}\mathtt{SEM}:~~~~
\Ais^* & = \arg \max_{\Ais}\ \mathbb{E}[W] =\arg \max_{\Ais}\  \sum_{m=1}^M \mathbb{E}[\ws_m] + \sum_{n=1}^N \Big( \tau_n \cdot \mathbb{E}[\wc_n] + (1-\tau_n) \cdot \mathbb{E}[\wcx_n] \Big) \\
s.t. &\textstyle
\quad \ais_{i}(\thxi ) \in [0,1],\ \forall \theta \in \THETA , \forall \xi \in \{0,1\}, \forall i \in \{1,...,I\},\\
&\textstyle
\quad \sum_{i = 1}^I \ais_i (\thxi ) \leq \xi,\ \forall \theta \in \THETA ,\forall \xi \in \{0,1\}, \qquad \mathtt{(SIC)}\\
& \textstyle
\quad D_n \geq \mathbb{E}[d_n],\ {\forall n\in \N.}
\end{aligned}
\end{equation}

Obviously, the  E-SEM problem (\ref{eq:obj-incomp}) is an \textit{infinite-dimensional} optimization problem (see \cite{Fattorini1999}), in which the solution (or strategy) is not a number or a vector, but rather a continuous quantity.
By the second  constraint (SIC), we immediately have: $\ais_i (\theta, \xi) \equiv 0$ if $\xi=0$, $\forall \theta\in \THETA, \forall i = 1,...,I$.
Therefore, in the rest of this paper, we will focus on solving $\ais_i (\theta, 1), \forall \theta\in \THETA, \forall i = 1,...,I$, i.e., the allocation of idle spectrums, denoted by $\Aisi \triangleq \{\vecais(\theta, 1) \}_{ \theta \in \THETA}$.
\textit{For notational convenience, we will write the tuple $\{\theta, 1\}$ as $\theta$, and thus denote $\Aisi \triangleq \{\vecais(\theta) \}_{ \theta \in \THETA}$.}


%% file: Section3-Symmetry-OR.tex

\section{Off-line Optimal Policy}\label{sec:symm}

%

In this section, we solve the E-SEM problem given in  (\ref{eq:obj-incomp}).
We first reduce the strategy space by removing some  \textit{irrelevant} strategies, so as to reduce the computational complexity.\footnote{The ``irrelevant strategies'' denote those that never emerge in an optimal solution.
Therefore, ignoring the irrelevant strategies does not affect the optimality of the solution. We will show this in Proposition \ref{prop:restriction}.}
Then we derive the optimal solution systemically using the primal-dual method.
Due to space limits, we present all proofs in \cite{Techrpt}.

\subsection{Allocation Strategy Reduction}\label{sec:incomp-reduction}

Now we simplify the E-SEM problem (\ref{eq:obj-incomp}) by restricting our attention to a particular subset of the   strategy space.
Define the \textit{weight} of each vertex in the graph $\G$ as the utility of the associated SU.
Notice that the welfare generated by a spot market user is exactly his or her utility, and thus we have: if an idle spectrum $ \theta$ (or spectrum $\{\theta, 1\}$) is allocated to an independent contract user set $\boldsymbol{n} \in \IS(\Gc)$, it must be allocated to an MWIS of the side market $\Gs_{\boldsymbol{n}}$ at the same time.
This observation helps us to eliminate  some strategies without affecting the optimality of the solution.

Let $\IS_{i}^{1} \eq \IS_i \bigcap \mathcal{M}$ and $\IS_{i}^{2} \eq \IS_i \bigcap \mathcal{N}$
denote the spot market user set and contract user set in  $\IS_i$ (i.e., the $i$-th independent set of graph $\G$), respectively.
Formally, we have:


\begin{proposition}\label{prop:restriction}
Suppose $\Aisi^* =\{\vecais^*(\theta)\}_{\theta\in \THETA} $, where $\vecais^*(\theta) = \big( {\ais_1}^* (\theta) , ...,  {\ais_I}^* (\theta)\big) $, is an optimal solution of  (\ref{eq:obj-incomp}). Then for any $\theta\in \THETA$ we have:
for every independent set $\IS_i  \in \IS(\G)$,
 \begin{itemize}
\item[\textit{(a)}] If $\IS_{i}^{1} $ is \textit{not} an MWIS of the graph $\Gs_{\IS_{i}^{2}}$ (i.e., the side market (see Definition \ref{def:subg-side}) of the contract user set $\IS_{i}^{2}$), then  ${\ais_i}^* (\theta) = 0$; or equivalently,
\item[\textit{(b)}] If ${\ais_i}^* (\theta)> 0$, then $\IS_{i}^{1} $ must be an MWIS of the graph $\Gs_{\IS_{i}^{2}}$.
 \end{itemize}
 \end{proposition}

%

\yyy{Proposition \ref{prop:restriction} implies that in an optimal solution, an idle spectrum $\theta$ will \textit{never} be allocated to an independent set $\IS_i = \IS_{i}^{1} \bigcup \IS_{i}^{2}$ where $\IS_{i}^{1}$ is \textit{not} an MWIS of the graph $\Gs_{\IS_{i}^{2}}$.
We refer to such an independent set $\IS_i$ as an \textit{irrelevant independent set}.
Obviously, we can   ignore those allocation strategies 
that allocate spectrums to the irrelevant independent sets, and focus only on the allocation among the ``relevant'' independent sets $\IS_i$ (where $\IS_{i}^{1}$ is an MWIS of the graph $\Gs_{\IS_{i}^{2}}$).
In effect, we can restrict our attention to the independent  sets of the futures market graph $\Gc$, instead of the independent sets of the whole graph $\G$: \textit{if a spectrum is allocated to an independent contract user set (say $\boldsymbol{n} $) of $\Gc$, it must be allocated to an MWIS of the side market graph $\Gs_{\boldsymbol{n}}$ at the same time.}
Consider the example in Fig.~\ref{system-illu}. Any idle spectrum will \emph{not} be allocated to the independent set $\IS_i = \{1,\ 6\}$, since $\{6\}$ is not an MWIS of $\Gs_{\{1\}}$; otherwise, we can immediately increase the total welfare by reallocating the spectrum to another independent set $\IS_j = \{1,\ 6,\ 8\}$.}


The above result allows us to simplify the E-SEM problem.
Denote $\ISc_i \in \IS(\Gc)$ as the $i$-th independent contract user set of the futures market graph $\Gc$, and $\Ic \triangleq |\IS(\Gc )|$ as the total number of independent sets of the graph $\Gc$.
Let $ \aic_i(\theta) \in [0,1]$ denote the allocation probability of an idle spectrum $ \theta $ to the $i$-th independent contract user set $\ISc_i$.
\footnote{Here the superscript ``$\textsc{ISc}$'' in $\aic_i(\theta)$ is used to distinguish the allocation probability for each independent contract user set (of the graph $\Gc$) from those for each individual SU or each independent set of graph $\G$.}
Formally, we can define the \textit{reduced} allocation strategy as $ \Aic \triangleq  \{ \vecaic (\theta) \}_{\theta \in \THETA}$, where
$$
 \vecaic (\theta) \triangleq \big(
\aic_{0}(\theta),\aic_{1}(\theta),...,\aic_{\Ic}(\theta) \big)
$$
is the reduced allocation probability vector of an idle spectrum $ \theta $, consisting of the allocation probability of spectrum $\{\theta ,1\}$ to every \textit{independent contract user set} of the graph $\Gc $.
Note that the 0-th independent contract user set is simply defined as the empty set, i.e.,  $\ISc_0 = \emptyset$.

Similar to (\ref{eq:feasible-i}), a reduced allocation strategy $\Aic$ is feasible, if and only if for every information realization $\{\theta, 1\}$, the reduced allocation probability vector $\vecaic(\theta)$ satisfies:
\begin{equation}\label{eq:feasible-ic}
\textstyle
\texttt{SIC}:~~~~  \sum_{i = 0}^{\Ic} \aic_i (\theta) \leq 1.
\end{equation}


By Proposition \ref{prop:restriction}, if an idle spectrum $\theta$ is allocated to the $i$-th independent contract user set $\ISc_i$, it must be allocated to an MWIS  of the side market graph $\Gs_{\ISc_i}$ (denoted by $\ISs_i$) at the same time.
Let $z_i(\theta)$ denote the total weight of $\ISs_i$. Obviously, $z_i(\theta)$ is the maximum total welfare generated by spot market users in the side market $\Gs_{\ISc_i}$ (over an idle spectrum $ \theta $), referred to as the \textit{side welfare} of $\ISc_i$.\footnote{Here the prefix ``side'' follows the fact that such a welfare is generated by spot market users along with the usage of contract user set $\ISc_i$, i.e., it is a side effect of the allocation of contract user set $\ISc_i$.}
Thus, given a feasible strategy $\Aic$, the expected side welfare of $\ISc_i$ is:
\begin{equation}\label{eq:EZk-11}
\begin{aligned}
\mathbb{E}[W^{\textsc{s}}_i]  &  = \sum_{k=1}^K \sum_{t=1}^T \sum_{\xi_{kt} = 0}^1 f_{\boldsymbol{\xi}} (\xi_{kt}) \cdot \xi_{kt}   \int_{\theta_{kt}}
  \aic_i(\thxikt)  \cdot z_i(\theta_{kt})  \cdot f_{\THETA}(\theta_{kt}) \mathrm{d} \theta_{kt}
  \\
 & \textstyle = \rho   S\cdot \int_{\theta}
  \aic_i(\theta)  \cdot z_i(\theta) \cdot f_{\THETA}(\theta) \mathrm{d} \theta.
\end{aligned}
\end{equation}

The expected social welfare generated by the spot market is the sum of the expected side welfare of all independent contract user sets (including the empty set), that is:
\begin{equation}\label{eq:EZk-1}
\begin{aligned}
\textstyle
\mathbb{E}[W^{\textsc{s}}] = \sum_{i=0}^{\Ic} \mathbb{E}[W^{\textsc{s}}_i]   =  \sum_{i=0}^{\Ic} \Big(  \rho   S\cdot \int_{\theta}
  \aic_i(\theta)  \cdot z_i(\theta) \cdot f_{\THETA}(\theta) \mathrm{d} \theta\Big).
\end{aligned}
\end{equation}

Similar to (\ref{eq:anam}), the allocation probability of an idle spectrum $\theta$ to a contract users $n$ is:
\begin{equation}
\textstyle
\ac_n (\theta) = \sum_{i=0}^{\Ic} l_n^i\cdot \aic_i(\theta) = \sum_{i=1}^{\Ic} l_n^i\cdot \aic_i(\theta).
\end{equation}
The 2nd equality follows because $l_n^0 \equiv 0$, $\forall n \in \mathcal{N}$ (since $\ISc_0 = \emptyset$).
Furthermore, the expected number of spectrums allocated to contract user $n$ (i.e., $\mathbb{E}[d_n]$) and the expected utility generated by contract user $n$ (i.e., $\mathbb{E}[ {w}_n]$ and $\mathbb{E}[\widetilde{w}_n]$) are given by (\ref{eq:EDn})-(\ref{eq:ERn-social}) directly.
The expected spectrum efficiency is:
\begin{equation}\label{eq:EUA-1}
\begin{aligned}
\textstyle
\mathbb{E}[W]  =  \sum_{i=0}^{\Ic} \mathbb{E}[W^{\textsc{s}}_i]   + \sum_{n=1}^N \Big( \tau_n \cdot \mathbb{E}[\wc_n] + (1-\tau_n) \cdot \mathbb{E}[\wcx_n] \Big),
\end{aligned}
\end{equation}

Based on the reduced allocation strategy $\Aic$ and the associated expected spectrum efficiency formula in (\ref{eq:EUA-1}), we can rewrite the E-SEM in (\ref{eq:obj-incomp}) into the following equivalent form:
\begin{equation}\label{eq:obj-incomp-1}
\begin{aligned}
\mathtt{E}\mbox{-}\mathtt{SEM(EQ)}:~~~~
\Aic^* & = \arg \max_{\Aic } \ \mathbb{E}[W] = \arg \max_{\Aic } \ \sum_{i=0}^{\Ic} \mathbb{E}[W^{\mathrm{s}}_i]  + \sum_{n=1}^N \Big( \tau_n \cdot \mathbb{E}[\wc_n] + (1-\tau_n) \cdot \mathbb{E}[\wcx_n] \Big)   \\
s.t. &\textstyle
\quad \mbox{(i)} \quad \aic_{i}(\theta) \in [0,1],\ \forall \theta \in \THETA ,\forall i \in \{0,1,...,{\Ic}\},\\
&\textstyle
\quad \mbox{(ii)} \quad \sum_{i = 0}^{\Ic} \aic_i (\theta ) \leq 1,\ \forall \theta \in \THETA ,\qquad \mathtt{(SIC)}\\
& \textstyle
\quad \mbox{(iii)} \quad D_n \geq \mathbb{E}[d_n],\ {\forall n\in \N.}
\end{aligned}
\end{equation}

Notice that the independent sets in the futures market graph $\Gc$ are much less than in the whole graph $\G$, since the number of independent sets increases exponentially with the number of vertices.
Thus, by (\ref{eq:obj-incomp-1}), we can greatly reduce the complexity of the E-SEM problem in (\ref{eq:obj-incomp}).
Recall the example in Fig.~\ref{system-illu}, there are a total of 26 independent sets in graph $\G$, while only $6$ in graph $\Gc$.

\subsection{Necessary Conditions}\label{sec:incomp-nece}

For exposition convenience,  {we will omit the superscript ``$ {\textsc{Isc}}$'' in $ {\aic_i(\theta)}$, as long as there is no confusion caused}.
Suppose $\Aic^* = \{ \vecaic^*(\theta) \}_{\theta \in \THETA} $ where $\vecaic^*(\theta)=
\left( {a}_0^*(\theta),  {a}_1^*(\theta), ...,  {a}_{\Ic}^*(\theta)\right)$ is the optimal solution to  (\ref{eq:obj-incomp-1}). The following necessary conditions hold.

\begin{lemma}[Necessary Condition I]\label{lemma:nc1}
For any $\theta \in \THETA$, we have:
$\sum_{i=0}^{\Ic} {a}_i^*(\theta) = 1$.
\end{lemma}

Intuitively, Lemma \ref{lemma:nc1} states that every idle spectrum will be allocated to SUs. This is because it can always produce certain positive utility when used by some SUs, e.g., spot market users.

\begin{lemma}[Necessary Condition II]\label{lemma:nc2}
For any contract user $n $, we have:
$\mathbb{E}[d_n]  \leq D_n$.
\end{lemma}

Intuitively, Lemma \ref{lemma:nc2} states that none of the contract users will get spectrums (on average) more than his or her demand.
This is because spectrum over-allocation to a contract user brings trivial utility gain for the contract user (see the definition of contract user's utility in Eq.~[\ref{eq:ERn}]).

%


\subsection{Optimal Solution}
\label{sec:incomp2}

We focus on the optimal solution for soft contracts (with unit utility loss for every under-allocated spectrum). The analysis for hard contracts (with a fixed total utility loss for the unsatisfaction in demand) is similar and shown in \cite{Techrpt}. 


By Lemma \ref{lemma:nc1}, we can replace the decision variable $a_0(\theta)$ by $a_0(\theta) = 1-\sum_{i=1}^{\Ic} {a}_i(\theta)$.
By Lemma \ref{lemma:nc2}, we can rewrite the potential welfare loss in (\ref{eq:contract-welfare-loss}) as
$\mathbb{P}( d_n  ,D_n) = (D_n- \mathbb{E}[d_n] ) \cdot \widehat{P}_n$ (for soft contracts).
Furthermore, by the 2nd constraint of the problem (\ref{eq:obj-incomp-1}), we can simplify the 1st constraint as $a_i(\theta)\geq 0$, $\forall i$.
Thus, the optimization problem (\ref{eq:obj-incomp-1}) can be rewritten as:
\begin{equation}\label{eq:obj-incomp2-soft}
\begin{aligned}
\Aicx^* & =
\arg \max_{\Aicx }\  {\textstyle \left(
\rho   S  \int_{\theta} \big({\textstyle 1-\sum_{i=1}^{\Ic}
  a_i(\theta) }\big)    z_0(\theta)    f_{\THETA}(\theta) \mathrm{d} \theta +
\sum_{i=1}^{\Ic} \rho   S  \int_{\theta}
  a_i(\theta)    z_i(\theta)   f_{\THETA}(\theta) \mathrm{d} \theta
\right.}
\\
& \textstyle  \left. \qquad \qquad  + \sum_{n=1}^N \big[ \tau_n \cdot \big( B_n  - (D_n- \mathbb{E}[d_n] )  \cdot \widehat{P}_n \big) +
(1-\tau_n) \cdot \rho   S  \int_{\theta} \ac_n(\theta) u_n(\theta )   f_{\THETA}(\theta) \mathrm{d} \theta \big] \right)
\\
&   = \arg \max_{\Aicx }\ {\textstyle \left( F + \rho S \cdot \sum_{i=1}^{\Ic} \int_{\theta} H_i(\theta) a_i(\theta) f_{\THETA}(\theta) \mathrm{d} \theta\right)},
\\
s.t. &\textstyle \quad \mbox{(i)} \quad  a_i(\theta)\geq 0,\  \forall \theta \in \THETA, \forall i \in \{1,...,{\Ic}\}; \\
&\textstyle \quad \mbox{(ii)}\quad \sum_{i=1}^{\Ic} a_i(\theta) \leq 1,\ \forall \theta\in \THETA;\\
&\textstyle \quad \mbox{(iii)}\quad \mathbb{E}[d_n]   \leq D_n,\ \forall n\in \mathcal{N};
\end{aligned}
 \end{equation}
where 
\begin{itemize}
\itemsep=-1mm
 \item $\Aicx  = \{\vecaicx (\theta)\}_{\theta \in \THETA}$ with $\vecaicx (\theta) = \big(a_1(\theta),...,a_{\Ic}(\theta)\big)$ is the reduced allocation strategy;
  \item $\mathbb{E}[d_n] =\rho   S  \int_{\theta} \ac_n(\theta)   f_{\THETA}(\theta) \mathrm{d} \theta  =  \sum_{i=1}^{\Ic} l_n^i\cdot \rho S \int_{\theta} a_i(\theta) \cdot f_{\THETA}(\theta) \mathrm{d} \theta$ is the expected number of spectrums allocated to the contract user $n$;
 \item $F  = \rho S\cdot \int_{\theta} z_0(\theta) f_{\THETA}(\theta) \mathrm{d} \theta + \sum_{n=1}^N \tau_n \cdot \big(B_n - \widehat{P}_n \cdot   D_n\big)$ is a constant independent of  the allocation;
  \item $H_i(\theta)=-z_0(\theta) +z_i(\theta)  +  \sum_{n=1}^N l_n^i \cdot \big(\tau_n \cdot \widehat{P}_n + (1-\tau_n)\cdot u_n(\theta)\big) $ is an information-specific variable.
\end{itemize}

To analytically solve the optimization problem (\ref{eq:obj-incomp2-soft}), we first study the dual problem of (\ref{eq:obj-incomp2-soft}) using the primal-dual method (see \cite{linear}~for details).
Then we determine the optimal dual variables, from which we further derive the optimal primal solution.

\subsubsection{{Primal-Dual Method}}

We introduce Lagrange multipliers (also called \textit{dual variables} or \textit{shadow prices}) $\mu_i(\theta) $,  $\eta(\theta) $, and $\lambda_n $ for three constraints in (\ref{eq:obj-incomp2-soft}).
The Lagrangian is:
\begin{equation*}\label{eq:obj-incomp2-soft-L}
\begin{aligned}
\textsc{Lag} &  \textstyle   = F + \rho S \cdot \sum_{i=1}^{\Ic} \int_{\theta} H_i(\theta) a_i(\theta) f_{\THETA}(\theta) \mathrm{d} \theta    + \sum_{i=1}^{\Ic}  \int_{\theta} \reve{\mu_i(\theta)}  a_i(\theta) f_{\THETA}(\theta) \mathrm{d} \theta  \\
 & \textstyle   \quad\quad    + \int_{\theta} \reve{\eta(\theta)}   \Big(1 - \sum_{i=1}^{\Ic}  a_i(\theta) \Big) f_{\THETA}(\theta) \mathrm{d} \theta   + \sum_{n=1}^N \reve{\lambda_n}   \Big(D_n - \sum_{i=1}^{\Ic}  l_n^i \cdot \rho S \int_{\theta} a_i(\theta) \cdot f_{\THETA}(\theta) \mathrm{d} \theta \Big), \\
\end{aligned}
\end{equation*}
which can be written as
$\textsc{Lag} \triangleq \int_{\theta} \mathcal{W}(  \theta)   f_{\THETA}(\theta) \mathrm{d} \theta$,
where $ \mathcal{W}( \theta)$, called Sub-Lagrangian, is given by
\begin{equation*}\label{eq:obj-incomp2-soft-LAb}
\begin{aligned}
\mathcal{W}( \theta)
&\textstyle  \triangleq F + \eta(\theta) + \sum_{n=1}^N \lambda_n  D_n
 + \sum_{i=1}^{\Ic} \Big(\rho S\cdot H_i(\theta) +  \mu_i(\theta) - \eta(\theta)  - \rho S \cdot \sum_{n=1}^N l_n^i  \lambda_n 	\Big)\cdot a_i(\theta) .
\end{aligned}
\end{equation*}

\begin{definition}[{Marginal Welfare -- MW}]\label{def:mp}
We define the \textit{marginal welfare} (MW) as the first partial derivative of $ \mathcal{W}( \theta)$ over each independent contract user set's allocation probability $a_i(\theta)$: 
 \begin{equation}\label{eq:def:mp}
\textstyle
   \MW_i( \theta) \triangleq   \frac{\partial  \mathcal{W}( \theta)}{\partial a_i(\theta)} = \rho S \cdot H_i(\theta)  +  \mu_i(\theta)  -  \eta(\theta) - \rho S\cdot \sum_{n=1}^N l_n^i\cdot \lambda_n.
 \end{equation}
 \end{definition}

By Euler-Lagrange conditions for optimality (see \cite{linear}), if an inner optimal solution exists, the optimal solution $\Aicx^*(\theta)$ must satisfy the conditions: $ \MW_i(\theta) = 0,\forall i$; otherwise, the optimal solution occurs at the boundaries.
From Definition \ref{def:mp}, we can see that the MW $ \MW_i( \theta)$ is independent of $\Aicx(\theta)$. Thus, the optimal solution $\Aicx^*(\theta)$ must occur at the boundaries, that is:
 \begin{equation}\label{eq:obj-incomp2-soft-opt}
a_i^*(\theta) = \left\{
\begin{aligned}
0,~~~~ & \textstyle \qquad \MW_i(\theta)<0 \\
1,~~~~ &\textstyle \qquad \MW_i(\theta)>0 \\
\end{aligned}
\right.
 \end{equation}

By (\ref{eq:def:mp}) and (\ref{eq:obj-incomp2-soft-opt}), the optimal solution is determined by the dual variables $\mu_i(\theta), \eta(\theta) $, and $\lambda_n $. By the primal-dual method (\cite{linear}), every dual variable must be nonnegative and is positive only when the associated constraint is tight, i.e., the following dual constraints hold:
\begin{equation*}
 \left\{
 \begin{aligned}
 \mbox{(D.1)} & \textstyle \quad \mu_i^*(\theta)\geq 0,\ a_i^*(\theta)\geq 0,\ \mu_i^*(\theta)\cdot a_i^*(\theta) = 0,\quad  \forall i \in \{1,...,\Ic\}, \forall \theta \in \THETA;
 \\
 \mbox{(D.2)} &\textstyle \quad \eta^*(\theta)\geq 0,\  1 - \sum_{i=1}^{\Ic} a_i^*(\theta)\geq 0,\ \eta^*(\theta)\cdot \big(1 - \sum_{i=1}^{\Ic} a_i^*(\theta) \big)=0, \quad \forall \theta\in \THETA;
  \\
 \mbox{(D.3)} &\textstyle \quad \lambda_n^*\geq 0, \  D_n -  \sum_{i=1}^{\Ic} l_n^i \cdot\rho S \int_{\theta} a_i^*(\theta) \cdot f_{\THETA}(\theta) \mathrm{d} \theta \geq 0,
     \\
   & \textstyle  \quad   \lambda_n^* \cdot \big(D_n - \sum_{i=1}^{\Ic} l_n^i \cdot \rho S \int_{\theta} a_i^*(\theta) \cdot f_{\THETA}(\theta) \mathrm{d} \theta \big)  =0, \quad \forall n\in \mathcal{N}.
 \end{aligned}\right.
\end{equation*}

%
%

By the duality principle, the primal problem (\ref{eq:obj-incomp2-soft}) is equivalent to the problem of finding a set of
optimal dual variables $\mu_i^*(\theta), \eta^*(\theta)$, and $\lambda_n^* $, such that all dual constraints are satisfied (dual problem).
The optimality of
this approach is guaranteed as the objective and constraints in (\ref{eq:obj-incomp2-soft}) are linear, which
implies that (\ref{eq:obj-incomp2-soft}) is a convex optimization problem.

For convenience, we further introduce the concepts of \textit{inner marginal welfare} (I-MW) and \textit{core marginal welfare} (C-MW), which will be frequently used in the later discussion.
 \begin{definition}[{Inner Marginal Welfare -- I-MW}]\label{def:mp1}
 \begin{equation}
\textstyle
  \IMW_i(\theta) \triangleq  \rho S \cdot H_i(\theta)  -  \eta(\theta) - \rho S\cdot \sum_{n=1}^N l_n^i\cdot \lambda_n.
\end{equation}
 \end{definition}
 \begin{definition}[{Core Marginal Welfare -- C-MW}]\label{def:mp2}
 \begin{equation}
\textstyle
  \CMW_i(\theta)   \triangleq  \rho S \cdot H_i(\theta)  - \rho S\cdot \sum_{n=1}^N l_n^i\cdot \lambda_n .
 \end{equation}
 \end{definition}

It is easy to see that $   \IMW_i(\theta)  =   \MW_i( \theta)  - \mu_i(\theta)
$ and $ \CMW_i(\theta) = \MW_i( \theta) -  \mu_i(\theta)  +  \eta(\theta)$. That is, both I-MW and C-MW are variations of MW by ignoring certain non-critical dual variables in the MW. This is also the reason that we refer to  them as ``inner'' and ``core'' marginal welfares.

\vspace{-3mm}

\subsubsection{{Optimal Shadow Prices and Optimal Solution}}

Now we study the optimal shadow prices (dual variables) and optimal primal  solution.

We start by showing the necessary conditions for the optimal  $\mu_i^*(\theta)$. Specifically,

 \begin{lemma}[Feasible Range of $\mu_i^*(\theta)$]\label{lemma:mu-multi}
The optimal shadow prices $\mu_i^*(\theta)$ must satisfy:
\begin{itemize}
\itemsep=-1mm
\item[\emph{(a)}] $\mu_i^*(\theta)=0$, if $\IMW_i( \theta)\geq 0$, and \emph{(b)} $\mu_i^*(\theta)\in[0,\ |\IMW_i( \theta)|]$, if $\IMW_i( \theta)<0$.
\end{itemize}
 \end{lemma}

\begin{figure*}
\hspace{-5mm}
   \begin{minipage}[t]{0.5\linewidth}
    \centering
    \includegraphics[scale=.4]{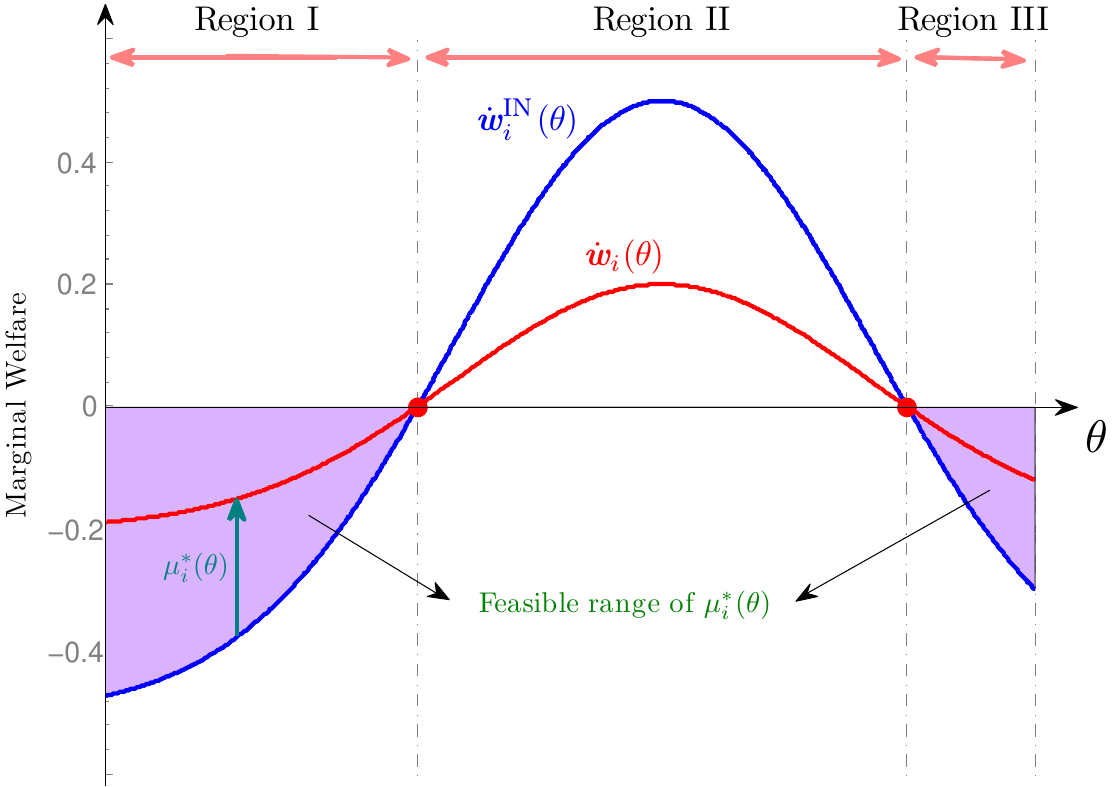}
     \caption{Illustration of $\mu_i^*(\theta)  $ in Lemma \ref{lemma:mu-multi}}
\label{f-mu-eta}
    \end{minipage}
  \begin{minipage}[t]{0.02\linewidth}~
  \end{minipage}
 \begin{minipage}[t]{0.5\linewidth}
    \centering
    \includegraphics[scale=.4]{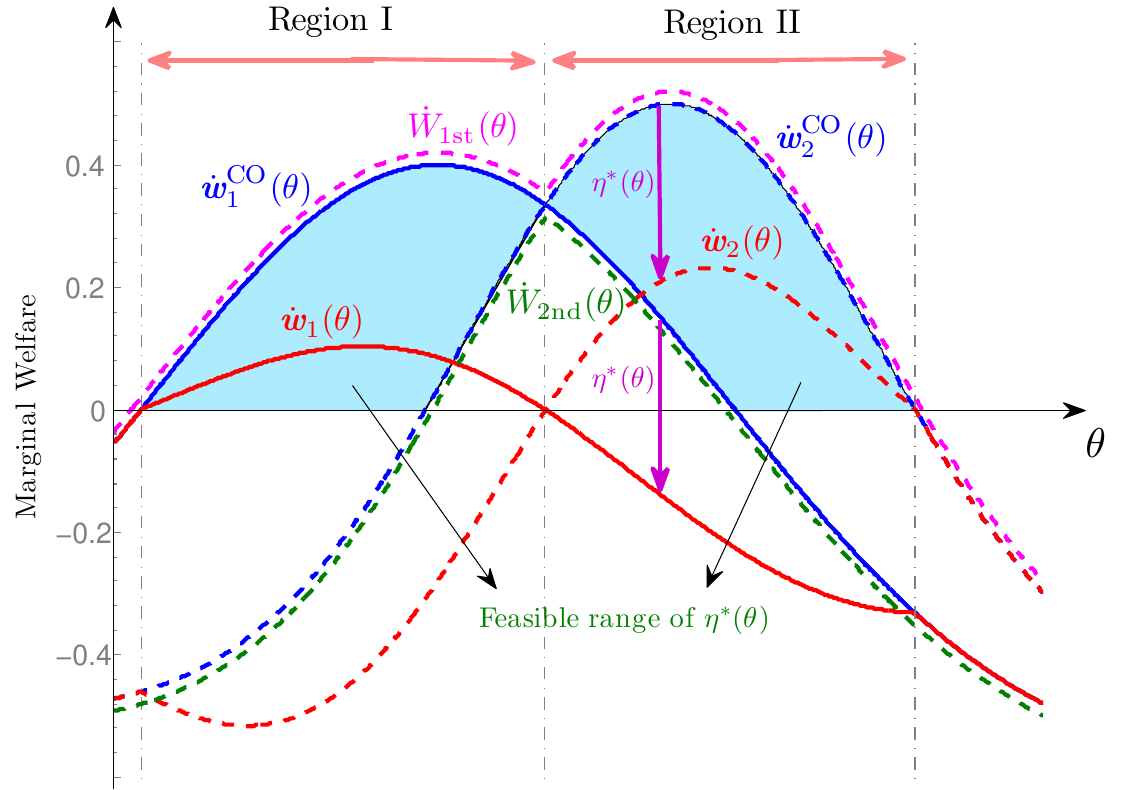}
     \caption{Illustration of $\eta^*(\theta) $ in Lemma \ref{lemma:eta-multi}}
    \label{f-eta-M}
    \end{minipage}
\end{figure*}

Notice that $\MW_i( \theta) = \IMW_i( \theta) + \mu_i^*(\theta)  $. Lemma \ref{lemma:mu-multi} implies that (a) $\IMW_i( \theta)\geq 0$ if and only if $\MW_i( \theta) \geq 0$, and (b) $\IMW_i( \theta) < 0$ if and only if $\MW_i( \theta)   \leq 0$. That is,
$\mu_i^*(\theta)$ never changes the sign of the MW $\MW_i(\theta)$, which implies that $\mu_i^*(\theta) $ has no impact on the optimal solution  (\ref{eq:obj-incomp2-soft-opt}).
Intuitively, this is because the optimal solution given by (\ref{eq:obj-incomp2-soft-opt}) never violates the first constraint (associated with shadow prices $\mu_i (\theta)$) of the primal problem (\ref{eq:obj-incomp2-soft}). 
To better understand Lemma \ref{lemma:mu-multi}, we provide an illustration of $\IMW_i( \theta)$, $\MW_i( \theta)$, and the feasible range of $\mu_i^*(\theta)$ (shown as the shadow area) in Fig.~\ref{f-mu-eta}, in which Region II corresponds to case (a), and Regions I and III correspond to case (b) in Lemma \ref{lemma:mu-multi}.
In Region II, the optimal $\mu_i^*(\theta)$ is $0$, and thus $\MW_i( \theta) = \IMW_i( \theta) \geq 0$.
In Region I or III, the optimal $\mu_i^*(\theta)$ is in
the shadow area, and thus $\MW_i( \theta) =\IMW_i( \theta) + \mu_i^*(\theta) \leq 0$.

Next we show the necessary conditions for the optimal $\eta^*(\theta)$.
Denote:
$$
\CMWmax(\theta) \triangleq \max_{i\in \{1,...,\Ic \} } \CMW_i( \theta)
\mbox{~~and~~}
\CMWsec(\theta) \triangleq \max_{i\in \{1,...,\Ic \}/\{i_1\} }\CMW_i( \theta)
$$
as the highest and second-highest C-MWs among all independent contract user sets, respectively, where $i_1 \triangleq\arg \max_{i\in \{1,...,\Ic \} }\CMW_i( \theta)$.
The feasible range of $\eta^*(\theta)$ is shown as follows.

\begin{lemma}[Feasible Range of $\eta^*(\theta)$]\label{lemma:eta-multi}
The optimal shadow prices $\eta^*(\theta)$ must satisfy:
\begin{itemize}
\itemsep=-1mm
\item[\emph{(a)}] $\eta^*(\theta) \in [\max(0,\CMWsec  (\theta)),\ \CMWmax (\theta)]$, if $\CMWmax ( \theta)>0$, and \emph{(b)} $\eta^*(\theta)=0$, if $\CMWmax (\theta) \leq 0$.
\end{itemize}
 \end{lemma}

\rev{Notice  that $  \MW_i( \theta) =  \CMW_i( \theta)  +   \mu_i^*(\theta) - \eta^*(\theta)$ and $  \IMW_i( \theta) =  \CMW_i( \theta) - \eta^*(\theta)$.
That is, the shadow price $\eta^*(\theta)$ can be viewed as an \textit{identical}
reduction on all C-MWs (at a particular $\theta$).
Lemma \ref{lemma:eta-multi} suggests that a feasible $\eta^*(\theta)$ would lie between $\CMWmax(\theta)$ and $\max(0,\CMWsec(\theta))$,} with which there is, at most, one independent contract user set (the $i_1$-th one $\ISc_{i_1}$, if exists) having {positive I-MW}, i.e., $\IMW_i( \theta) \leq 0$, $\forall i \neq i_1$. We therefore have $\MW_i( \theta) \leq 0 $ and $a_i^*(\theta)=0$, $\forall i\neq i_1$ by Lemma \ref{lemma:mu-multi} and (\ref{eq:obj-incomp2-soft-opt}).
Thus, the coupling constraint (ii)  will be satisfied.
To better understand Lemma \ref{lemma:eta-multi}, we provide an illustration of $\CMW_i( \theta)$, $\MW_i( \theta)$, and  $\eta^*(\theta)$ {(shown as the shadow areas)} in Fig.~\ref{f-eta-M}. {Here we have two independent contract user sets, $i=1,2$.}
The upper dash curve denotes $\CMWmax(\theta)$, the highest C-MW of two independent sets. The lower dash curve denotes $\CMWsec(\theta)$, the second highest (i.e., lowest) C-MW of two independent sets.
\rev{Regions I and II correspond to case (a) where $\CMWmax(\theta)> 0$, and the remaining  regions correspond to case (b) where $\CMWmax(\theta)\leq 0$. It is easy to see that in Region I (or II), the $1$st (or $2$nd) independent set realizes the highest positive C-MW, i.e., $\CMWmax(\theta) = \CMW_1( \theta)>0$ (or $\CMWmax(\theta) = \CMW_2( \theta)>0$), and thus given any feasible $\eta^*(\theta)$ in the shadow area, the 1st (or 2nd) independent set has the unique positive marginal welfare.}

%

By Lemmas \ref{lemma:mu-multi} and   \ref{lemma:eta-multi}, we can find that the optimal solution given by (\ref{eq:obj-incomp2-soft-opt}) is equivalent to such a solution that allocates spectrums to the independent contract user set with the highest and positive C-MW (and meanwhile the associated side market).
Denote $\boldsymbol{i}_1$ $ \triangleq $ $\{i \in \{1,...,\Ic\} \ |\ \CMW_i(\theta)=\CMWmax(\theta)\}$ as the collection of independent  sets with the highest C-MW (over a spectrum $\{\theta, 1\}$). Formally, we have the following optimal solution.

 \begin{lemma}[Optimal Allocation Strategy]\label{lemma:opt-multi}
For any idle spectrum $\theta$,
the optimal allocation strategy $\Aicx^*(\theta)$ is given by: \emph{(i)} if $i\notin \boldsymbol{i}_1$, then $a_{i}^*(\theta)=0$; and \emph{(ii)} if $ i \in \boldsymbol{i}_1$, then
 \begin{itemize}
\itemsep=-1mm
\item[\emph{(ii.a)}]
$a_{i}^*(\theta) \in [0,1]$ such that $\sum_{i\in \boldsymbol{i}_1} a^*_i(\theta)=1$, if $\CMWmax (\theta)>0$;
\item[\emph{(ii.b)}]
$a_{i}^*(\theta)\in [0,1]$ such that $\sum_{i\in \boldsymbol{i}_1} a^*_i(\theta) \leq 1$,  if $\CMWmax (\theta)=0$;
\item[\emph{(ii.c)}]
$a_{i}^*(\theta)=0$,  if $\CMWmax (\theta)<0$;
 \end{itemize}
 \end{lemma}

Lemma \ref{lemma:opt-multi} states that in the optimal solution, only the independent contract user sets with the highest positive C-MW may obtain an idle spectrum $\theta$.
If the C-MW is negative for all independent contract user sets (i.e., $\CMWmax(\theta)<0$), the spectrum is allocated to the $0$-th independent contract user set (empty set) and the associated side market (the whole spot market). \rev{For example, in Fig.~\ref{f-eta-M}, we have $\boldsymbol{i}_1 = \{1\}$ in Regions I and I$^{\prime}$ (on the left side of Region I) and $\boldsymbol{i}_1 = \{2\}$ in Regions II and II$^{\prime}$ (on the right side of Region II). Thus, we have $a_{1}^*(\theta) = a_{2}^*(\theta) = 0$ in Regions I$^{\prime}$ and II$^{\prime}$ since $\CMWmax(\theta)<0$; $a_{1}^*(\theta) = 1, a_{2}^*(\theta) = 0$ in Region I; and $a_{1}^*(\theta) = 0, a_{2}^*(\theta) = 1$ in Region II.}

When SUs have continuous and heterogeneous utilities, the realization of $\theta$ such that $\CMWmax(\theta)=0$ (i.e., case (b) in Lemma \ref{lemma:opt-multi}) or $|\boldsymbol{i}_1|>1$ (i.e., multiple independent sets {having} the same highest C-MP) has a zero size support, and therefore can be ignored.
Thus, the optimal allocation given by Lemma \ref{lemma:opt-multi} is equivalent to:
 \begin{equation}\label{eq:xxx}
a^*_i(\theta) = 1 \mbox{~~if and only if~~} \CMW_i(\theta) >0 \mbox{ and }  \CMW_i(\theta) > \max_{j\neq i} \CMW_j(\theta).
 \end{equation}
Recall the example in Figure \ref{f-eta-M}. There are two distinct realizations of $\theta$ such that $\CMWmax(\theta)=0$, i.e., the boundary points between Region I and I$^{\prime}$ and between Region II and II$^{\prime}$. In addition, there is only one distinct realization of $\theta$  such that $|\boldsymbol{i}_1|>1$, i.e., the boundary point between Region I and II. Obviously, these realizations have a zero size support.

Now let us look into  (\ref{eq:xxx})  to gain further insights. First, notice that:
$$
\textstyle
\CMW_i(\theta)   =  \rho S \cdot \Big(-z_0(\theta) +z_i(\theta)  +  \sum_{n=1}^N l_n^i \cdot \big(\tau_n \cdot \widehat{P}_n + (1-\tau_n)\cdot u_n(\theta)\big) - \sum_{n=1}^N l_n^i \cdot \lambda_n \Big).
$$
The first term $z_0(\theta)$ is the
side welfare (from a spectrum $\{\theta, 1\}$) of the 0-th independent contract user set (empty set).
That is, it is the welfare generated by the spot market users, when the spectrum is not allocated to any contract user.
In this sense, it can be viewed as an \textit{outside option} with regard to the futures market.
The second term $z_i(\theta)$ is the side welfare of the $i$-th independent contract user set. That is, it is the welfare generated by the spot market users (in the associated side market), when the spectrum is allocated to the $i$-th independent contract user set.
The third term $\sum_{n=1}^N l_n^i \cdot \big(\tau_n \cdot \widehat{P}_n + (1-\tau_n)\cdot u_n(\theta)\big)$ denotes the direct welfare \textit{improvement} (also called \textit{welfare gain}) generated by the contract users in the $i$-th independent contract user set from a spectrum $\{\theta, 1\}$.\footnote{\label{aaaaa}
Specifically, if the SR allocates a spectrum $\{\theta, 1\}$ to the $i$-th independent contract user set, the total welfare generated by those contract users is $\sum_{n=1}^N l_n^i \cdot (\tau_n \cdot \frac{B_n}{D_n} + (1-\tau_n)\cdot u_n(\theta) )$; whereas, if the SR does not allocate the spectrum  $\{\theta,1\}$ to the $i$-th independent set, the total welfare is $\sum_{n=1}^N l_n^i \cdot (\tau_n \cdot ( \frac{B_n}{D_n} -\widehat{P}_n)  )$. That is, there is no utilization-based welfare and every contract user in the independent set suffers a unit welfare loss $\widehat{P}_n$.
Here, the welfare loss $\widehat{P}_n$ is implied by Lemma \ref{lemma:nc2}, which states that each contract user can never get spectrums more than his or her demand. Thus, by allocating one less spectrum to the contract user, he or she must suffer a unit welfare loss $\widehat{P}_n$. Based on above, the welfare gain is $\sum_{n=1}^N l_n^i \cdot \big(\tau_n \cdot \widehat{P}_n + (1-\tau_n)\cdot u_n(\theta)\big)$.}
The last term $\sum_{n=1}^N l_i^n \cdot \lambda_n^*$ is the shadow prices of the $i$-th independent contract user set, which imposes an identical ``shade'' on the C-MW of the $i$-th independent set over all spectrums.
It is used to rule out the relatively low welfare spectrums and assign the high welfare spectrums to the $i$-th independent contract user set.

Let $\THETA_i^+ \triangleq
\big\{\theta\ |\  \CMW_i(\theta) >0\ \&\ \CMW_i(\theta) > \max_{j\neq i} \CMW_j(\theta)  \big\}$ denote the set of spectrum information $\theta$ with which the $i$-th independent contract user set having the unique highest and positive C-MP.
Since $\THETA_i^+$ is determined by the shadow prices (vector) $\boldsymbol{\lambda}^* \triangleq (\lambda_1^*,...,\lambda_N^*)$, we can write $\THETA_i^+$ as a function of $\boldsymbol{\lambda}^* $, denoted by $\THETA_i^+(\boldsymbol{\lambda}^*)$.
The optimal solution is therefore equivalent to: $a_i^*(\theta)=1$ if and only if $\theta \in \THETA_i^+(\boldsymbol{\lambda}^*)$.
Accordingly, the set of spectrums allocated to a contract user $n$ can be written as: $\THETA_n^+ = \bigcup_{i=1}^{\Ic} l_n^i \odot \THETA_i^+$, where $l_n^i \odot \THETA_i^+ = \emptyset$ if $l_n^i=0$ and $l_n^i \odot \THETA_i^+ = \THETA_i^+ $ if $l_n^i=1$.
\revh{Furthermore, the expected number of spectrums allocated to a contract user $n$ can be written as: $\mathbb{E}[d_n] =\rho S \int_{\theta\in \THETA_n^+(\boldsymbol{\lambda}^*)} f_{\THETA}(\theta) \mathrm{d} \theta$.}
According to the dual constraints (D.3), the optimal shadow price $\lambda_n^*$ is either $0$ if the associated constraint (iii) is not tight, or otherwise a non-negative value that makes the associated constraint tight. Formally,

 \begin{lemma}[Optimal Shadow Price $\lambda_n^*$]\label{lemma:lambda-multi}
The optimal shadow price
$\lambda_n^*$ is given by:
 \begin{equation*}
\textstyle
\lambda_n^* = \max
\left\{0,\ \
\arg_{\lambda_n } \left( \rho S \int_{\theta\in \THETA_n^+(\boldsymbol{\lambda}_{-n}^*,\lambda_n)} f_{\THETA}(\theta) \mathrm{d} \theta = D_n \right) \right\},\quad \forall n\in \mathcal{N},
 \end{equation*}
where $\boldsymbol{\lambda}_{-n}^*  $ is the vector of all independent contract user sets' shadow prices except that of the $i$-th   independent set, i.e., $\boldsymbol{\lambda}_{-n}^*  \triangleq (\lambda_1^*,...,\lambda_{n-1}^*,\lambda_{n+1}^* ,...,\lambda_N^*)$.
\end{lemma}

After obtaining the optimal shadow price vector $\boldsymbol{\lambda}^*$ and substituting the optimal $\boldsymbol{\lambda}^*$ into the C-MWs of all independent contract user sets,
we can easily determine the optimal allocation of any spectrum under any information $\{\theta, 1\}$ according to Lemma \ref{lemma:opt-multi} or  (\ref{eq:xxx}).
Intuitively, the shadow price $\lambda_n^*$ imposes a \textit{vertical} (downward) shift on the C-MPs of those independent sets containing the contract user $n$, such that $\mathbb{E}[d_n]$ meets the demand $D_n$.

\subsection{The Optimal Policy}

We now summarize the \textit{optimal policy} based on the optimal solution derived above. Simply speaking, a policy is a set of pre-defined \textit{allocation rules}, specifying explicitly or implicitly the allocation of any spectrum under every possible information realization.
We will show that the optimal solution in Lemma \ref{lemma:opt-multi} is equivalent to such a policy that allocates every spectrum to an MWIS of the graph $\G$ with well-defined weights. Formally, we show this in the following theorem.
\begin{theorem}[Optimal Policy]\label{solu:policy}
For any idle spectrum with information realization $\{\theta, 1\}$,
\begin{itemize}
  \item Define the \textit{weight} of each vertex of graph $\G$ in the following way:
    \begin{itemize}
            \item The {weight} of each spot market user $m$ is: $v_m$;
            \item The {weight} of each contract user $n$ is: $\tau_n \cdot \widehat{P}_n + (1-\tau_n)\cdot u_n(\theta) -  \lambda_n^*$;
      \end{itemize}
  \item Allocate the  spectrum $\{\theta, 1\}$ to an \textit{MWIS} of the weighted graph $\G$ defined above.
\end{itemize}
\end{theorem}


By the optimal policy in Theorem \ref{solu:policy}, every spectrum will be allocated to an MWIS of $\G$, where the weight of each vertex (SU) is defined as his \textit{current} valuation directly for spot market users or a function of his \textit{current} valuation for contract users.
That is, the weight is related to the realized actual information of the current spectrum and the optimal policy only, but independent of the information of future or past spectrums.
\reee{In this sense, we disentangle the time-coupled spectrum
allocation problem (which determines the spectrum allocations in all time slots jointly based on the complete network information) into a set of sub-problems.
Each sub-problem determines the spectrum allocation in a particular time slot based on the currently realized information and the pre-derived optimal shadow prices.
The coupling relationship of allocations in different time slots is captured by the pre-derived optimal shadow prices (in Lemma \ref{lemma:lambda-multi}).}

%% file: Section4-VCG-OR.tex
\section{On-line VCG Mechanisms}\label{sec:vcg}


In the previous section, we have shown in Theorem \ref{solu:policy} that the E-SEM problem can be disentangled {into a set of sub-problems}, each determining the allocation of one particular spectrum. More specifically, every spectrum is allocated to an MWIS of the graph $\G$, in which the weight of each vertex is related only to the utility of the associated SU over the spectrum, while independent of his or her utilities over other spectrums or the utility of other SUs.

Obviously, if the SR  can observe all of the realized information of one spectrum, it is straightforward to determine the optimal allocation of that spectrum, based on Theorem \ref{solu:policy}.
As mentioned previously, however, the SU's valuation is private information, and cannot be observed by the SR.
That is, the network information is asymmetric between the SR  and SUs.
In this case, the real-time allocation of every spectrum essentially depends on the SUs' reports (also called \textit{bids}) about their valuations.
SUs may have incentives to falsely report their valuations to enlarge their welfare. Therefore, an incentive compatible mechanism is necessary for credibly eliciting the SUs' private valuations and then optimally allocating every spectrum.


\subsection{VCG Mechanism}

We first propose a {VCG auction}, which is {incentive-compatible} for SUs to truthfully report their private valuations, and is {optimal} for the real-time allocation of every spectrum.
Essentially, a VCG auction charges each SU a fee if his or her report is \textit{pivotal}, i.e., if his or her report changes the optimal allocation so as to harm other SUs.
Let $\widetilde{v}_m$ and $\widetilde{u}_n$ denote the reports (i.e., bids) of spot market user $m$ and contract user $n$. Denote  $\widetilde{\theta} \triangleq \big( \widetilde{v}_{1 } , $ $ ...,$ $ \widetilde{v}_{M } ; $ $  \widetilde{u}_{1 } ,..., \widetilde{u}_{N } \big)$ as the bid vector of all SUs.
Formally, the VCG based spectrum allocation and charging mechanism is shown as follows.

%

\begin{mechanism}[VCG Auction.]\label{solu:vcg}
For any idle spectrum with a bid vector $\widetilde{\theta}$,
\begin{itemize}
  \item Define the \textit{weight} of each vertex of graph $\G$ in the following way:
      \begin{itemize}
            \item The {weight} of each spot market user $m$ is: $x_m \triangleq \widetilde{v}_m$;
            \item The {weight} of each contract user $n$ is: $x_n \triangleq \tau_n \cdot \widehat{P}_n + (1-\tau_n)\cdot \widetilde{u}_n(\theta) -  \lambda_n^*$;
      \end{itemize}
  \item Allocate the spectrum to an \textit{MWIS} of the above weighted graph $\G$:
            \begin{itemize}
            \item Denote $\IS^* \in \IS(\G)$ as the allocated independent set, i.e., $\sum_{j\in \IS^*} x_j \geq \sum_{j\in \IS_i} x_j,\ \forall \IS_i \in \IS(\G)$;
      \end{itemize}
  \item Charge each SU $k $ in the following way:
            \begin{itemize}
            \item Denote $\IS_{-k}^* \in \IS(\G_{-k})$ as an MWIS of $\G_{-k}$, the graph consisting of all SUs except $k$;
            \item Charge the  SU $k $ a critical price defined as: $p_k \triangleq \sum_{j\in \IS_{-k}^*} x_j - \sum_{j\in \IS^*/\{k\}} x_j$.
      \end{itemize}
\end{itemize}
\end{mechanism}

The above mechanism is a standard VCG auction (see \cite{Revelation}), thus leads to truth-telling. The novelty of our VCG auction is that the bid of each contract user $n$ is not exogenously given, but rather related to the optimal shadow prices $\lambda_n^*$ derived in the previous section.
{Hence, such a solution (i.e., the off-line policy and the on-line VCG auction) allows us to optimally allocate every spectrum in an on-line manner under stochastic and asymmetric network information.}

\reee{It is notable that the above spectrum auction with spatial spectrum reuse is mathematically equivalent to
the auction with single-minded buyers proposed in \cite{Nisan}.
Specifically, in an auction with single-minded buyers, each bidder requests a specific bundle of items, and achieves a positive revenue if obtaining this bundle of items, and zero otherwise. We can view any two bidders with non-conflicting requirements (i.e., there is no intersection between their requested bundles of items) as non-conflicting (independent) buyers, based on which we can construct such a conflict graph in which conflicting bidders are connected by an edge and non-conflicting bidders are disconnected. Accordingly, a set of independent bidders (i.e., those not connected by edges) can be chosen as auction winners at the same time. This is equivalent to the single-slot spectrum auction in our work.
We would like to emphasize that our work is not restricted to the study of the single-slot spectrum auction (in a particular time slot). In fact, we focus on the auction of spectrum in a period of $T$ time slots. This multi-slot spectrum allocation problem is essentially a combination of $T$ coupled single-slot spectrum auction problems (which are coupled by the optimal shadow prices derived in advance). This is also the key difference between our work and the single-minded auction in \cite{Nisan}.}

Next we provide some comments on this mechanism.
First, as shown in Section \ref{sec:model:contract}, the contract user's actual payment is pre-defined, and determined only by the number of spectrums that he or she obtains, while independent of the utility. In the above VCG auction, however, the contract user's payment depends on his or her valuation.
Actually, we can view a contract user's payment in the VCG auction as a subsidy that encourages the contract user to report truthfully.
Second, if an SU $k$ is not allocated a spectrum (i.e., $k\notin \IS^*$), 
then we have $\IS_{-k}^* = \IS^*$, and thus his payment is $p_k= \sum_{j\in \IS_{-k}^*} x_j - \sum_{j\in \IS^*/\{k\}} x_j = 0$.
Third, the above VCG auction relies on the solving of MWIS problems, which are well-known to be NP-hard. Thus, the VCG auction may not be suitable for large networks. This motivates us to further study a low-complexity mechanism instead.

\subsection{VCG-like Mechanism}

Due to the spatial spectrum reuse, the VCG auction in Mechanism \ref{solu:vcg} requires a solution to the MWIS problem, which is NP-hard.
Thus, the above VCG auction is not suitable for on-line implementation.
To this end, we further propose a low-complexity \textit{VCG-like mechanism}, based on approximate MWIS algorithms with polynomial-time complexity.
Formally, the VCG-like spectrum allocation and charging mechanism works in the following way.

\begin{mechanism}[VCG-Like Mechanism.]\label{solu:vcg-like}
For any idle spectrum with a bid vector $\widetilde{\theta}$,
\begin{itemize}
  \item Define the \textit{weight} of each vertex in the graph $\G$ in the following way:
      \begin{itemize}
            \item The {weight} of each spot market user $m$ is: $x_m \triangleq \widetilde{v}_m$;
            \item The {weight} of each contract user $n$ is: $x_n \triangleq \tau_n \cdot \widehat{P}_n + (1-\tau_n)\cdot \widetilde{u}_n(\theta) -  \lambda_n^*$;
      \end{itemize}
  \item Allocate each spectrum in the following greedy way: Define the survival SU set $\N^{\dag} = \N$
            \begin{itemize}
			\item Do until $\N^{\dag} = \emptyset$: Allocate the spectrum to the maximum weight SU in $\N^{\dag}$, and
			\item[]  ~~~~~~~~~~~~~~~~~~~~~~~~Remove this SU as well as the neighboring SUs from $\N^{\dag}$;
            \item Denote $\IS^{\dag} \in \IS(\G)$ as the allocated independent SU set;
      \end{itemize}
  \item Charge each SU $k $ in the following way:
            \begin{itemize}
            \item Denote $\IS_{-k}^{\dag} \in \IS(\G_{-k}) $ as the allocated SU set using the above allocation mechanism on $\G_{-k}$;
            \item Charge the  SU $k $ a critical price defined as: $p_k \triangleq \sum_{j\in \IS_{-k}^{\dag}} x_j - \sum_{j\in \IS^{\dag}/\{k\}} x_j$.
      \end{itemize}
\end{itemize}
\end{mechanism}
%


The key difference between this VCG-like mechanism and the previous VCG auction lies in the allocation strategy. Specifically, in the VCG auction, every spectrum is allocated to an MWIS of SUs; whereas, in the VCG-like mechanism, every spectrum is allocated to a set of SUs based on a greedy algorithm.
Some important observations are as follows.
First, the above VCG-like mechanism has a polynomial-time complexity.
Second, the above VCG-like mechanism leads to truth-telling (under this greedy allocation mechanism).
This can be easily proved by the monotonicity property and bid-independent property (see \cite{Auction}).
Third, the above VCG-like mechanism is \textit{not} optimal due to the lack of optimal algorithm for MWIS calculations.
The welfare loss caused by this non-optimal VCG-like mechanism will be studied in the next section.


%% file: Section5-Loss-OR.tex

\section{Welfare Loss Analysis}\label{sec:pl}

Now we study the welfare loss induced by the approximate MWIS algorithms.
We first notice that the E-SEM problem in (\ref{eq:obj-incomp-1}) depends only on the stochastic distribution information, and the optimal solution (policy) can be derived in an \textit{off-line} manner.
That is, we can enumerate all possible scenarios and calculate the corresponding solution beforehand.
In this sense, we can always apply the precise MWIS algorithm to compute the off-line optimal policy exactly.
Hence, the approximate MWIS algorithms only affect the real-time allocation of every spectrum (which must be determined in an \textit{on-line} manner).


In the previous section, we have proposed a specific on-line VCG-like mechanism (based on a greedy approximate MWIS algorithm) for the real-time allocation of every spectrum.
Now we characterize the {welfare loss} under \textit{generic} VCG-like mechanisms, or equivalently, under generic approximate MWIS algorithms.
This quantitative analysis helps us to achieve a desirable tradeoff between   performance and   complexity.

\subsection{Precise and Approximate C-MWs}

We first examine how an approximate MWIS algorithm affects the achieved performance.
By Lemma \ref{lemma:opt-multi}, the optimal allocation of every spectrum $\theta$ relies on the \textit{accurate calculations} of $z_i(\theta),\forall i = 0,1,...,\Ic $, i.e., the side welfare of the $i$-th independent contract user set.
With the above VCG-like mechanism, however, we can only get an approximate (degraded) side welfare $\tilde{z}_i(\theta)$ with certain performance guarantee.
This degradation will potentially change the relationship between C-MWs, and further change the spectrum allocation.
For example, a relatively larger (smaller) degradation on the side welfare ${z}_i(\theta)$ will lead to a smaller (larger) C-MW, and therefore a smaller (larger) allocation probability to the $i$-th independent contract user set.

Without loss of generality, we treat the performance ratio (i.e., the ratio of the achieved performance and the optimal performance) of the adopted approximate MWIS algorithm as a random variable drawn from $[0,1]$.
Let random variable $\boldsymbol{\epsilon}_i$ denote the performance ratio of
the MWIS problem of calculating $ {z}_i(\theta)$, and $\boldsymbol{\epsilon} \triangleq (\boldsymbol{\epsilon}_0 , \boldsymbol{\epsilon}_1, ..., \boldsymbol{\epsilon}_{\Ic})$ denote the random vector consisting of the performance ratios of all $\Ic+1$ MWIS problems.
Let $\epsilon_i\in[0,1]$ denote a particular realization of the random variable $\boldsymbol{\epsilon}_i$, and $ {\epsilon} \triangleq ( {\epsilon}_0 ,  {\epsilon}_1, ...,  {\epsilon}_{\Ic})$ denote a particular realization of the random vector $\boldsymbol{\epsilon}$. For simplicity and without loss of generality, we assume that all performance ratios in vector $\boldsymbol{\epsilon} $ are independent of each other, but they are not necessarily identically distributed.


%
%

For any spectrum $\{\theta, 1\}$, the \textit{precise} C-MW of the $i$-th independent contract user set is
 \begin{equation}\label{eq:precise-cmp}
\textstyle
 \CMW_i(\theta)   = \rho S \cdot   \Big(- z_0(\theta) + z_i(\theta)  +  \sum_{n=1}^N l_n^i \cdot \big( \tau_n \cdot \widehat{P}_n + (1-\tau_n)\cdot u_n(\theta) -  \lambda_n^* \big) \Big),
 \end{equation}
while the \textit{approximate} C-MW under  certain performance ratio realization of $ \boldsymbol{\epsilon}$ is:
 \begin{equation}\label{eq:app-cmp}
 \begin{aligned}
 \CMWx_i(\theta)
&\textstyle   = \rho S \cdot   \Big(- \epsilon_0 \cdot z_0(\theta) + \epsilon_i \cdot z_i(\theta)  +  \sum_{n=1}^N l_n^i \cdot \big( \tau_n \cdot \widehat{P}_n + (1-\tau_n)\cdot u_n(\theta) -  \lambda_n^* \big) \Big).
 \end{aligned}
 \end{equation}

\subsection{Welfare Loss and Performance Ratio}

Now we study the welfare loss and the achievable performance ratio.
We can divide the whole spectrum range $\THETA$ into $\Ic+1$ parts based on  the allocated independent contract user set under the optimal allocation policy: (i) $\THETA_1$, $i=1,...,\Ic$, the spectrums intended for the $i$-th independent contract user set; and (ii) $\THETA_0$, the spectrums intended for  the spot market only. Formally,
\begin{equation*}
\left\{
\begin{aligned}
\THETA_i &  = \{\theta\in\THETA \ |\ \CMW_i(\theta) > 0\ \&\ \CMW_i(\theta) > \max_{j\neq i} \CMW_j(\theta) \},\quad i=1,...,\Ic. \\
\THETA_0 &  = \{\theta\in\THETA \ |\ \CMW_i(\theta)< 0, \forall i=1,...,\Ic \}.
\end{aligned}
\right.
\end{equation*}

In the following, we will first derive the welfare loss (upper-bound) in each part of $\THETA$. \revh{Then we   characterize the achievable performance ratio (lower-bound).}

\subsubsection{Welfare Loss in $\THETA_i$}

We first consider the welfare loss on spectrums in $\THETA_i$. Under the optimal allocation policy based on precise C-MWs, every idle spectrum $\theta $ in $ \THETA_i$ is allocated to the $i$-th independent contract user set (and the associated side market) with an instantaneous welfare:
$$
\textstyle
z_i(\theta)  +  \sum_{n=1}^N l_n^i \cdot  (1-\tau_n)\cdot u_n(\theta).
$$
The first term denotes the side welfare, and the second term denotes the (quality-based) utility of the $i$-th independent contract user set.
Note that we ignore the demand-related utility of contract user $n$, since it is constant and not affected by the mis-allocation; we also ignore the potential utility loss induced by under-allocation here, which will be analyzed specifically later.

Obviously, with precise C-MWs (using optimal MWIS algorithms), the $i$-th independent contract user set has the highest positive C-MW among all independent sets.
With approximate C-MWs (using approximate MWIS algorithms), however, there are possibly three different outcomes, each leading to a different allocation and thus a different welfare loss. Specifically, we have the following:
 \begin{itemize}
\itemsep=-1mm
  \item[(1)]
   {The $i$-th independent contract user set itself has the highest and positive approximate C-MW}, i.e., $ \CMWx_i(\theta) > 0$ and $\CMWx_i(\theta) > \max_{j\neq i} \CMWx_j(\theta)$. In this case, the spectrum will still be allocated to the  $i$-th independent contract user set with a degraded instantaneous welfare $\epsilon_{i}\cdot z_i(\theta)  +  \sum_{n=1}^N l_n^i \cdot  (1-\tau_n)\cdot u_n(\theta) $. Thus, the welfare loss can be written as:
  $$
\textstyle
    \psi_i(\theta) \triangleq (1 - \epsilon_{i})\cdot  z_{i}(\theta).
$$

  \item[(2)]
   {The $i'$-th (with $i' \neq i$) independent contract user set  has the highest and positive approximate C-MW}, i.e., $ \CMWx_{i'}(\theta) > 0$ and $\CMWx_{i'}(\theta) > \max_{j\neq i'} \CMWx_j(\theta)$. In this case, the spectrum will be mistakenly allocated to the  $i'$-th independent set  instead with an instantaneous welfare $\epsilon_{i'}\cdot z_{i'}(\theta)  +  \sum_{n=1}^N l_n^{i'} \cdot  (1-\tau_n)\cdot u_n(\theta) $. Thus, the welfare loss can be written as:
$$
\textstyle
\psi_{i'}(\theta) \triangleq z_{i}(\theta) - \epsilon_{i'}\cdot z_{i'}(\theta)  + \sum_{n=1}^N (l_n^{i} - l_n^{i'}) \cdot  (1-\tau_n)\cdot u_n(\theta).
$$

  \item[(3)]
   {All independent contract user sets have the negative approximate C-MWs}, i.e., $\CMWx_i(\theta)< 0, \forall i=1,...,\Ic$. In this case, the spectrum will be mistakenly allocated to the spot market only with an instantaneous welfare $\epsilon_0\cdot z_0(\theta)$. Thus, the welfare loss can be written as:
$$
\textstyle
\psi_0(\theta) \triangleq z_{i}(\theta) - \epsilon_0\cdot z_0(\theta) + \sum_{n=1}^N l_n^{i} \cdot  (1-\tau_n)\cdot u_n(\theta).
$$
 \end{itemize}

Denote $\mathbf{e}_i^{i}(\theta) \triangleq \{\epsilon \in \boldsymbol{\epsilon} \ |\  \CMWx_i(\theta) > 0\ \&\ \CMWx_i(\theta) > \max_{j\neq i} \CMWx_j(\theta) \}$ as the set of performance ratio $\epsilon $ such that the  $i$-th independent contract user set itself has the highest and positive approximate C-MW (i.e., case [1]),
$\mathbf{e}_i^{i'}(\theta) \triangleq \{\epsilon\in \boldsymbol{\epsilon} \ |\  \CMWx_{i'}(\theta) > 0\ \&\ \CMWx_{i'}(\theta) > \max_{j\neq i'} \CMWx_{j}(\theta) \}$ as the set of $\epsilon  $ such that the  $i'$-th (where $i'\neq i$) independent contract user set has the highest and positive approximate C-MW (i.e., case [2]),
and ${\mathbf{e}}_i^0(\theta) \triangleq \{ \epsilon\in \boldsymbol{\epsilon} \ |\ \CMWx_{i}(\theta)< 0, \forall i=1,...,\Ic\}$ as the set of $\epsilon $ such that all independent contract user sets have negative approximate C-MWs (i.e., case [3]).\footnote{Similar to $\boldsymbol{\Theta}$, here we use the same notation $\boldsymbol{\epsilon}$ to denote the whole range of $\epsilon$. For example, if $\epsilon_i$ is randomly drawn from $[0,1]$, then $\boldsymbol{\epsilon}$ is the Cartesian product of $\Ic+1$ intervals $[0,1]$.}
It is easy to see that $\bigcup_{j=0}^{\Ic} \mathbf{e}_i^j $ equals $\boldsymbol{\epsilon}$, i.e., the whole range of $\epsilon $.
Here, the subscript and superscript in $\mathbf{e}_i^{j}(\theta)$ denote the indexes of allocated independent contract user sets under precise C-MWs and under approximate C-MWs, respectively.
For convenience, we will omit the parameter $\theta$ in ${\mathbf{e}}_i^j(\theta)$, as long as there is no confusion caused.

Based on the above, the average welfare loss on every spectrum in $\THETA_i$ is formally given by
\begin{equation*}
\begin{aligned}
 & ~~~~ \textstyle  \int_{\theta \in \THETA_i} \left( \int_{\epsilon  \in \mathbf{e}_i^i} \psi_i(\theta) \cdot f_{\boldsymbol{\epsilon}}(\epsilon) \mathrm{d} \epsilon +   \int_{\epsilon \in \mathbf{e}_i^0}  \psi_0(\theta) \cdot f_{\boldsymbol{\epsilon}}(\epsilon) \mathrm{d} \epsilon
 + \sum_{i' \neq i} \int_{\epsilon \in \mathbf{e}_i^{i'}} \psi_{i'}(\theta)  \cdot f_{\boldsymbol{\epsilon}}(\epsilon) \mathrm{d} \epsilon
 \right) \cdot f_{\THETA}(\theta) \mathrm{d} \theta
\\
& \textstyle = \int_{\theta \in\THETA_i} \left( \int_{\epsilon \in \boldsymbol{\epsilon} } \psi_i(\theta)   f_{\boldsymbol{\epsilon}}(\epsilon) \mathrm{d} \epsilon +   \int_{\epsilon \in \mathbf{e}_i^0}  \big( \psi_0(\theta) -\psi_i(\theta) \big)  f_{\boldsymbol{\epsilon}}(\epsilon) \mathrm{d} \epsilon
 + \sum_{i' \neq i} \int_{\epsilon\in \mathbf{e}_i^{i'}} \big(\psi_{i'}(\theta) - \psi_i(\theta)\big)   f_{\boldsymbol{\epsilon}}(\epsilon) \mathrm{d} \epsilon
 \right)   f_{\THETA}(\theta) \mathrm{d} \theta
\\
& \textstyle \leq \int_{\theta \in\THETA_i} \left( (1-\bar{\epsilon}_i)
\cdot z_i(\theta)
-
\int_{\epsilon\in \mathbf{e}_i^0}
  Y_i  \cdot f_{\boldsymbol{\epsilon}}(\epsilon) \mathrm{d} \epsilon
+   \sum_{i' \neq i}  \int_{\epsilon\in \mathbf{e}_i^{i'}} (Y_{i'}-Y_i) \cdot f_{\boldsymbol{\epsilon}}(\epsilon) \mathrm{d} \epsilon
  \right)\cdot f_{\THETA}(\theta) \mathrm{d} \theta
\\
& \textstyle = \int_{\theta \in \THETA_i} (1-\bar{\epsilon}_i)
\cdot z_i(\theta) \cdot f_{\THETA}(\theta) \mathrm{d} \theta - Y_i  \cdot \beta_i^0 + \sum_{i'=1}^{\Ic} (Y_{i'}-Y_i) \cdot \beta_i^{i'},
\end{aligned}
\end{equation*}
where
\begin{itemize}
\itemsep=-1mm


\item
$\bar{\epsilon}_i = \mathbb{E}[\epsilon_i]  \triangleq \int_{\epsilon \in \boldsymbol{\epsilon} } \epsilon_i  f_{\boldsymbol{\epsilon}}(\epsilon) \mathrm{d} \epsilon$   is the average performance ratio of the approximate MWIS algorithm in calculating $z_i(\theta)$, i.e., the expectation of random variable $\boldsymbol{\epsilon}_i$;

\item
$Y_i = \sum_{n=1}^N l_n^i \cdot X_n$ is the sum of $X_n$ of all contract users in the  $i$-th independent set, where $X_n \triangleq \widehat{P}_n - \lambda_n^* $ is  a constant associated with contract user $n$ only;

\item
$ \int_{\epsilon \in \mathbf{e}_i^0} f_{\boldsymbol{\epsilon}}(\epsilon) \mathrm{d} \epsilon$ is the probability of $\epsilon \in \mathbf{e}_i^0(\theta)$, i.e., the probability of {mis-allocating} a spectrum $\theta \in \THETA_i$ to the 0-th independent contract user set (or the spot market only) under $\CMWx_i(\theta), \forall i $. 

\item
$\beta_i^0 = \int_{\theta \in \THETA_i}
\int_{\epsilon \in \mathbf{e}_i^0} f_{\boldsymbol{\epsilon}}(\epsilon) f_{\THETA}(\theta)\mathrm{d} \epsilon \mathrm{d} \theta$ is the average mis-allocation probability of spectrums intended for the  $i$-th independent contract user set to the 0-th one. That is, $\beta_i^0$ is the average probability of transferring spectrums from the $i$-th independent set to the 0-th one;

\item
$\int_{\epsilon \in \mathbf{e}_i^{i'}} f_{\boldsymbol{\epsilon}}(\epsilon) \mathrm{d} \epsilon$ is the probability of $\epsilon \in \mathbf{e}_i^{i'}(\theta)$, i.e., the probability of {mis-allocating} a spectrum $\theta \in \THETA_i$ to the $i'$-th independent set under approximate C-MWs $\CMWx_i(\theta), \forall i$. 

\item
$\beta_i^{i'} = \int_{\theta \in \THETA_i}
\int_{\epsilon \in \mathbf{e}_i^{i'}} f_{\boldsymbol{\epsilon}}(\epsilon) f_{\THETA}(\theta)\mathrm{d} \epsilon \mathrm{d} \theta$ is the average mis-allocation probability of spectrum  intended for the $i$-th independent contract user set to the $i'$-th one. That is, $\beta_i^{i'}$ is the average probability of transferring spectrums from the $i$-th independent set to the $i'$-th one;

\item
The inequality follows because: (i) $\CMWx_i(\theta)< 0$ if  $\epsilon \in \mathbf{e}_i^0$, and $\CMWx_i(\theta)< \CMWx_{i'}(\theta)$ if $\epsilon \in \mathbf{e}_i^{i'}$; and (ii) $\CMWx_i(\theta) = \psi_0(\theta) -\psi_i(\theta) + Y_i$. Thus, we have: $\psi_0(\theta) -\psi_i(\theta) < -Y_i$, and $\psi_{i'}(\theta) - \psi_i(\theta) < Y_{i'}-Y_i$.
\end{itemize}

Thus, we can obtain an upper-bound of the expected welfare loss on every spectrum in $\THETA_i$ as
 \begin{equation} \label{eq:eps-RII-PL1}
\begin{aligned}
\textstyle
  \textsc{Loss}^{(i)} \leq     \Phi_i   - Y_i  \cdot \beta_i^0 + \sum_{i'=1}^{\Ic} (Y_{i'}-Y_i) \cdot \beta_i^{i'},
\end{aligned}
 \end{equation}
where $\Phi_i \triangleq \int_{\theta \in\THETA_i} (1-\bar{\epsilon}_i)\cdot
z_i(\theta)  f_{\THETA}(\theta) \mathrm{d} \theta$ is the average side welfare loss of the $i$-th independent contract user set (over every   spectrum in $\THETA_i$).

\subsubsection{Welfare Loss in $\THETA_0$}

Using a similar analysis to that in (\ref{eq:eps-RII-PL1}), we can obtain an upper-bound of the expected welfare loss on every spectrum in $\THETA_0$ as:
 \begin{equation} \label{eq:eps-RII-PL2}
\begin{aligned}
\textstyle
\textsc{Loss}^{(0)}
\leq  \Phi_0 + \sum_{i=1}^{\Ic} Y_i  \cdot \beta_0^i.
\end{aligned}
 \end{equation}
where $\Phi_0 \triangleq \int_{\theta \in \THETA_0} (1-\bar{\epsilon}_0)\cdot
z_0(\theta) f_{\THETA}(\theta) \mathrm{d} \theta$ is the average side welfare loss of the $0$-th independent contract user set (over every spectrum in $\THETA_0$), and $\beta_0^i = \int_{\theta \in \THETA_0} \int_{\epsilon \in \mathbf{e}_0^i} f_{\boldsymbol{\epsilon}}(\epsilon) f_{\THETA}(\theta)\mathrm{d} \epsilon \mathrm{d} \theta$ is the average
probability of transferring spectrums from the 0-th independent contract user set to the  $i$-th one.


\revh{By (\ref{eq:eps-RII-PL1}) and (\ref{eq:eps-RII-PL2}), the average welfare loss on every spectrum is bounded by:
\begin{equation} \label{eq:eps-RII-PL1-all}
\begin{aligned}
\textstyle
 \sum_{i=0}^{\Ic} \textsc{Loss}^{(i)}
\leq &\textstyle   \sum_{i=0}^{\Ic} \Phi_i  + \sum_{i=1}^{\Ic} \left( \sum_{i'=1}^{\Ic} (Y_{i'}-Y_i) \cdot \beta_i^{i'} - Y_i  \cdot \beta_i^0 \right)+ \sum_{i=1}^{\Ic} Y_i  \cdot \beta_0^i  \\
= & \textstyle   \sum_{i=0}^{\Ic} \Phi_i + \sum_{i=1}^{\Ic} \left( \sum_{i'=1}^{\Ic} (Y_{i}\cdot \beta_{i'}^i -Y_i \cdot \beta_i^{i'}) - Y_i  \cdot \beta_i^0 \right)+ \sum_{i=1}^{\Ic} Y_i  \cdot \beta_0^i  \\
= & \textstyle \sum_{i=0}^{\Ic} \Phi_i - \sum_{i=1}^{\Ic} Y_i  \cdot \Delta_{i}
\triangleq \sum_{i=0}^{\Ic} \Phi_i - \sum_{n=1}^N X_n \cdot \gamma_n  ,
\end{aligned}
\end{equation}
where
\begin{itemize}
\itemsep=-1mm

\item
$ \Delta_{i} \triangleq \sum_{i'=1}^{\Ic}  \beta_i^{i'}  + \beta_i^0 -\sum_{i'=1}^{\Ic}  \beta_{i'}^i -  \beta_0^i $ is the \textit{change} on average allocation probability of a spectrum to the $i$-th independent contract user set, in which the first two terms denote the allocation probability decrease (induced by mis-allocating the spectrums intended for the $i$-th independent set to others), and the last two terms denote the allocation probability increase;
\item
$\gamma_n \triangleq \sum_{i=1}^{\Ic}  l_n^i \cdot \Delta_{i}$ is the \textit{change} on   average allocation probability of a spectrum to contract user $n$, which is the sum of the probability changes of all related independent sets;
\item
The second line follows because $ \sum_{i=1}^{\Ic} \sum_{i'=1}^{\Ic}  Y_{i'}   \beta_i^{i'} = \sum_{i'=1}^{\Ic} \sum_{i=1}^{\Ic}  Y_{i}    \beta_{i'}^i  $ by exchanging the indices of two summing operations, and
the last line follows   because $ Y_i = \sum_{n=1}^N  l_n^i \cdot X_n$.
\end{itemize}}

\subsubsection{Welfare Loss by Spectrum Under-Allocation}

It is important to note that the welfare loss in (\ref{eq:eps-RII-PL1-all}) is induced by the \textit{non-optimal utilization} of spectrum due to the mis-allocation, without including the potential utility loss of contract users (defined in Eq.~[\ref{eq:contract-welfare-loss}]) induced by the spectrum \textit{under-allocation} due to the mis-allocation.

Now we study the welfare loss of contract users induced by the spectrum under-allocation. From (\ref{eq:contract-welfare-loss}) we can easily find  that such a potential welfare loss is determined by the expected number of spectrums that  each contract user obtains.
By Lemma \ref{lemma:nc2}, none of the contract users can obtain spectrums more than his or her demand. Thus, any negative change on a contract user $n$'s average allocation probability $\gamma_n$ would impose certain welfare loss for contract user $n$, while any positive change on  $\gamma_n$  may or may not bring welfare gain for contract user $n$ (depending on whether the contract user already obtained the demanded spectrums).
Formally, the average welfare loss induced by the spectrum under-allocation  (over every spectrum) is bounded by:
 \begin{equation}\label{eq:eps-RII-PL3}
\begin{aligned}
\textstyle
\textsc{Loss}^{\textsc{wl}}
&\textstyle \leq \sum_{n=1}^N \widehat{P}_n \cdot \Big[
\sum_{i=1}^{\Ic} l_n^i \cdot \Delta_{i}  \Big]^+  =  \sum_{n=1}^N \widehat{P}_n \cdot \big[ \gamma_n \big]^+.
\end{aligned}
 \end{equation}

\subsubsection{Overall Welfare Loss}

Combining (\ref{eq:eps-RII-PL1-all}) and (\ref{eq:eps-RII-PL3}), the expected overall welfare loss on every spectrum is bounded by:
 \begin{equation}\label{eq:eps-RII-PL123}
\begin{aligned}
\textstyle
\textsc{Loss}^{\textsc{All}} = & \textstyle    \sum_{i=0}^{\Ic} \textsc{Loss}^{(i)} + \textsc{Loss}^{\textsc{wl}} \\
 \leq  & \textstyle  \sum_{i=0}^{\Ic} \Phi_i - \sum_{n=1}^N X_n \cdot \gamma_n + \sum_{n=1}^N \widehat{P}_n \cdot \left[ \gamma_n \right]^+ \\
=  &\textstyle \sum_{i=0}^{\Ic}  (1-\bar{\epsilon}_i)\cdot \mathbb{E}[Z_i]^* - \sum_{n=1}^N X_n \cdot \gamma_n + \sum_{n=1}^N \widehat{P}_n \cdot \left[ \gamma_n \right]^+,
\end{aligned}
 \end{equation}
where
$\mathbb{E}[Z_i]^* \triangleq	 \int_{\theta\in\THETA_i} z_i(\theta) f_{\THETA}(\theta) \mathrm{d} \theta $. Since $\mathbb{E}[Z_i]^* = \frac{ 1}{\rho \cdot S }\cdot \mathbb{E}[W_i^{\textsc{s}}]^*$, where $\mathbb{E}[W_i^{\textsc{s}}]^*$ is the expected side welfare of the $i$-th independent set over all $S$ spectrums (defined in Eq.~[\ref{eq:EZk-1}])  under the optimal allocation, $\mathbb{E}[Z_i]^*$ represents the expected side welfare of the $i$-th independent set over every spectrum.

\subsubsection{Achievable Performance Ratio}

\revf{Let $\bar{\epsilon} \triangleq \min_{i} \bar{\epsilon}_i $ denote the minimum expected performance ratio of all MWIS algorithms applied.}
Subtracting the overall welfare loss from the optimal welfare (defined in Eq.~[\ref{eq:EUA-1}]), we have the following average achievable welfare
\begin{equation}\label{eq:eps-RII-AP}
\begin{aligned}
 &\textstyle
\textsc{Wf}^{\textsc{Achi}} =  \textsc{Wf}^{\textsc{Opt}} - \rho S \cdot \textsc{Loss}^{\textsc{All}}
\\
 & \textstyle \quad = \sum_{i=0}^{\Ic}  \mathbb{E}[W^{\textsc{s}}_i]^*   +   \sum_{n=1}^N  ( \tau_n   \mathbb{E}[{\wc_n} ]^* + (1-\tau_n)   \mathbb{E}[\wcx_n ]^*  ) -  \rho S \cdot \textsc{Loss}^{\textsc{All}}
 \\
 &\textstyle
\quad \geq
 \sum_{i=0}^{\Ic} \bar{\epsilon} \cdot \mathbb{E}[W^{\textsc{s}}_i]^*   +   \sum_{n=1}^N  \big( \tau_n   \mathbb{E}[{\wc_n} ]^* + (1-\tau_n)   \mathbb{E}[\wcx_n ]^*  \big)
+ \rho S \cdot \sum_{n=1}^N \Big( X_n   \gamma_n - \widehat{P}_n   \left[ \gamma_n \right]^+ \Big),
\end{aligned}
 \end{equation}
where  $\rho S \cdot \textsc{Loss}^{\textsc{All}}$ is the overall welfare loss over all $S$ spectrums, and
$\textsc{Wf}^{\textsc{Opt}} \triangleq   \mathbb{E}[W ]^* =
 \sum_{i=0}^{\Ic}  \mathbb{E}[W^{\textsc{s}}_i]^*   +   \sum_{n=1}^N  ( \tau_n   \mathbb{E}[{\wc_n} ]^* + (1-\tau_n)   \mathbb{E}[\wcx_n ]^*  )
$ is the   overall welfare (spectrum efficiency) under the optimal allocation (using precise C-MWs), wherein $(\tau_n   \mathbb{E}[{\wc_n} ]^* + (1-\tau_n)   \mathbb{E}[\wcx_n ]^*)$ is the expected welfare of the contract user $n$ under the optimal allocation. Note that $\rho S \cdot   \gamma_n$ is the negative change of the expected number of spectrums allocated to contract user $n$.

Let us elaborate on the insights behind (\ref{eq:eps-RII-AP}). \rev{First, there is a welfare loss ratio $(1-\bar{\epsilon})$ on the spot market (either the whole spot market, or the side spot market of a particular independent contract user set), shown by the first term in  (\ref{eq:eps-RII-AP}). This welfare loss ratio is the same as the performance loss ratio of the adopted approximate MWIS algorithm.
Second, there is an average welfare loss $t_n  \triangleq X_n \cdot \gamma_n - \widehat{P}_n \cdot \left[\gamma_n\right]^+ $ for each contract user $n$.} 
\revh{Obviously, if $\gamma_n > 0$, which implies that the number of spectrums for contract user $n$ is decreased (under approximate C-MWs), we have: $t_n = X_n \cdot \gamma_n - \widehat{P}_n \cdot \gamma_n = -\lambda_n^* \cdot \gamma_n \leq 0$; and if $\gamma_n < 0$, which implies the number of spectrum for contract user $n$ is increased (under approximate C-MWs), we also have: $t_n = X_n \cdot \gamma_n < 0$.}

Finally, we have the following bound for the relative welfare ratio (WR):
 \begin{equation}\label{eq:eps-RII-RP}
\begin{aligned}
\textstyle
\textsc{WR} = \frac{\textsc{Wf}^{\textsc{Achi}}}{  \textsc{Wf}^{\textsc{Opt}} }
 \geq
\bar{\epsilon} +  \sum_{n=1}^N
\frac{ (1-\bar{\epsilon})\cdot   \big( \tau_n   \mathbb{E}[{\wc_n} ]^* + (1-\tau_n)   \mathbb{E}[\wcx_n ]^*  \big) +  \rho S
\cdot t_n }{ \mathbb{E}[W ]^* }.
\end{aligned}
 \end{equation}

Note that the above discussion concerns the theoretical \textit{lower-bound} of the achieved performance ratio. Even if this bound is poor, the actual welfare may still be maintained at a good level. We will show by our simulation results 
in Section \ref{sec:simu2} 
that even with the worst MWIS algorithm in our simulations, the actual welfare ratio is more than 70\%; whereas, the theoretical bound is lower than $\overline{\epsilon}$ (i.e., 50\%).
\revth{Moreover, in \cite{Techrpt}, we will show that the above welfare ratio is \emph{conservatively} tight, in the sense that there does not exist another bound that is always better than our proposed one in any case.}

%% file: Section6-Simu-OR.tex

\section{Simulation Results}\label{sec:simu}

We perform simulations in MATLAB. Unless otherwise stated, we will use the following default setting: (i) all   spot market SUs are randomly distributed in a square area with 1000m$\times$1000m; (ii) all contract SUs are placed   on some predetermined  locations in the square area;  and (iii) all SUs' valuations are uniformly distributed in $[0,1]$, i.e., $F_{\boldsymbol{v}_m}(x) = F_{\boldsymbol{u}_n}(x) = x, \forall x\in [0,1]$.
Let IRs and IRc denote the interference ranges of the spot market users and contract users, respectively.

\subsection{Optimal Social Welfare $\mathbb{E}[W]^*$}


\begin{figure*}
\hspace{-5mm}
  \begin{minipage}[t]{0.5\linewidth}
    \centering
    \includegraphics[scale=.4]{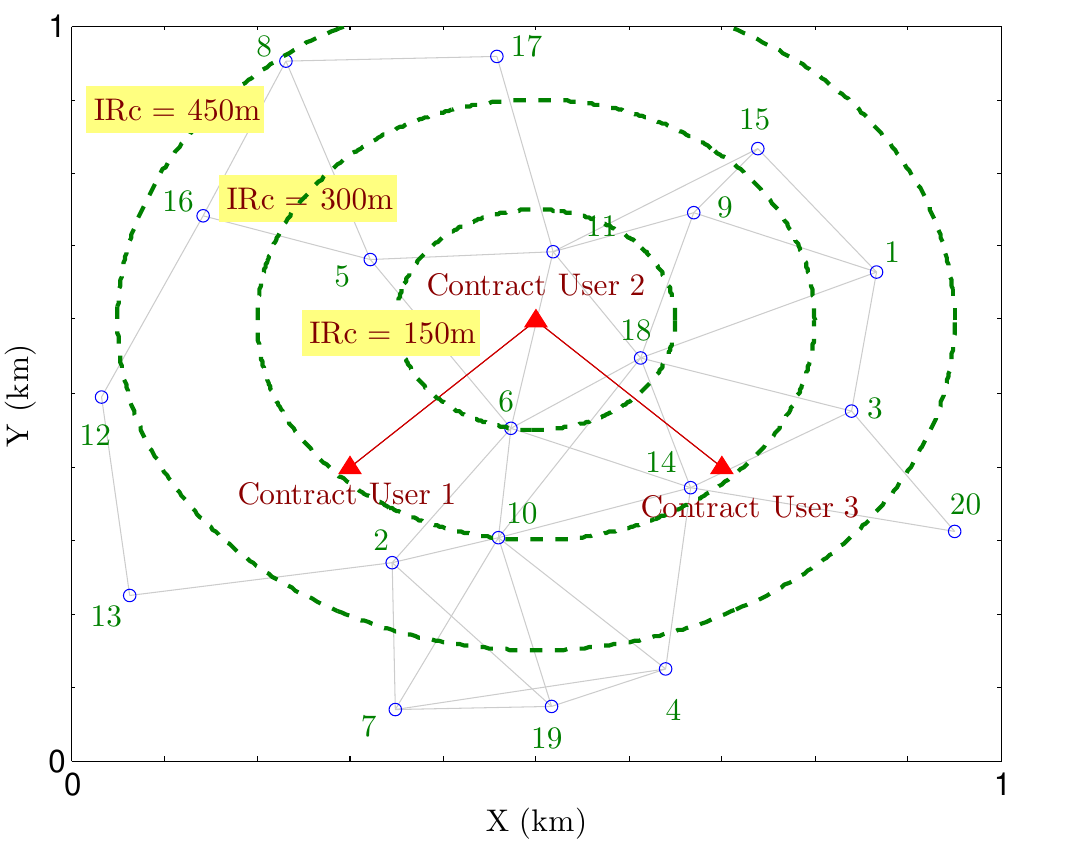}
    \caption{Network topology.} \label{f-topo}
    \end{minipage}
%
  \begin{minipage}[t]{0.005\linewidth}~
  \end{minipage}
  \begin{minipage}[t]{0.5\linewidth}
    \centering
    \includegraphics[scale=.4]{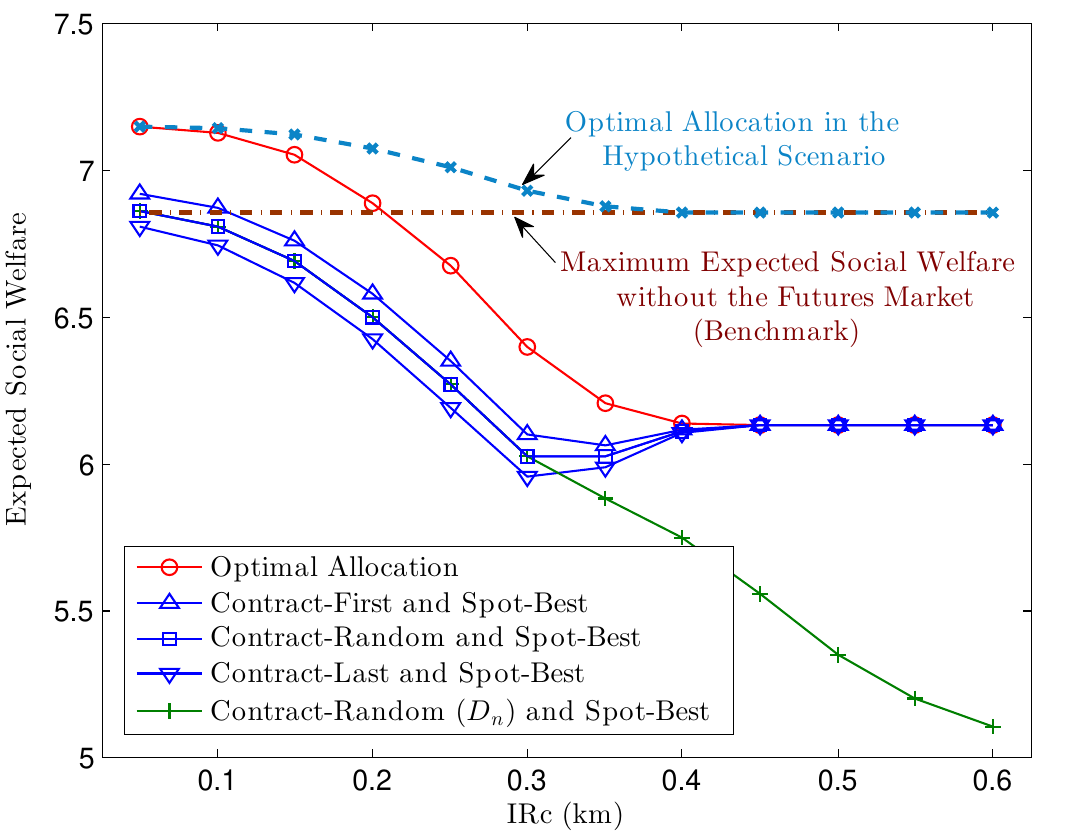}
    \caption{Expected social welfare.} \label{f-topo-profit}
  \end{minipage}
\end{figure*}

Fig.~\ref{f-topo} presents a realization of the random network topology with 20 randomly distributed spot market users and 3 contract users located at (300m, 400m), (500m, 600m) and (700m, 400m), respectively.
The interference range of spot market users is IRs = 300m.
The ellipses denote different values of IRc, i.e., the interference range of contract users.
Given a particular IRc, the spot market users outside the ellipse(s) of a contract user (set) form the corresponding side market.
The edges between contract users denote their interference relationships under a certain IRc (e.g., when IRc=300m). In Fig.~\ref{f-topo}'s example, contract users 1 and 2 (or 2 and 3) interfere with each other, while contract users 1 and 3 do not. Therefore, there are totally four independent contract user sets: \{1\}, \{2\}, \{3\}, and \{1, 3\}.

Fig.~\ref{f-topo-profit} presents the expected social welfare~\emph{vs}~IRc under different allocation strategies. Note that each value in the figure is the average of 1,000 simulations, and each simulation corresponds to a particular realization of the random network topology.
The dashed straight line denotes the maximum expected social welfare \textit{without} the futures market (benchmark), which is achieved by optimally allocating each spectrum to the spot market (i.e., to an MWIS of the spot market).
Obviously, this benchmark is independent of the interference range of contract users (IRc).
\revth{The upper dash curve (with mark ``$\times$'') denotes the maximum expected social welfare in a hypothetical hybrid market, which consists of both the spot market and the futures market, but ignores the potential
utility loss of contract users (due to the spectrum under-allocation) in the futures market}.
The red curve (with mark ``$\bigcirc$'') denotes the maximum expected social welfare under the optimal allocation strategy in our proposed hybrid market.
\reee{The lower three blue curves denote the expected social welfares under allocation strategies ``Contract-Last and Spot-Best'' (with mark ``$\bigtriangleup$''), ``Contract-Random and Spot-Best'' (with mark ``$\square$''), and ``Contract-Last and Spot-Best'' (with mark ``$\bigtriangledown$''), respectively.
Under the ``Contract-Last/Random/First and Spot-Best'' strategies, the SR allocates $\mathbb{E}[D_n]$ spectrums with the lowest,  arbitrary, and highest utilities to each contract user $n$,  respectively.}~Here, $\mathbb{E}[D_n]$ is the expected number of spectrum contract user $n$ achieves under the optimal allocation (derived in our paper).
The green curve (with mark ``$+$'') denotes the expected social welfare under such an allocation strategy that randomly allocates $ D_n $ (i.e., the demand of contract user $n$) spectrums to each contract user $n$, called ``Contract-Random ($D_n$) and Spot-Best''.
Note that under all of the above four strategies, when a spectrum is allocated to a contract user (or a set of independent contract users) in the futures market, it will also be allocated to an MWIS of the associated side (spot) market at the same time.
This is the reason that we call them ``Spot-Best''.~~~~~~

Fig.~\ref{f-topo-profit} shows that our proposed optimal allocation strategy outperforms other strategies
significantly.
The maximum expected social welfare under the proposed optimal strategy is on average 20\% higher than that under ``Contract-Random ($D_n$) and Spot-Best''.
This performance gap becomes even larger with more contract users.
\reee{Moreover, the performances under ``Contract-First/Random/Last and Spot-Best'' are very close to each other.
This is because different SUs' valuations for the same spectrum are independent of each other; hence,  a
spectrum that brings a high or low valuation for a contract user may not bring high or low valuations for spot market users.
When IRc is small, the performance under ``Contract-Last and Spot-Best'' is a bit higher than that under
``Contract-Random and Spot-Best'', which is further a bit higher than that under ``Contract-First and Spot-Best''.
This is because allocating a spectrum to a spot market
user can generate a higher effective welfare than allocating a spectrum (with the same valuation) to a contract user (as the latter is
discounted by $\tau$).
When IRc is large, the performances under ``Contract-First/Random/Last and Spot-Best'' are identical, and all converge to the maximum expected social welfare achieved in the proposed optimal allocation.
This is because $\mathbb{E}[D_n] \rightarrow 0$ in this case; thus, all spectrums are optimally allocated to the spot market as in our proposed optimal allocation.
Notice that this maximum expected social welfare is
less than the benchmark (without the futures market), due to the utility loss of contract users in our hybrid market.}
Finally, the performance under
``Contract-Random ($D_n$) and Spot-Best'' is same as that under ``Contract-Random and Spot-Best'' when IRc is small; whereas, it is worse than that under ``Contract-Random and Spot-Best'' when IRc is large.
This reason is as follows.
When IRc is small, $\mathbb{E}[D_n] = D_n$ (i.e., the contract user demand is fully satisfied) due to the weak conflict between spot market users and contract users);
when IRc is large, $\mathbb{E}[D_n] \rightarrow 0 $ due to the severe conflict between spot market users and contract users.~~~~~~~~~~~~~~


Fig.~\ref{f-topo-profit} further shows that the maximum expected social welfare with the futures market (under the proposed optimal strategy) may be smaller than the maximum expected social welfare without the futures market (benchmark), when, for example, IRc is large.
That is, accepting contract users in the futures market may reduce the total achieved social welfare,  compared to that under a pure spot market.
Notice that this welfare gap can serve
as a guidance used to decide whether or not to accept futures  contracts prior to the spectrum
allocation optimization considered in this work.
For example, in this figure, it is more efficient to accept futures contracts when IRc is smaller than 0.2, while it is more efficient in a pure spot market when Irc is larger than 0.2.
The reason that  a hybrid market may be worse than a pure spot market is as follows.
\revth{As long as a contract is accepted, the SR has the responsibility to fulfill the contract; otherwise, the contract user will suffer certain utility loss, which harms the entire social welfare.
In the benchmark case, however, we consider the spot market only, while removing all contract users as well as their impacts on the social welfare. That is, in this benchmark case, the SR will not allocate any spectrum to contract users and there will be no penalty for this.}

\revth{For a fair and more meaningful comparison, we propose the above hypothetical hybrid market scenario in our simulation.
Specifically, in the hypothetical scenario, there is no utility loss for contract users when their demands are not satisfied. Clearly, the benchmark solution (ignoring both the futures market and utility loss for contract users) is a feasible solution of operating such a hypothetical scenario (considering the futures market but ignoring the utility loss for contract users), but not a feasible solution of operating our proposed hybrid market (considering both the futures market and utility loss for contract users).
We can see that the maximum social welfare in the hypothetical scenario is always higher than the
maximum that can be achieved in the our proposed hybrid market and the maximum that can
be achieved in the benchmark case. More specifically, when the IRc is small, the maximum social
welfare in the hypothetical scenario is close to that in our proposed hybrid market. This is because
with a small IRc, the needs of contract users are likely to be fulfilled in both the proposed and the
hypothetical hybrid markets, hence leading to a small difference between these two market scenarios.
When the IRc is large, however, the maximum social welfare in the hypothetical scenario is close to that in the benchmark case. This is because with a large IRc, the needs of contract users are likely
to be violated (i.e., assigned few or even no spectrum by the SR) in both the proposed and the
hypothetical hybrid markets, hence leading to a large difference between these two market scenarios
(due to the welfare loss in our proposed scenarios), but a small difference between the hypothetical
scenario and the benchmark case (as in these two cases, spectrums are allocated to the spot market
only, without any welfare loss from contract users in the futures market).
}


\subsection{Welfare Loss by Approximate MWIS Algorithms}
\label{sec:simu2}

Now we study the welfare loss and the relative welfare ratio (WR) under the approximate MWIS algorithms.
We take the performance ratio of the adopted approximate MWIS algorithm as a random variable $\epsilon$ drawn from $[e_0,1]$ uniformly and independently, where $e_0\in [0,1]$. Accordingly, the average performance ratio is $\overline{\epsilon} = 0.5 \cdot (1+e_0)$.

\begin{figure*}
\hspace{-5mm}
  \begin{minipage}[t]{0.5\linewidth}
    \centering
    \includegraphics[scale=.4]{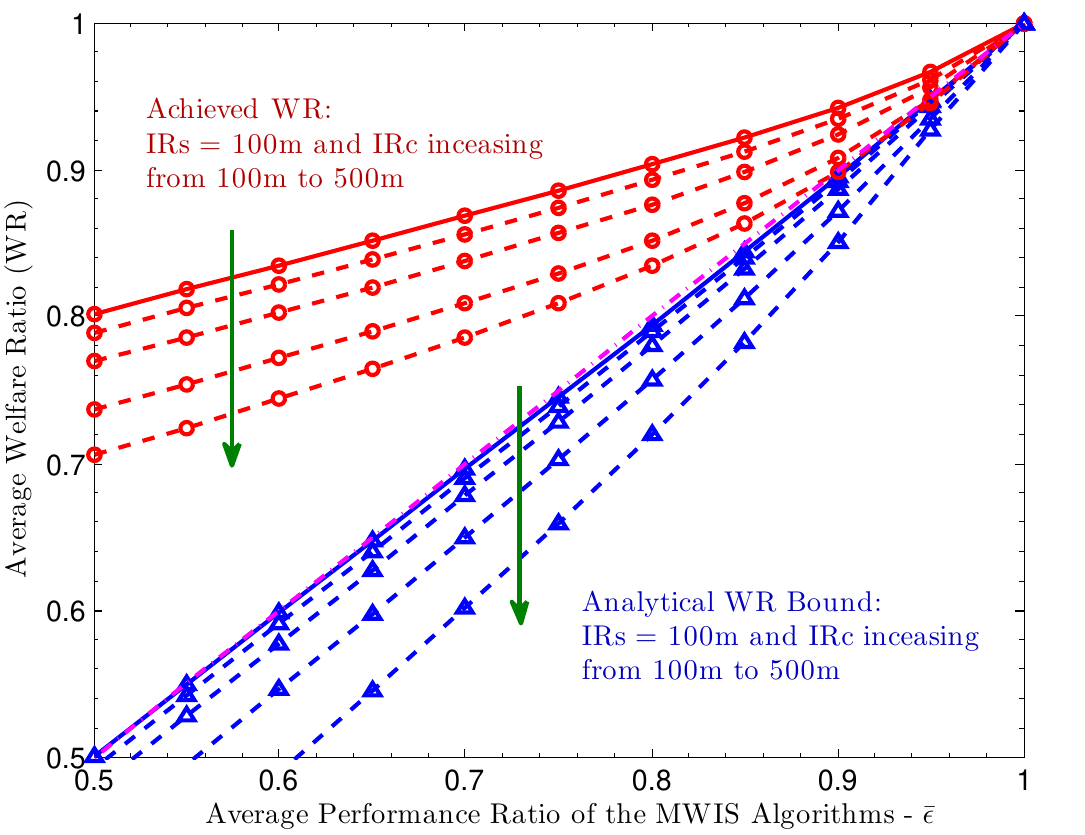}
    \caption{The achieved WR and analytical WR bound with IRs=100m.} \label{f-topo-PL}
    \end{minipage}
  \begin{minipage}[t]{0.005\linewidth}~
  \end{minipage}
  \begin{minipage}[t]{0.5\linewidth}
    \centering
    \includegraphics[scale=.4]{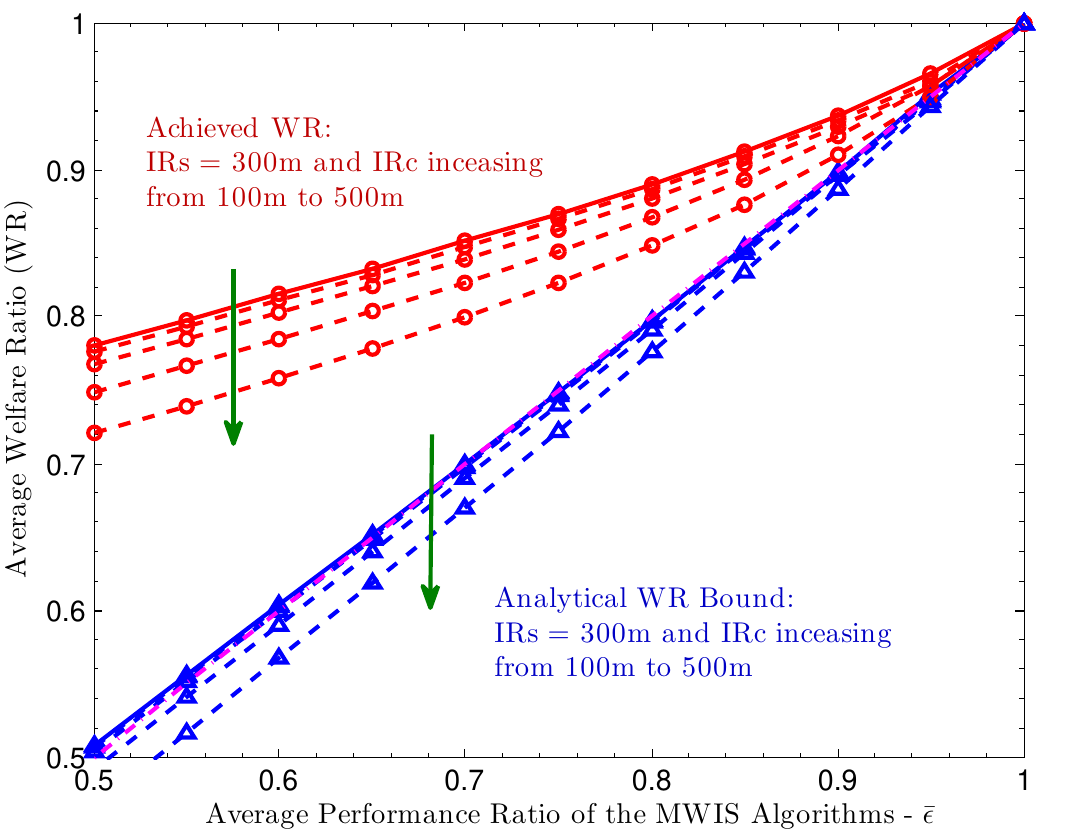}
    \caption{The achieved WR and analytical WR bound with IRs=300m.} \label{f-topo-PL3}
  \end{minipage}
%
\vspace{-5mm}
\end{figure*}

Figs.~\ref{f-topo-PL} and \ref{f-topo-PL3} present the achieved WR and  the analytical WR bound given in (\ref{eq:eps-RII-RP}) under different approximate MWIS algorithms (with different average performance ratio $\overline{\epsilon}$).
\revh{The upper (red) five curves denote the achieved WR when IRc increases from 100m to 500m, and the lower (blue) five curves denote the analytical WR bounds when IRc increases from 100m to 500m.}
From Figs.~\ref{f-topo-PL} and \ref{f-topo-PL3}
we can see that WR decreases with IRc. This is because a larger IRc means a smaller side welfare $z_i(\theta)$ for each independent contract user set $\ISc_i$, and the side welfare $z_0(\theta)$ becomes relatively more significant.
Thus, the C-MW $\CMW_i(\theta)$ defined in Definition \ref{def:mp2} is mainly determined by $z_0(\theta)$, and the welfare loss is mainly due to the degradation of $z_0(\theta)$. Additionally, with certain degradation of $z_0(\theta)$, the spectrum intended for the 0-th independent contract user set (i.e., the spot market only) may be mis-allocated to another independent set (e.g., the $i$-th one) with a welfare loss
$ z_{0}(\theta) -  \epsilon_{i}\cdot z_{i}(\theta) + \sum_{n=1}^N l_{n}^i (1-\tau_n) u_n(\theta) $. Obviously, the smaller the side welfare $  z_{i}(\theta)$, the higher the possible welfare loss.
We can also see that the gap between the relative WR (either the achieved WR or the analytical WR bound) under different values of IRc becomes smaller, when IRs becomes larger. \revh{For example, when IRs=100m (in Fig.~\ref{f-topo-PL}), the gap between the achieved WR under IRc=100m and IRc=500m is $80\%-70\%=10\%$ in the worst case ($\overline{\epsilon}=0.5$); while when IRs = 300m (in Fig.~\ref{f-topo-PL3}), this gap is only $78\% - 72\% =6\%$.} This is because a larger IRs implies a lower level of $z_i(\theta)$; thus, the impact of IRc on $z_i(\theta)$ becomes less significant on the C-MWs.~~~~~~~~~~~

\rev{Figs.~\ref{f-topo-PL} and \ref{f-topo-PL3}   show that even with the worst approximate MWIS algorithm in our simulations (i.e., $\overline{\epsilon} =0.5$ when $\epsilon$ drawn from $[0,1]$), the achieved social welfare can reach more than 70\% of the optimal social welfare. This implies that the optimal allocation (policy) proposed in this work is robust to the performance degradation of the approximate MWIS algorithms.}
\revth{We will further show in \cite{Techrpt}, that the above welfare ratio bound is conservatively tight, in the sense that there does not exist another bound that is \textit{always} better than our proposed one in any case.}

%% file: Section9-Conclusion-OR.tex
\section{Conclusion}\label{sec:conclu}

We study the short-term secondary spectrum trading in a {hybrid} spectrum  market, and focus on the {spectrum efficiency maximization} with spatial spectrum reuse under {stochastic} and asymmetric information.
We first compute an off-line optimal policy that specifies the allocation of any spectrum under every possible information realization, and then design an on-line VCG auction that elicits SUs' private information and determines the real allocation of every  spectrum.
Such a solution technique allows us to optimally allocate every spectrum in an
on-line manner, based on the policy derived and the information elicited.
We further propose a heuristics solution
based on a low complexity VCG-like mechanism, and quantify the associated spectrum efficiency loss.

\revf{We would like to point out that the problem considered in this work (i.e., spectrum
allocation in an exogenous hybrid market  in which every SU is associated and fixed in a particular market) can be viewed as a sub-problem of the whole market optimization
problem, which consists of both the hybrid market formulation and the spectrum allocation.
Our analysis of spectrum allocation in a given market provides a very important first
step towards the ``social optimal design'' for the whole model.
In our future work, we will study the problem of joint market formulation and spectrum allocation, in which every SU can make strategic decisions about the market selection, and can further negotiate with the SR for the contract details if choosing the futures market.}

%% file: Section-Ref-OR.tex

\bibliographystyle{nonumber}

%% file: Section-Appendix.tex

\newpage

\section{Appendix}

This file serves as the online technical report for the paper  ``Combining Spot and Futures Markets: A Hybrid Market Approach to Dynamic Spectrum Access'' by Lin~Gao, Biying~Shou, Ying-Ju~Chen, and  Jianwei~Huang, which is published in INFORMS Operations Research.

\subsection*{Outline of This Technical Report}

\begin{itemize}
\item \ref{app:truthful}. Model Justification: Truthful Revelation of SU Demand Type
\item \ref{app:expected}. Model Justification: Impact of Expected Contract Demand
\item \ref{app:small}. Model Justification: Problem of ``Small Allocation''
\item \ref{sec:sic-nece}. Model Justification: Some Necessary Conditions for SIC


\item \ref{sec:opt-hard}. Model Extension: Optimal Solution with Hard Contracts

\item \ref{app:WR}. Performance Justification: Tightness of Welfare Ratio Bound

\item \ref{prop:feasible-proof}. Proof for Proposition \ref{prop:feasible}
\item \ref{prop:restriction-proof}. {Proof for Proposition \ref{prop:restriction}}
\item \ref{lemma:nc1-proof}. {Proof for Lemma \ref{lemma:nc1}}
\item \ref{lemma:nc2-proof}. {Proof for Lemma \ref{lemma:nc2}}
\item  \ref{lemma:nc2-cor-proof}. {Proof for Corollary \ref{lemma:nc2-cor}}
\item \ref{lemma:mu-multi-proof}. {Proof for Lemma \ref{lemma:mu-multi}}
\item \ref{lemma:eta-multi-proof}. {Proof for Lemma \ref{lemma:eta-multi}}
\item \ref{lemma:opt-multi-proof}. {Proof for Lemma \ref{lemma:opt-multi}}
\item \ref{lemma:lambda-multi-proof}. {Proof for Lemma \ref{lemma:lambda-multi}}
\end{itemize}

\input{Section-Appendix-Response-OR}

\input{Section-Appendix-ProbFormu-OR}

\input{Section-Appendix-Proofs-OR}

%% file: Section-Appendix-Response-OR.tex

\subsection{Truthful Revelation of SU Demand Type}\label{app:truthful}

We first elaborate on an SU's demand type (elastic or inelastic), which mainly depends on the QoS requirement of the associated application/service.
For example, applications, such as FTP downloading, data backup, and cloud synchronization have elastic demand for spectrum, in the sense that the tasks are not urgent in terms of time.
Thus, the QoS will not be significantly affected, even if the spectrum resource is limited and the transmission rate is low for a substantial amount of time.
On the other hand, applications, such as VoIP, video streaming, and real-time data collection have inelastic demand for spectrum.
In other words, they can only function well when the data rate is guaranteed to be above certain thresholds such that the delay requirements are met, otherwise they will suffer significant performance loss.
It is important to note  that the information of application type is usually explicitly represented in the headers of data packets, and can be easily extracted by the network operators through deep packet inspection. Hence, it is reasonable to assume that SUs cannot fake their application types arbitrarily.

We further note that even if an SU has the capability of faking his or her application type, he or she does not have the incentive to do so if the hybrid market is properly designed.
Specifically, the futures market insures SUs against uncertainties of future spectrum acquisition  through pre-defined contracts; whereas, the spot market allows SUs to compete for spectrum based on their real-time demands and preferences.
Therefore, {with a proper design of the contracts in the futures market}, a rational SU with elastic demand will generally prefer
 spot transactions to achieve the flexible resource-price tradeoff.
 In contrast, a rational SU with inelastic demand requiring minimum data rates will prefer the certainty of contract in the futures market.

For example, consider an SU $n$ with an inelastic demand $D_n$, a fixed total welfare ${B}_n$ (when the demand is satisfied), and a potential welfare loss scheme $\mathbf{J}_n$ (when the demand is not satisfied). Suppose the SR offers the following contract: $\textsc{Ctr}_n = \{\widetilde{B}_n, D_n, \mathbf{J}_n\}$.
Then the  SU's surplus when choosing such a contract is:\footnote{Here, we suppose that the SU's welfare loss due to the unmet demand is    exactly compensated by the penalty $\mathbf{J}_n $ specified in the contract.}
$$
S_n^{\textsc{f}} =  {B}_n - \widetilde{B}_n.
$$
On the other hand, when  SU $n$ chooses the spot market, the expected surplus is:
$$
S_n^{\textsc{s}} = {B}_n - Loss(D_n - \bar{d}_n) - \bar{d}_n \cdot \bar{p}_n ,
$$
where $\bar{d}_n$ is the average number of spectrum obtained through the spot market, and $\bar{p}_n$ is the average payment for every spectrum.
As long as the welfare loss $Loss (D_n - \bar{d}_n)$ is sufficiently large, the SU will bid in such a way that the achieved demand $\bar{d}_n$ equals the required demand $D_n$.
Thus, the expected surplus under such a bidding strategy in the spot market  is:
$$
S_n^{\textsc{s}} = {B}_n- D_n \cdot \bar{p}_n ,
$$
where $\bar{p}_n $ is the average payment for every spectrum.
By comparing the SU's surpluses in both market markets, we can easily find that if the payment specified in the contract is no larger than the actual payment in the spot market (i.e., $\widetilde{B}_n \leq D_n \cdot \bar{p}_n $),  the SU can achieve a larger surplus from the contract in the futures market.
Note that such a contract design is also desirable by a social-planning spectrum regulator, as truth-telling of SU is the prerequisite for the spectrum utilization maximization.

Similarly, consider another example in which SU $n$ has the elastic demand.
When choosing the spot market, he or she bids truthfully  (under our proposed auction mechanism) and achieves an expected surplus:
$$
S_n^{\textsc{s}} =  \bar{d}_n\cdot (\bar{w}_n - \bar{p}_n) .
$$
Here, $\bar{d}_n$ is the average number of spectrum obtained through the spot market, $\bar{p}_n$ is the average payment for every spectrum, and $\bar{w}_n$ is the average welfare achieved from every spectrum.
Suppose that the SR offers the following contract $\textsc{Ctr}_n = \{\widetilde{B}_n, D_n, \mathbf{J}_n\}$, where $D_n = \bar{d}_n$.
When choosing the contract, the SU's  expected surplus is:
$$
S_n^{\textsc{f}} =  \widetilde{d}_n\cdot \widetilde{w}_n  + Penalty(D_n,\widetilde{d}_n ) -  \widetilde{B}_n ,
$$
where $\widetilde{d}_n\leq D_n$ is the average number of spectrum obtained through the contract, and $\widetilde{p}_n$ is the average welfare achieved from every spectrum.
We further suppose that the penalty is defined in the following way: $Penalty(D_n,\widetilde{d}_n ) = (D_n - \widetilde{d}_n) \cdot \frac{\widetilde{B}_n}{D_n}$, where $\frac{\widetilde{B}_n}{D_n}$ is the average payment for every spectrum. That is, the SR will pay back  all of the unmet demand to the SU.
Then, the SU's expected surplus when choosing the contract is:
$$
\textstyle
S_n^{\textsc{f}} =  \widetilde{d}_n\cdot \widetilde{w}_n  - \widetilde{d}_n\cdot \frac{\widetilde{B}_n}{D_n}.
$$
Notice that $\widetilde{d}_n\leq D_n = \bar{d}_n$ and $\widetilde{w}_n \leq \bar{w}_n$ (as the SU is more likely to obtain the high welfare spectrum through the auction). Thus, if the average payment in the contract is no smaller than the average payment in the spot market, i.e., $\frac{\widetilde{B}_n}{D_n} \geq \bar{p}_n$, the SU can achieve a larger surplus from spot market.~~~~~~~

From the above discussions, we show that through a proper design of contracts in the futures market, we can prevent an SU with   inelastic demand from choosing the spot market, and an SU with  elastic demand from choosing the futures market. Nevertheless, how to fully characterize the future contract such that any SU will truthfully reveal his or her demand type and any other required parameters (such as the demand/welfare loss of an SU with inelastic demand, or the average number of spectrum achieved in the spot market for an SU with elastic demand) is a challenging problem
in general.
Thus, to maintain analytical tractability, in this paper we focus on  the optimal spectrum allocation problem in an \emph{exogenously given} hybrid market, in which all SUs are assumed to truthfully reveal their demand types (market selections). In other words, the spot market and the futures market are already given and fixed.

\subsection{Impact of Expected Contract Demand}\label{app:expected}

In our model, we assume that a contract user cares about the expected number of spectrum (reflecting the expected data rate) that he or she will obtain, while not the strict number of spectrum (reflecting the strict data rate). The key reason for this assumption is as follows.

The main practical  motivation for such an assumption is that most wireless applications in practice require an \emph{expected/average} data rate during a certain time period, while only very few applications is required to satisfy a \emph{strict} data rate constraint at all   times.
For example, video streaming concerns the average downloading rate in every minute, and VoIP concerns the average downloading/uploading rate in every second.
When the actual data rate is occasionally less than the average rate, various coding and error concealment technologies can be employed so that the end-users will not feel significant performance degradation.
Thus, with a proper choice of the length of allocation period (e.g., 1 minute for video streaming, or 1 second for VoIP), such a contract formulation (with the expected spectrum demand) is suitable for most wireless applications.
In fact, with the uncertainties of
spectrum availability and SU utility for each spectrum, it is practically impossible  to guarantee a strict spectrum supply for each contract user. To our knowledge, no such guarantee has been provided even in the latest communication standard.

To make our analysis more complete, we also provide the detailed theoretical  evaluation for the impact of this assumption on the achieved performance.
That is, we evaluate the potential welfare loss induced by this assumption, i.e., the gap between the maximum social welfare under the expected contract demand and under the strict contract demand.
The key idea and details are given as follows.
First, we denote $W_{\textsc{e}}$ as the expected social welfare computed by the expected contract demand, and $\boldsymbol{A}^{*}_{\textsc{e}}$ as the optimal allocation that maximizes the expected social welfare $W_{\textsc{e}}$ (i.e., those studied in this paper). That is:
$$
\boldsymbol{A}^{*}_{\textsc{e}} = \arg \max_{\boldsymbol{A}} \  W_{\textsc{e}} (\boldsymbol{A}).
$$
Similarly, we denote $W_{\textsc{s}}$ as the strict social welfare computed by the strict contract demand, and $\boldsymbol{A}_{\textsc{s}}^{*}$ as the optimal allocation that maximizes the strict social welfare $W_{\textsc{s}}$. That is:
$$
\boldsymbol{A}^{*}_{\textsc{s}} = \arg \max_{\boldsymbol{A}}\  W_{\textsc{s}} (\boldsymbol{A}).
$$
Then, using basic optimization principles, we have the following observation:
$$
W_{\textsc{e}}^* \geq W_{\textsc{s}}^* \geq W_{\textsc{s}}(\boldsymbol{A}^{*}_{\textsc{e}}),
$$
where $W_{\textsc{e}}^* \triangleq W_{\textsc{e}}(\boldsymbol{A}^*_{\textsc{e}})$ is the maximum social welfare under the expected contract demand constraint, and $W_{\textsc{s}}^*  \triangleq W_{\textsc{s}} (\boldsymbol{A}^*_{\textsc{s}})$ is the maximum strict social welfare  under the strict contract demand constraint.
The first inequality follows because the expected contract demand constraint is looser than the strict contract demand constraint; hence, the maximum social welfare $W_{\textsc{s}}^* $ under the strict contract demand constraint is upper-bounded by the maximum social welfare $W_{\textsc{e}}^*$ under the expected contract demand constraint.
The second inequality follows because $\boldsymbol{A}^*_{\textsc{s}}$ is the optimal allocation that maximizes $W_{\textsc{s}}$; hence, $W_{\textsc{s}}^*  =W_{\textsc{s}}(\boldsymbol{A}^*_{\textsc{s}}) $ must be no smaller than $ W_{\textsc{s}}^*(\boldsymbol{A}^{*}_{\textsc{e}})$.

The above observation  further implies that the gap between $W_{\textsc{e}}^* $ and $ W_{\textsc{s}}^* $ is bounded by the gap between $W_{\textsc{e}}^* $ and $ W_{\textsc{s}}(\boldsymbol{A}^{*}_{\textsc{e}})$. That is:
$$
W_{\textsc{e}}^* - W_{\textsc{s}}^* \geq W_{\textsc{e}}^* - W_{\textsc{s}}(\boldsymbol{A}^{*}_{\textsc{e}}).
$$
The above formula implies that we can use the gap $W_{\textsc{e}}^* $ and $ W_{\textsc{s}}(\boldsymbol{A}^{*}_{\textsc{e}})$ to characterize an effective upper-bound of the gap between $W_{\textsc{e}}^* $ and $ W_{\textsc{s}}^* $.
Note that it is difficult to compute the maximum strict social welfare $W_{\textsc{s}}^*$ directly; whereas, it is easy to compute the strict social welfare $ W_{\textsc{s}}(\boldsymbol{A}^{*}_{\textsc{e}})$ under the allocation $\boldsymbol{A}^{*}_{\textsc{e}}$.
Thus, the above observation is useful for estimating the potential welfare loss induced by the assumption of expected contract demand.

\begin{figure}
\centering
\includegraphics[scale=0.35]{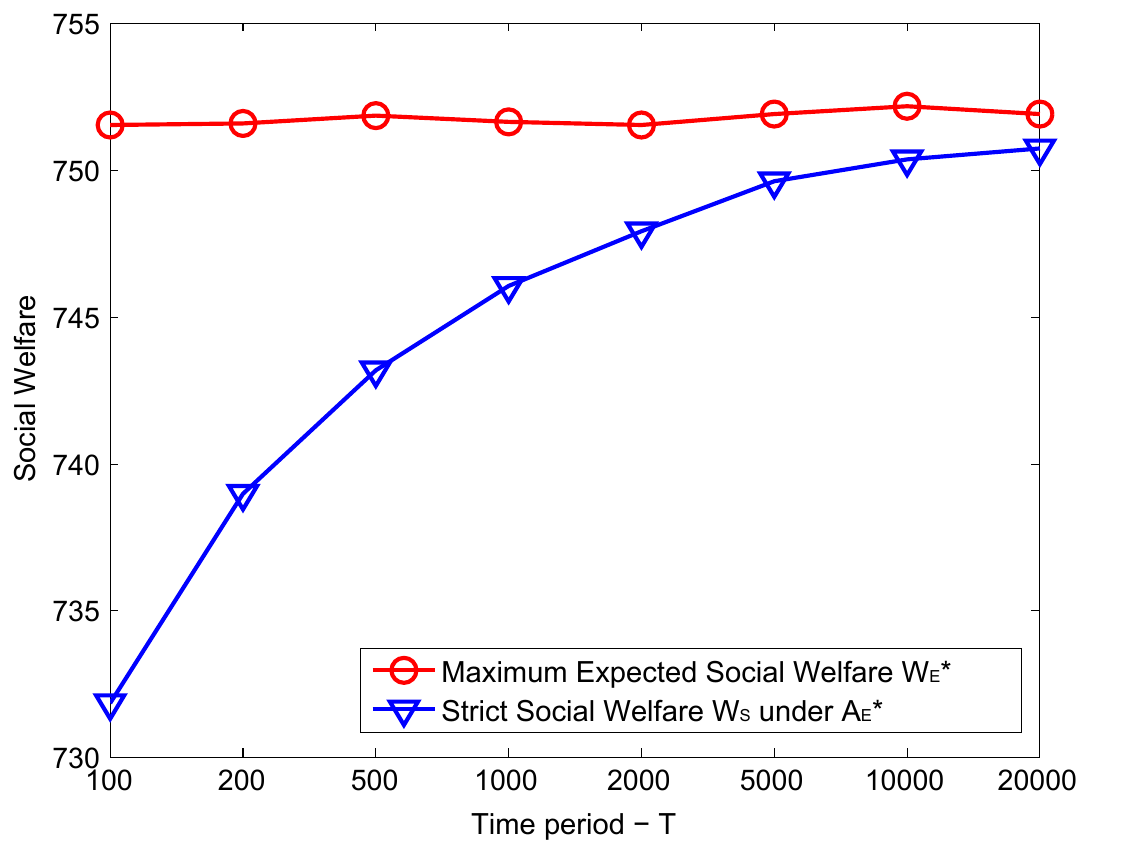}
\includegraphics[scale=0.35]{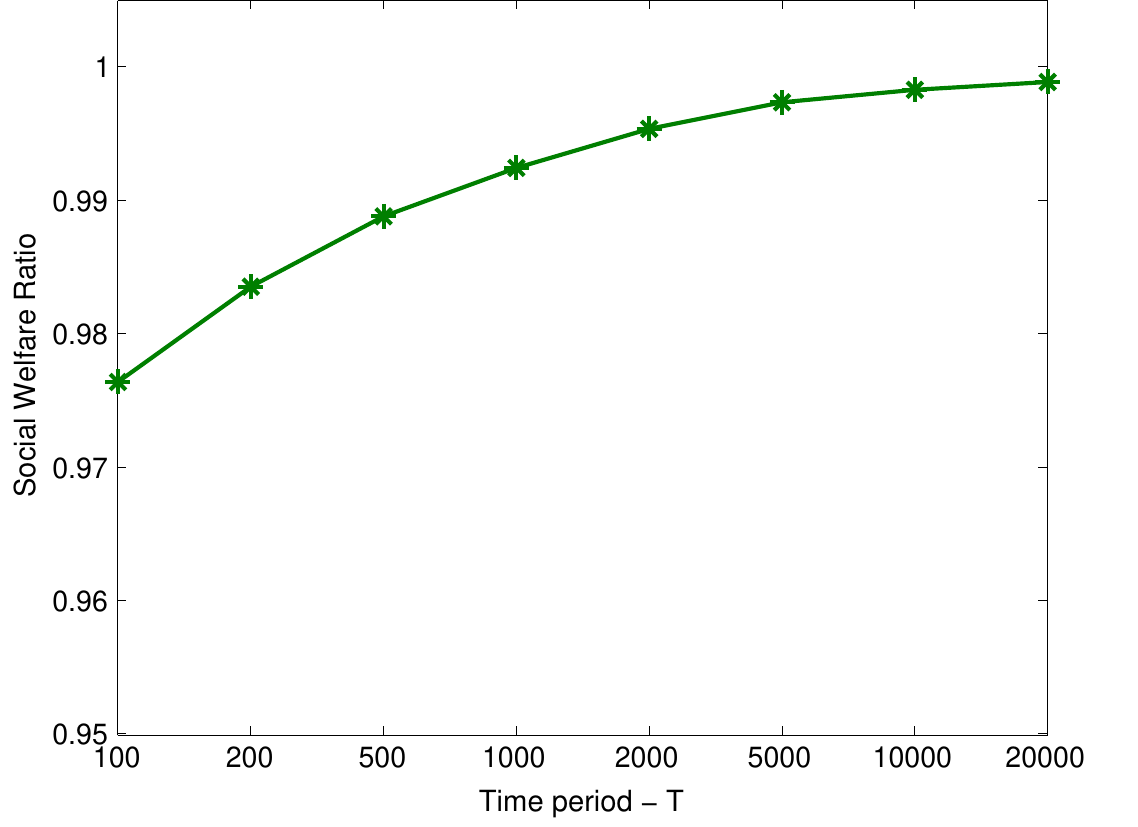}
\caption{(a) The maximum expected social welfare $W_{\textsc{e}}^*$ and the strict social welfare $W_{\textsc{s}}(\boldsymbol{A}^{*}_{\textsc{e}})$ under $\boldsymbol{A}^*_{\textsc{e}}$;
(b) The ratio of $W_{\textsc{s}}(\boldsymbol{A}^{*}_{\textsc{e}})$ and $W_{\textsc{e}}^*$.}\label{fig:gap}
\vspace{-5mm}
\end{figure}

Finally, we provide numerical results to illustrate the gap of $W_{\textsc{e}}^* $ and $ W_{\textsc{s}}(\boldsymbol{A}^{*}_{\textsc{e}})$.
Numerical results in Fig.~\ref{fig:gap} show that the gap is lower than 3\% when the scheduling  period $T$ is 100, and further decreases with the increasing of the scheduling period $T$.
\revv{Take the widely-used 3G LTE (FDD) system as an example. The time length of each slot is 1 ms, and thus a scheduling period of $T$ (slots) corresponds to a total time length of $T$ ms.
Then, the above numerical results indicate that the performance loss is no larger than 3\% when scheduling spectrum every $100$ ms, and is no larger than 1\% when scheduling spectrum every second (i.e., $T=1000$).
These results indicate that if the spectrum regulator jointly schedules the spectrum in a relatively long time period (e.g., larger than $100$ ms), the gap between the maximum expected social welfare and the maximum strict social welfare is very small (e.g., smaller than 3\%).}

\subsection{Problem of ``Small Allocation''}\label{app:small}

\revth{
It is important to note that our proposed analytical framework is available only when each contract user requests a considerable number of spectrum in each period, such that the time average approaches to the true expectation (of allocated spectrum).
In a counter-example, if a user requests a small allocation, say 1 channel for 1 period, he'll either get it or not, hence the time average won't be the expectation. Next we discuss such a ``small allocation'' problem.

We would like to clarify that such a ``small allocation'' is unlikely to happen in a practical wireless communication system. In
fact, each contract user requires a considerable number of spectrum in each period to support a desirable QoS. This is mainly due to two reasons. First, the length of each time slot in a wireless
system is often quite small, e.g., in milliseconds or even in microseconds, which corresponds to the
typical length of frame in many wireless communication systems (such as the 4G LTE networks
and Wi-Fi networks). In fact, the physical limit of choosing the time sloth length is the so called
\emph{coherence time}, which is the time within which the channel condition does not change. In wireless
communications, such coherence time is usually very small due to fast small scale multi-path fading.
Second, the time scale of each time period is relatively large, e.g., in minutes or even hours, which
corresponds to the validity period of contracts. Hence, the number of slots in each time period
(i.e., $T$) is very large, e.g., $T = 6\cdot 10^5$ when the length of slot is 1 millisecond and the length
of period is 10 minutes. Due to the short time slot and long time period, a contract user has to
require a considerable number of spectrum in each time period to achieve a certain data rate (so
as to achieve the desirable QoS). For example, when the length of slot is 1 millisecond, the length
of period is 10 minutes (i.e., $T = 6\cdot 10^5$), and the average data rate provided in each spectrum is
1Mbps, then a contract user with a data rate requirement of 264Kbps (i.e., the typical data rate
of the low resolution ($320\times 240$) video streaming service) requires a total of $1.584\cdot 10^5$ time slots in each time period.
}

%% file: Section-Appendix-ProbFormu-OR.tex

\subsection{Some Necessary Conditions for SIC}\label{sec:sic-nece}

Before providing the problem formulation, we first present the formal definition for \emph{allocation strategy} and study the corresponding spatial interference constraints (SIC). \rev{An allocation strategy specifies the allocation probability of any spectrum (to every SU) under any possible information realization $\theta$}.\footnote{Here we omit the time index $t$ of $\theta(t)$ for brevity, as we are not talking about a particular spectrum $t$.}
More specifically,
 \begin{definition}[Allocation Strategy]\label{def:alloc-strategy}
An allocation strategy $\boldsymbol{A}(\theta)$ is a mapping from every information realization $\theta$ to a vector consisting of the allocation probability for every SU,\footnote{Note that we use the same notation $\boldsymbol{\Theta}$ to denote the information space, i.e., the set of all possible  $\theta$.}
 \begin{equation*}
\boldsymbol{A}(\theta)\triangleq( {a}_1^{\mathrm{s}}(\theta), ..., {a}_M^{\mathrm{s}}(\theta);  {a}_1^{\mathrm{c}}(\theta), ...,  {a}_N^{\mathrm{c}}(\theta)),\quad \forall \theta \in \boldsymbol{\Theta},
 \end{equation*}
where $ {a}_m^{\mathrm{s}}(\theta) \in  [0,1]$ and $ {a}_n^{\mathrm{c}}(\theta) \in  [0,1]$ denote the allocation probabilities of a spectrum with information $\theta$ to spot market user $m$ and contract user $n$, respectively.
 \end{definition}

Now we study the necessary and sufficient conditions for a feasible allocation strategy $\boldsymbol{A}(\theta)$, referred to as the \emph{spatial interference constraints (SIC)}.

Since the SUs in the same clique can never use the same spectrum at the same time, we have the following necessary conditions for a feasible $\boldsymbol{A}(\theta)$:
 \begin{equation}\label{eq:feasible}
\textstyle \sum_{i \in Q} a_i (\theta) \leq 1, \quad \forall \theta \in \boldsymbol{\Theta},\ \forall Q \in \mathcal{Q}(G),
 \end{equation}
where $a_i(\theta) = a_i^{\mathrm{c}}(\theta)$ if SU $i$ is a contract user, and $a_i(\theta) = a_i^{\mathrm{s}}(\theta)$ if SU $i$ is a spot market user.

\begin{figure}
   \centering
    \includegraphics[scale=.6]{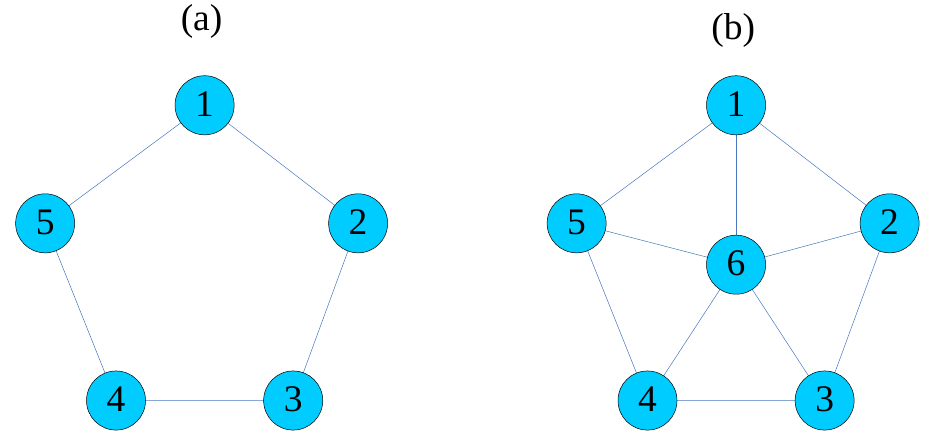}
    \caption{An illustration of graphs with odd holes.} \label{fig:SIC-examples}
\end{figure}

Unfortunately, the conditions in (\ref{eq:feasible}) are \emph{not} sufficient for satisfying the spatial interference constraints. This can be illustrated by a simple graph $G^{\mathrm{(a)}}$ with 5 SUs forming a ring, {as shown in Figure \ref{fig:SIC-examples} (a)}. \revh{We can easily see that the set of all cliques in graph $G^{\mathrm{(a)}}$ is $\mathcal{Q}(G^{\mathrm{(a)}}) = \{\{1\},\{2\},\{3\},\{4\},\{5\},\{1,2\},\{2,3\}, \{3,4\}, \{4,5\},\{5,1\}\}$.}
Consider an allocation strategy with $a_k(\theta) = 0.5, \forall k=1,...,5$. It satisfies the conditions in (\ref{eq:feasible}). But there is no feasible schedule to achieve this allocation, \revh{that is, it is not a feasible allocation.}
The same problem arises in any graph with \emph{odd holes}, i.e., a loop formed by an odd number of edges without a chord in between. \revh{This suggests that in addition to the conditions in (\ref{eq:feasible}), we need to add additional constraints corresponding to the odd holes. Specifically, referring to \cite{graph2001}, the constraint corresponding to an odd hole with $K$ vertices is:
\begin{equation}\label{eq:odd}
\textstyle
\sum_{k=1}^K a_k(\theta) \leq \frac{K-1}{2}, \quad \forall \theta \in \boldsymbol{\Theta}.
\end{equation}
Due to space limitations, we do not present the detailed proof here. Recall the example in Figure \ref{fig:SIC-examples} (a), an allocation strategy with $a_k(\theta) = 0.4, \forall k=1,...,5$ is feasible since $\sum_{k=1}^K a_k(\theta)\leq 2 $, and we can achieve such an allocation by the following schedule: scheduling/assigning the spectrums (with information $\theta$) to the user sets \{1,3\}, \{1,4\}, \{2,4\}, \{2,5\} and \{3,5\} with equal fraction (probability), i.e., 20\%. Obviously, each user achieves an allocation probability of 40\%.}

\revh{The worse thing is that even if we consider the constraints in (\ref{eq:odd}) imposed by the odd holes in addition to those in (\ref{eq:feasible}) imposed by the cliques, we are \emph{not} guaranteed to have a feasible allocation.
This can be further illustrated by the example shown in Figure \ref{fig:SIC-examples} (b).
We can see that there are 3 types of cliques in the graph $G^{\mathrm{(b)}}$: the cliques formed by each SU (e.g., \{1\}), the cliques formed by the adjacent pairs of SUs (e.g., \{1,2\}), and the cliques formed by the 3 SUs in the same triangles (e.g., \{1,2,6\}). We can also see there are five $3$-SUs odd holes each corresponding to one triangle and one $5$-SUs odd hole $\{1,2,3,4,5\}$.
Consider an allocation strategy with $a_k(\theta) = \frac{1}{3}, \forall k=1,...,6$. Obviously, it satisfies  the conditions  in (\ref{eq:feasible}) and (\ref{eq:odd}). But there is no feasible schedule to achieve this allocation.}

\subsection{Optimal Solution with Hard Contracts}\label{sec:opt-hard}

Similar to Lemma \ref{lemma:nc1} and Lemma \ref{lemma:nc2} in Section \ref{sec:incomp-nece}, we have the following necessary condition for hard contracts.

\begin{corollary}[Necessary Condition]\label{lemma:nc2-cor}
For any contract  user $n $ with the hard contract, we have:
$\mathbb{E}[d_n]   = 0$ or $ D_n$.
 \end{corollary}
Corollary \ref{lemma:nc2-cor} states that a hard contract user either gets spectrums (on average) equals to his demand, or nothing.
The reason is that once a hard contract is violated, the welfare loss is a constant $\widehat{B}_n$ regardless of the actual number of spectrums the contract user $n$ obtains.

According to Corollary \ref{lemma:nc2-cor}, a hard contract is either fully satisfied, i.e., $\mathbb{E}[d_n] = D_n$, or fully violated, i.e., $\mathbb{E}[d_n]   = 0$.
Thus, for a future market with $N$ hard contracts, there are $2^N$ possible outcomes based on the state (satisfied or violated) of each contract.
Each outcome corresponds to a potential best solution, and the global optimal solution is the one that achieves the highest profit within the $2^N$ potential best solutions.

Consider a particular outcome where a subset of hard contracts $\mathcal{N}^{+}$ are satisfied and others are violated, i.e., $\mathbb{E}[d_n]   = D_n, \forall n\in \mathcal{N}^{+}$ and $\mathbb{E}[d_n]   = 0, \forall n\notin \mathcal{N}^{+}$. Obviously, we have $a_n(\theta) \equiv 0, \forall \theta \in \boldsymbol{\Theta} $ for all contract users $ n \notin \mathcal{N}^{+}$, and thus we can ignore these contracts directly. For the reduced market (by removing those contract users not in $\mathcal{N}^{+}$), the PO's objective function in (\ref{eq:obj-incomp2-soft}) can be rewritten as follows:
\begin{equation}\label{eq:obj-incomp2-hard}
\begin{aligned}
\boldsymbol{A}_{0}^*(\theta) &\textstyle  = \arg \max_{\boldsymbol{A}_{0}(\theta)}\ \left( F + S \cdot \sum_{k=1}^K \int_{\theta} H_k(\theta) a_k(\theta) f_{\boldsymbol{\Theta}}(\theta) \mathrm{d} \theta\right),
\\
s.t. &\textstyle \quad \mbox{(i)} \quad  a_k(\theta)\geq 0,\ \forall k =1,...,K, \forall \theta \in \boldsymbol{\Theta}; \\
&\textstyle \quad \mbox{(ii)}\quad \sum_{k=1}^K a_k(\theta) \leq 1,\ \forall \theta\in \boldsymbol{\Theta};\\
&\textstyle \quad \mbox{(iii)}\quad \mathbb{E}[d_n]  = D_n,\ \forall n\in \mathcal{N}^{+};
\end{aligned}
 \end{equation}
Note that the notations in (\ref{eq:obj-incomp2-hard}) are defined in a same way as those in (\ref{eq:obj-incomp2-soft}).

Compare the objective function (\ref{eq:obj-incomp2-hard}) for hard contracts to (\ref{eq:obj-incomp2-soft}) for soft contracts, we can find that both objective functions have the  {same structure} \emph{except} for the third constraint. Specifically, in (\ref{eq:obj-incomp2-soft}) the third constraint is an \emph{inequality} constraint, while in  (\ref{eq:obj-incomp2-hard}) the third constraint is an \emph{equality} constraint. To solve (\ref{eq:obj-incomp2-hard}), we similarly introduce Lagrange multipliers $\mu_k(\theta) $ for constraint (i), $\eta(\theta) $ for constraint (ii) and $\lambda_n $ for constraint (iii).
The dual constraints for   $\mu_n(\theta) $ and $\eta(\theta) $ are same as those for (\ref{eq:obj-incomp2-soft}), while the dual constraint for  $\lambda_n $ becomes:
\begin{equation*}
\begin{aligned}
 \mbox{(D.3)} &\textstyle \quad D_n = \mathbb{E}[d_n]  , \mbox{ and } \lambda_n^*  \cdot \big(D_n  - \mathbb{E}[d_n] \big)  =0, \quad \forall n\in \mathcal{N}^+.
 \end{aligned}
\end{equation*}
By the Euler-Lagrange conditions for optimality and duality principle, we can derive the best solution for (\ref{eq:obj-incomp2-hard}) in a similar way to compute the optimal solution for (\ref{eq:obj-incomp2-soft}). Specifically, the feasible ranges of $\mu_k^*(\theta)$ and $\eta^*(\theta)$ are totally same as those for soft contracts given by Lemmas \ref{lemma:mu-multi} and \ref{lemma:eta-multi}. By the dual constraint for $\lambda_n $ mentioned above, the optimal $\lambda_n^*$ is a bit different to that for soft contracts given by Lemma \ref{lemma:lambda-multi}:
 \begin{equation*}
\textstyle
\lambda_n^* =\left(
\arg_{\lambda_n }  S \int_{\theta\in \boldsymbol{\Theta}_n^+(\boldsymbol{\Lambda}_{-n}^*,\lambda_n)} f_{\boldsymbol{\Theta}}(\theta) \mathrm{d} \theta = D_n \right),\quad \forall n\in \mathcal{N}.
 \end{equation*}
That is, for hard contracts, we do not need to restrict the shadow price $\lambda_n$ to be non-negative.
Given the optimal shadow prices, the optimal solution can be characterized by Lemma \ref{lemma:opt-multi}.

\subsection{Tightness of Welfare Ratio Bound}\label{app:WR}

In this section, we analyze the tightness of our proposed welfare ratio bound.
We first explain what factors affect the tightness of the welfare ratio (WR). We provide three examples to show when the bounds can be conservatively tight.
We then explain why an approximate algorithm and the WR analysis are  important in this problem, although the optimal policy can be solved off-line.~~~~~~~~

We first show that the tightness of the welfare ratio ($\mathrm{WR}$) in Eq.~(40) depends on the performance of the adopted approximate MWIS algorithm.
This is illustrated by our simulation results in Figs.~7 and 8, which show that $\mathrm{WR}$ is tighter
with better approximate MWIS algorithms (i.e., those with higher $\bar{\epsilon}$), and looser with worse approximate MWIS algorithms (i.e., those with lower $\bar{\epsilon}$). In particular, when  the adopted approximate MWIS algorithm achieves a similar performance to the optimal  MWIS algorithm (i.e., $\bar{\epsilon}$ is close to 1), we   have $\mathrm{WR} \rightarrow 1$, which implies that the welfare ratio bound is  tight in this case.

We further show that the tightness of $\mathrm{WR}$ also depends on the market structure.
For example, we have $\mathrm{WR} \rightarrow \bar{\epsilon}$,
if $\sum_{m=1}^M \mathsf{E}(w_m^s)^* >> \sum_{n=1}^N \big(\tau_n \mathsf{E}(w_n^c)^* + (1-\tau_n) \mathsf{E}(\widetilde{w}_n^c)^* \big)$, i.e.,
the potential welfare from the spot market is much larger than that from the futures market under the optimal allocation.
This implies that the proposed  welfare ratio bound is tight, since $\bar{\epsilon}$ is the performance bound of the adopted approximate MWIS algorithms.
On the other hand, we have $t_n \rightarrow 0$ and $\mathrm{WR} \rightarrow \bar{\epsilon} + (1-\bar{\epsilon}) = 1$, if
$\sum_{m=1}^M \mathsf{E}(w_m^s)^* << \sum_{n=1}^N \big(\tau_n \mathsf{E}(w_n^c)^* + (1-\tau_n) \mathsf{E}(\widetilde{w}_n^c)^* \big)$, i.e.,  the potential welfare from the futures market is much larger than that from the spot market.\footnote{\revth{Here we impliedly assume that there is no error in computing the optimal allocation among contract users in the futures market. This is due to the assumption that the size of futures market is assumed to be much smaller than that of the spot market.}}
This also implies that the proposed welfare ratio bound is tight.

In the above discussion, we have already provided three illustrative examples in which the proposed welfare ratio bounds are tight.
In the first example, we use a close-to-optimal approximate MWIS algorithm, and achieve a $\mathrm{WR} $ close to 1.
In the second example, we consider a market scenario with a neglectable futures market, and  achieve a $\mathrm{WR} $ close to $\bar{\epsilon}$, where $\bar{\epsilon}$ is the performance bound of the adopted approximate MWIS algorithms.
In the last example, we consider a market scenario with a neglectable spot market, and achieve a $\mathrm{WR} $ close to 1.
\revth{Based on the above discussion, we conclude that our proposed bound is conservatively tight, in the sense that there does not exist another bound that is \emph{always} better than our proposed one.}

Now we explain why an approximate algorithm and the WR analysis is important in this problem.
The optimal policy  that maximizes the expected spectrum efficiency (in Theorem 1)  can be derived in an off-line manner based on the stochastic network information.
This indeed is one of the main contributions of this paper.
However, the actual allocation of each spectrum in each time slot  depends not only on the pre-derived optimal policy, but also on the real-time network information realization.
Thus, it must be determined in an on-line manner.
With spatial spectrum reuse, however, determining the optimal allocation of each spectrum  (in each time slot) via a VCG auction  involves solving NP-hard problems (i.e., the MWIS problems).
Hence, the corresponding optimal allocation cannot be easily determined in an on-line manner due to the high complexity of solving NP-hard problems.
This is the major reason why we introduce the approximate algorithms for the real-time spectrum allocation in each time slot.

%% file: Section-Appendix-Proofs-OR.tex

\subsection{Proof for Proposition \ref{prop:feasible}} \label{prop:feasible-proof}
 
\begin{proof}
A feasible allocation must be achievable in practice by a feasible schedule. A schedule essentially specifies the fraction of spectrums (with certain information $\theta$) to a particular set of independent SUs. 
Define the fraction of spectrums to a particular independent SUs set as the allocation probability to this independent set.

A feasible schedule implies that the total fraction of spectrums (with information $\theta$) to all independent SU sets does not exceed 100\%, which is exactly equivalent to the condition in (\ref{eq:feasible-i}). That is, for any feasible allocation or schedule, there is always an allocation strategy $\boldsymbol{A}(\theta)$ (with respect to the independent SU set) satisfying (\ref{eq:feasible-i}) and achieving the same allocation. 
\end{proof}

\subsection{Proof for Proposition \ref{prop:restriction}} \label{prop:restriction-proof}
 
\begin{proof}
We prove the proposition by contradiction. Suppose ${a}_k^{\mathrm{Ind*}} (\theta) > 0$ for certain independent set $I_k =I_k^{\mathrm{c}} \bigcup I_k^{\mathrm{s}}$, where $I_k^{\mathrm{s}}$ is not an MWIS of the side market $G^{\mathrm{s}}_{I_k^{\mathrm{c}}}$. Then the PO can improve his expected profit by transferring the spectrums for independent set $I_k =I_k^{\mathrm{c}} \bigcup I_k^{\mathrm{s}}$ to another independent set $ \{I_k^{\mathrm{c}}, I_k^{\mathrm{s*}}\} $, where $I_k^{\mathrm{s*}}$ is indeed an MWIS of the side market $G^{\mathrm{s}}_{I_k^{\mathrm{c}}}$. The reason is that the PO can achieve a higher revenue from the spot market user $I_k^{\mathrm{s*}}$ than that from set $I_k^{\mathrm{s}}$, while the revenue from the contract user set $I_k^{\mathrm{c}}$ does not change (since the allocation to contract users remains the same).  It follows that the ${a}_k^{\mathrm{Ind*}} (\theta) $ must be zero.
\end{proof}

\subsection{Proof for Lemma \ref{lemma:nc1}}\label{lemma:nc1-proof}
 
\begin{proof} We prove the lemma by contradiction. Suppose there exists any $\theta$ such that $\sum_{k=0}^K {a}_k^*(\theta) < 1$.
Obviously, by increasing ${a}_0^*(\theta)$ to ${a}_0^+(\theta)=1-\sum_{k=1}^K {a}_k^*(\theta)$, the PO can improve the expected revenue $\mathbb{E} [Z_0]$ from the spot market (which is given in (\ref{eq:EZk-1})) without affecting the revenue and cost from the contract users. This implies the PO can achieve a higher expected profit with a new strategy $\boldsymbol{A}^+(\theta)=\left( {a}_0^+(\theta),  {a}_1^*(\theta), ...,  {a}_K^*(\theta)\right)$.
\end{proof}

\subsection{Proof for Lemma \ref{lemma:nc2}}\label{lemma:nc2-proof}

\begin{proof} We prove the lemma by contradiction. Suppose there exists a contract user $n$ with $\mathbb{E}[d_n] = S\cdot \int_{\theta}  a_n (\theta) \cdot f_{\boldsymbol{\Theta}}(\theta) \mathrm{d} \theta > D_n$, where $a_n (\theta)  =\sum_{k=0}^K l_k^n\cdot a_k(\theta)$.
Let $\mathcal{K}_n$ denote the set of independent sets containing contract user $n$, i.e., $l_k^n=1, \forall k \in \mathcal{K}_n$. Let $\mathcal{K}_{n/\{n\}} $ denote   the set of independent sets each corresponding to one element in $\mathcal{K}_n$ by removing contract user $n$.\footnote{For example, in Figure \ref{system-illu}, we have $\mathcal{K}_1=\{\{1\},\{1,4\}\}$  and $\mathcal{K}_{1/\{1\}} =\{\emptyset, \{4\}\}$ for contract user 1, and $\mathcal{K}_4=\{\{4\},\{1,4\},\{2,4\}\}$ and $\mathcal{K}_{4/\{4\}} =\{\emptyset, \{1\}, \{2\}\}$ for contract user 4.}

Let $\boldsymbol{A}^+(\theta) = \left( {a}_0^+(\theta),  {a}_1^+(\theta), ..., {a}_K^+(\theta)\right)$ denote a new allocation strategy, by transferring certain spectrum for each independent set in $\mathcal{K}_n$ to the corresponding set in $\mathcal{K}_{n/\{n\}} $, i.e., (i) ${a}_k^+(\theta)= {a}_k^*(\theta)-\sigma, \forall k \in \mathcal{K}_n$, and ${a}_k^+(\theta)= {a}_k^*(\theta)+\sigma, \forall k \in \mathcal{K}_{n/\{n\}}$, and (ii) ${a}_k^+(\theta)= {a}_k^*(\theta) , \forall k \notin \mathcal{K}_{n }\bigcup \mathcal{K}_{n/\{n\}}$.
It is easy to see that with  $\boldsymbol{A}^+(\theta)$, the allocation probability to contract user $n$ decreases to $a_n^+ (\theta) = a_n (\theta) -  \sigma \cdot |\mathcal{K}_n|$, and thus
the expected number of spectrums for contract user $n$ decreases to $\mathbb{E}[d_n^+] = \mathbb{E}[d_n] - \sigma \cdot S\cdot |\mathcal{K}_n|$.  As long as $0< \sigma \leq \frac{\mathbb{E}[d_n]-D_n}{S\cdot |\mathcal{K}_n|}$, we have
$\mathbb{E}[d_n^+]= \mathbb{E}[d_n]-\sigma \cdot S\cdot |\mathcal{K}_n| \geq D_n$, i.e., the contract $n$ is still fulfilled. This implies that the PO suffers no short-term revenue loss from   contract user $n$.
Besides, the allocation probabilities to all other contract users remain the same, and thus the PO suffers no short-term revenue loss from other  contract users. On the other hand, the PO can achieve a higher expected side revenue from the side markets of the independent sets in $\mathcal{K}_{n/\{n\}} $ than those in $\mathcal{K}_{n} $.
It follows that the PO can achieve higher expected profit with new strategy $\boldsymbol{A}^+(\theta)$.    \end{proof}

\subsection{Proof for Corollary \ref{lemma:nc2-cor}}\label{lemma:nc2-cor-proof}
 
\begin{proof}
By Lemma \ref{lemma:nc2}, we have $\mathbb{E}[d_n] \leq D_n, \forall n \in \mathcal{N}$. Thus, we only need to show that there is no contract user $n$ with $0<\mathbb{E}[d_n] < D_n$. We prove the corollary by contradiction. Suppose there exists a contract user $n$ with $0<\mathbb{E}[d_n] < D_n$. According to the definitions of hard contract, the expected revenue achieved from contract user $n$ is $\mathbb{E}[R_n] = B_n  - \widehat{B}_n $. 
Let us define $\mathcal{K}_n$ and $\mathcal{K}_{n/\{n\}} $ in a same way as those in Appendix-\ref{lemma:nc2-proof}.

Let $\boldsymbol{A}^+(\theta) = \left( {a}_0^+(\theta),  {a}_1^+(\theta), ..., {a}_K^+(\theta)\right)$ denote a new allocation strategy, by transferring all spectrums for each independent set in $\mathcal{K}_n$ to the corresponding set in $\mathcal{K}_{n/\{n\}} $, i.e., (i) ${a}_k^+(\theta)= 0, \forall k \in \mathcal{K}_n$, and ${a}_k^+(\theta)= {a}_k^*(\theta)+{a}_{k'}^*(\theta), \forall k \in \mathcal{K}_{n/\{n\}}$, where $k'$ is the corresponding independent set in $\mathcal{K}_n$, and (ii) ${a}_k^+(\theta)= {a}_k^*(\theta) , \forall k \notin \mathcal{K}_{n }\bigcup \mathcal{K}_{n/\{n\}}$.
It is easy to see that with  $\boldsymbol{A}^+(\theta)$, the allocation probability to contract user $n$ decreases to $a_n^+ (\theta) = 0$, while the probabilities to all other contract users remain the same.
Obviously, the PO achieves the same expected revenue from all contract users. On the other hand, the PO can achieve a higher expected side revenue from the side markets of the independent sets in $\mathcal{K}_{n/\{n\}} $ than those in $\mathcal{K}_{n} $.
Thus, the PO can achieve higher expected profit with new strategy $\boldsymbol{A}^+(\theta)$. 
  \end{proof}

\subsection{Proof for Lemma \ref{lemma:mu-multi}}\label{lemma:mu-multi-proof}

\begin{proof}  We prove the lemma by \emph{duality principle} in particular the dual constraint (D.1), i.e., $ \mu_k^*(\theta)\geq 0$, $a_k^*(\theta)\geq 0 $ and $ \mu_k^*(\theta) \cdot a_k^*(\theta) = 0, \forall k\in \mathcal{K}, \theta \in \boldsymbol{\Theta}$. According to the Definitions \ref{def:mp} and \ref{def:mp1}, we have: $\mathcal{L}^{(k)}(\theta) = \mathcal{J}_1^{(k)}( \theta)   + \mu_k^*(\theta), \forall k$.

Proof for Condition (a). Suppose $\mu_k^*(\theta)>0$ by contradiction. Then we have: $\mathcal{L}^{(k)}(\theta) = \mathcal{J}_1^{(k)}( \theta) + \mu_k^*(\theta) > 0$, which implies $a_k^*(\theta) = 1$ by the Euler-Lagrange conditions for optimality or (\ref{eq:obj-incomp2-soft-opt}). Obviously, this violates the   constraint  $ \mu_n^*(\theta)  \cdot  a_n^*(\theta) = 0$.

Proof for Condition (b). Suppose $\mu_k^*(\theta)>|\mathcal{J}_1^{(k)}( \theta)| > 0$ by contradiction. Then we have: $\mathcal{L}^{(k)}(\theta) = \mathcal{J}_1^{(k)}( \theta) + \mu_k^*(\theta) > 0$, which implies $a_k^*(\theta) = 1$. Obviously, this violates the   constraint  $ \mu_k^*(\theta)  \cdot  a_k^*(\theta) = 0$. Thus, we have: $\mu_k^*(\theta)\in[0,\ |\mathcal{J}_1^{(k)}( \theta)|]$, since $\mu_k^*(\theta)\geq 0$.
\end{proof}

\subsection{Proof for Lemma \ref{lemma:eta-multi}}\label{lemma:eta-multi-proof}
 
\begin{proof}
We prove the lemma by \emph{duality principle} in particular the dual constraint (D.2), i.e., $ \eta^*(\theta)\geq 0$, $ \sum_{k=1}^K a_k^*(\theta)\leq 1$ and $ \eta^*(\theta) \cdot \big(1 - \sum_{k=1}^K a_k^*(\theta) \big)=0, \forall \theta\in \boldsymbol{\Theta}$. According to the Definitions \ref{def:mp1} and  \ref{def:mp2}, we have: $\mathcal{J}_1^{(k)}( \theta) = \mathcal{J}_2^{(k)}( \theta)  - \eta^*(\theta) $, $\forall k$.

Proof for Condition (a). We first prove that $\eta^*(\theta) \leq K_1( \theta)\triangleq \max_{k\in \mathcal{K}}\mathcal{J}_2^{(k)}( \theta)$. Suppose $\eta^*(\theta)>K_1( \theta)$ by contradiction. Then we have: $\mathcal{J}_1^{(k)}( \theta) =\mathcal{J}_2^{(k)}( \theta) - \eta^*(\theta)<0, \forall k $. Together with Lemma \ref{lemma:mu-multi}, we further have: $\mathcal{L}^{(k)}( \theta)<0, \forall k$, and thus $a_k^*(\theta)=0,\forall k$, which implies $\sum_{k=1}^{K} a_k^*(\theta)=0$.
Obviously, this violates the constraint $\eta^*(\theta)\cdot (1-\sum_{k=1}^{K} a_k^*(\theta))=0$.

We then prove that $\eta^*(\theta) \geq \max(0,K_2(\theta))$. The result is obvious if $K_2(\theta)\leq 0$ since $\eta^*(\theta)\geq 0$. Thus, we focus on the case of $K_2(\theta)>0$, where there exist at least two independent sets with positive core marginal profit $\mathcal{J}_2^{(k)}( \theta)$. Let $k_1 $ and $k_2 $ denote the independent sets with the highest and second-highest core marginal profits, respectively. That is, $\mathcal{J}_2^{(k_1)}( \theta) = K_1(\theta)>0$ and $\mathcal{J}_2^{(k_2)}( \theta) = K_2(\theta)>0$.
Suppose $ \eta^*(\theta)<K_2( \theta)$ by contradiction. Then there exist at least two independent sets, i.e., $k_1 $ and $k_2 $, with positive mental marginal profit, i.e., $\mathcal{J}_1^{(k_1)}( \theta)>0$ and $\mathcal{J}_1^{(k_2)}( \theta)>0$. Together with Lemma \ref{lemma:mu-multi}, we further have: $\mathcal{L}^{(k_1)}( \theta)>0$ and $\mathcal{L}^{(k_2)}( \theta)>0$, and thus $a_{k_1}^*(\theta)=1$ and $a_{k_2}^*(\theta)=1$, which implies $\sum_{k=1}^{K} a^*_k(\theta)>1$. This obviously violates the   constraint $\sum_{k=1}^{K} a^*_k(\theta)\leq 1$.

Proof for Condition (b). Suppose $\eta^*(\theta)>0$ by contradiction. Then we have: $\mathcal{J}_1^{(k)}(\theta)=\mathcal{J}_2^{(k)}(\theta)-\eta^*(\theta) < 0, \forall k$. Together with Lemma \ref{lemma:mu-multi}, we further have:  $\mathcal{L}^{(k)}( \theta)<0, \forall k$, and thus $a_k^*(\theta)=0,\forall k$, which implies $\sum_{k=1}^{K} a_k^*(\theta)=0$. Obviously, this violates the   constraint $\eta^*(\theta)\cdot (1-\sum_{k=1}^{K} a_k^*(\theta))=0$.
 \end{proof}

\subsection{Proof for Lemma \ref{lemma:opt-multi}}\label{lemma:opt-multi-proof}

\begin{proof}
We prove the lemma by the Euler-Lagrange conditions for optimality, i.e., (\ref{eq:obj-incomp2-soft-opt}).

We first prove $a_{k}^*(\theta)=0$ for all independent sets $k\notin \boldsymbol{k}^*$. By contradiction, we suppose there exists an independent set $k_0 \notin \boldsymbol{k}^*$ with $a_{k_0}^*(\theta)>0$. According to (\ref{eq:obj-incomp2-soft-opt}), we have:
$\mathcal{L}^{(k_0)}(\theta)\geq 0$. 
By Lemmas \ref{lemma:mu-multi} and \ref{lemma:eta-multi}, we further have: $\mathcal{J}^{(k_0)}_2(\theta)\geq 0$.
Further, according to the definition of $\boldsymbol{k}^*$, we have: $\mathcal{J}_2^{(k_0)}(\theta) < K_1(\theta) $, since $k_0 \notin \boldsymbol{k}^*$. This implies $\mathcal{J}^{(k)}_2(\theta) = K_1(\theta) > 0, \forall k \in \boldsymbol{k}^*$. By Lemmas \ref{lemma:mu-multi} and \ref{lemma:eta-multi}, we further have: $\mathcal{L}^{(k)}(\theta)>0$, and thus $a_{k}^*(\theta) =1, \forall k \in \boldsymbol{k}^*$, which implies that $\sum_{k=1}^K a_{k}^*(\theta) = a_{k_0}^*(\theta) + \sum_{k\in\boldsymbol{k}^*} a_{k}^*(\theta) + \sum_{k\notin \boldsymbol{k}^*\cup \{k_0\}} a_{k}^*(\theta) > 1$. This obviously violates the constraint $\sum_{k=1}^K a_{k}^*(\theta)  \leq 1$.

We then prove the optimal solution for $k\in \boldsymbol{k}^*$. Obviously, $\mathcal{J}^{(k)}_2(\theta) = K_1(\theta), \forall k\in \boldsymbol{k}^*$.

Proof for case (a). By Lemma \ref{lemma:eta-multi}, we have: $\eta^*(\theta) \leq K_1(\theta)$ if $K_1(\theta)>0$, which implies $\mathcal{J}^{(k)}_1(\theta) = \mathcal{J}^{(k)}_2(\theta) - \eta^*(\theta) = K_1(\theta) - \eta^*(\theta)\geq 0$, $\forall k\in \boldsymbol{k}^*$. By Lemma \ref{lemma:mu-multi}, we further have: $\mu_k^*(\theta)  = 0 $ if $\mathcal{J}^{(k)}_1(\theta) \geq 0$, which further implies $\mathcal{L}^{(k)}(\theta) = \mathcal{J}^{(k)}_1(\theta) - 0 \geq 0$, $\forall k\in \boldsymbol{k}^*$.

{Case} (a.1). Suppose $\eta^*(\theta) = K_1(\theta)>0$. By constraint $\eta^*(\theta)\cdot (1-\sum_{k=1}^{K} a^*_k(\theta))=0$, we have: $\sum_{k=1}^{K} a^*_k(\theta) = 1$, which implies $\sum_{k\in \boldsymbol{k}^*} a^*_k(\theta) = 1$, since $a_{k}^*(\theta)=0$, $\forall k\notin \boldsymbol{k}^*$.

{Case} (a.2). Suppose $\eta^*(\theta) < K_1(\theta) $. Then we must have: $K_2(\theta) < K_1(\theta)$, otherwise $\eta^*(\theta)$ must equal to $K_1(\theta)$ since $\eta^*(\theta) \in [\max(0, K_2(\theta)), K_1(\theta)]$. This implies that there is only one independent set, denoted by $ k_1 $, in the set $\boldsymbol{k}^*$.
Obviously, $\mathcal{L}^{(k_1)}(\theta) =K_1(\theta) -\eta^*(\theta) > 0$. According to (\ref{eq:obj-incomp2-soft-opt}), we have: $a^*_{k_1}(\theta) = 1$, which implies $\sum_{k\in \boldsymbol{k}^*} a^*_k(\theta) = 1$, since $\boldsymbol{k}^* = \{k_1\}$.

Proof for case (b). By Lemma \ref{lemma:eta-multi}, we have: $\eta^*(\theta) =0$ if $K_1(\theta)=0$, which implies $\mathcal{J}^{(k)}_1(\theta) = \mathcal{J}^{(k)}_2(\theta) - 0 = K_1(\theta)= 0$, $\forall k\in \boldsymbol{k}^*$. By Lemma \ref{lemma:mu-multi}, we further have: $\mu_k^*(\theta) = 0 $ if $\mathcal{J}^{(k)}_1(\theta) = 0$, which further implies $\mathcal{L}^{(k)}(\theta) = \mathcal{J}^{(k)}_1(\theta) - 0 = 0$, $\forall k\in \boldsymbol{k}^*$. According to the dual constraint (D.2), we have: $\sum_{k\in \boldsymbol{k}^*} a^*_k(\theta) \leq 1$.

Proof for case (c). By Lemma \ref{lemma:eta-multi}, we have: $\eta^*(\theta) =0$ if $K_1(\theta)<0$, which implies $\mathcal{J}^{(k)}_1(\theta) = \mathcal{J}^{(k)}_2(\theta) -0 = K_1(\theta)< 0$, $\forall k\in \boldsymbol{k}^*$. By Lemma \ref{lemma:mu-multi}, we further have: $\mu_k^*(\theta) \leq |\mathcal{J}^{(k)}_1(\theta)|$ if $\mathcal{J}^{(k)}_1(\theta) < 0$, which further implies $\mathcal{L}^{(k)}(\theta) = \mathcal{J}^{(k)}_1(\theta) + \mu_k^*(\theta) = K_1(\theta)+ \mu_k^*(\theta)   \leq 0$, $\forall k\in \boldsymbol{k}^*$.

{Case} (c.1). Suppose $\mu_k^*(\theta) = |\mathcal{J}^{(k)}_1(\theta)| >0$. By constraint $ \mu_k^*(\theta) \cdot a_k^*(\theta) = 0$, we have: $a_k^*(\theta) = 0$, $\forall k\in \boldsymbol{k}^*$.

{Case} (c.2). Suppose $\mu_k^*(\theta) < |\mathcal{J}^{(k)}_1(\theta)|$. Then we have: $ \mathcal{L}^{(k)}(\theta)  < 0$, $\forall k\in \boldsymbol{k}^*$. According to  (\ref{eq:obj-incomp2-soft-opt}), we have: $a^*_{k}(\theta) = 0$, $\forall k\in \boldsymbol{k}^*$.
 \end{proof}

\subsection{Proof for Lemma \ref{lemma:lambda-multi}}\label{lemma:lambda-multi-proof}

\begin{proof} We first note that $\mathbb{E}[d_n] \triangleq S \int_{\theta\in \boldsymbol{\Theta}_n^+(\boldsymbol{\Lambda}^*)} f_{\boldsymbol{\Theta}}(\theta) \mathrm{d} \theta$ decreases with the contract user $n$'s own shadow price $\lambda_n $, while increases with   other contract users' shadow prices $\lambda_{n'}, \forall n'\neq n$.
For convenience, we write $\mathbb{E}[d_n]$ as a function of $\lambda_n$, denoted by $\mathbb{E}[d_n( \lambda_n)] $.\footnote{Note that $\mathbb{E}[d_n( \lambda_n)] $ also depends on the shadow prices $\boldsymbol{\Lambda}_{-n}$ of other contract users.}
Let $\lambda_n^o $ denote the shadow price $\lambda_n$ such that $\mathbb{E}[d_n( \lambda_n)]= D_n$, i.e., $\lambda_n^o \triangleq \arg_{\lambda_n } (\mathbb{E}[d_n( \lambda_n)]= D_n)$. Thus, we can rewrite the optimal shadow price $\lambda_n^*$ in Lemma \ref{lemma:lambda-multi} as $ \lambda_n^* = \max
\big\{0, \lambda_n^o\},\forall n$.
We prove the lemma by \emph{duality principle} in particular the dual constraint (D.3), i.e., $\lambda_n^*\geq 0$,  $D_n - \mathbb{E}[d_n( \lambda_n^*)] \geq 0$ and  $\lambda_n^* \cdot  \big(D_n - \mathbb{E}[d_n( \lambda_n^*)] \big)  =0$, $\forall n\in \mathcal{N}$.

Consider the first case: $\mathbb{E}[d_n(0)] \leq D_n$. It is easy to see that $\lambda_n^o   < 0$, since $\mathbb{E}[d_n( \lambda_n)] $ decreases with $\lambda_n$. Thus, we only need to prove $ \lambda_n^*=\max
\big\{0, \lambda_n^o\big\} =0$.
Suppose $\lambda_n^*>0$ by contradiction. Then we have: $\mathbb{E}[d_n(\lambda_n^*)] <\mathbb{E}[d_n(0)]  \leq D_n$. This obviously violates the   constraint $\lambda_n^* \cdot \big(D_n - \mathbb{E}[d_n( \lambda_n^*)] \big)  =0$.

Consider the second case: $\mathbb{E}[d_n(0)] > D_n$. It is easy to see that $\lambda_n^o  > 0$.
Thus, we only need to prove $ \lambda_n^* = \big\{0, \lambda_n^o\big\} = \lambda_n^o >0$. Suppose $\lambda_n^* > \lambda_n^o$ by contradiction. Then we have: $\mathbb{E}[d_n(\lambda_n^*)] < \mathbb{E}[d_n(\lambda_n^o)] = D_n$. This obviously violates the   constraint $\lambda_n^* \cdot \big(D_n - \mathbb{E}[d_n( \lambda_n^*)] \big)  =0$. Suppose $\lambda_n^* < \lambda_n^o$ by contradiction. Then we have: $\mathbb{E}[d_n(\lambda_n^*)] > \mathbb{E}[d_n(\lambda_n^o)] = D_n$. This obviously violates the constraint $  \mathbb{E}[d_n( \lambda_n^*)]\leq  D_n $.
\end{proof}